\documentclass[preprint]{aastex61}
\usepackage{CJK}

\usepackage{amsmath}
\usepackage{amssymb}
\usepackage{epsfig}
\usepackage{apjfonts}
\usepackage{natbib}
\usepackage{hyperref}
\usepackage{epstopdf}
\usepackage{verbatim}
\usepackage{enumitem}
\usepackage{color}
\setlist[enumerate]{itemsep=-1mm}
\usepackage{soul, xcolor}
\setstcolor{blue}

\usepackage{mathtools}

\usepackage{bigints}

\usepackage{changepage}
\usepackage{tikz}

\makeatletter
\global\let\tikz@ensure@dollar@catcode=\relax
\makeatother

\begin{document}


\begin{CJK*}{UTF8}{gbsn}

\title{COol Companions ON Ultrawide orbiTS (COCONUTS).\\ I. A High-Gravity T4 Benchmark around an Old White Dwarf\\ and A Re-Examination of the Surface-Gravity Dependence of the L/T Transition}

\author{Zhoujian Zhang (张周健)}
\affiliation{Institute for Astronomy, University of Hawaii at Manoa, Honolulu, HI 96822, USA}
\affiliation{Visiting Astronomer at the Infrared Telescope Facility, which is operated by the University of Hawaii under contract NNH14CK55B with the National Aeronautics and Space Administration.}

\author{Michael C.\ Liu}
\affiliation{Institute for Astronomy, University of Hawaii at Manoa, Honolulu, HI 96822, USA}

\author{J. J.\ Hermes}
\affiliation{Department of Astronomy, Boston University, Boston, MA 02215, USA }

\author{Eugene A.\ Magnier}
\affiliation{Institute for Astronomy, University of Hawaii at Manoa, Honolulu, HI 96822, USA}

\author{Mark S.\ Marley}
\affiliation{NASA Ames Research Center, Mail Stop 245-3, Moffett Field, CA 94035, USA}

\author{Pier-Emmanuel Tremblay}
\affiliation{Department of Physics, University of Warwick, Coventry CV4~7AL, UK}

\author{Michael A.\ Tucker}
\affiliation{Institute for Astronomy, University of Hawaii at Manoa, Honolulu, HI 96822, USA}

\author{Aaron Do}
\affiliation{Institute for Astronomy, University of Hawaii at Manoa, Honolulu, HI 96822, USA}

\author{Anna V.\ Payne}
\affiliation{Institute for Astronomy, University of Hawaii at Manoa, Honolulu, HI 96822, USA}

\author{Benjamin J.\ Shappee}
\affiliation{Institute for Astronomy, University of Hawaii at Manoa, Honolulu, HI 96822, USA}

\begin{abstract}
We present the first discovery from the COol Companions ON Ultrawide orbiTS (COCONUTS) program, a large-scale survey for wide-orbit planetary and substellar companions. We have discovered a co-moving system COCONUTS-1, composed of a hydrogen-dominated white dwarf (PSO~J$058.9855+45.4184$; $d=31.5$~pc) and a T4 companion (PSO~J$058.9869+45.4296$) at a $40.6^{\prime\prime}$ ($1280$~au) projected separation. We derive physical properties for COCONUTS-1B from (1) its near-infrared spectrum using cloudless Sonora atmospheric models, and (2) its luminosity and the white dwarf's age ($7.3_{-1.6}^{+2.8}$~Gyr) using Sonora evolutionary models. The two methods give consistent temperatures and radii, but atmospheric models infer a lower surface gravity and therefore an unphysically young age. Assuming evolutionary model parameters ($T_{\rm eff}=1255^{+6}_{-8}$~K, $\log\,g=5.44^{+0.02}_{-0.03}$~dex, $R=0.789^{+0.011}_{-0.005}$~R$_{\rm Jup}$), we find cloudless model atmospheres have brighter $Y$- and $J$-band fluxes than the data, suggesting condensate clouds have not fully dispersed around $1300$~K. The W2 flux ($4.6$~$\mu$m) of COCONUTS-1B is fainter than models, suggesting non-equilibrium mixing of CO. To investigate the gravity dependence of the L/T transition, we compile all 60 known L6$-$T6 benchmarks and derive a homogeneous set of temperatures, surface gravities, and masses. As is well-known, young, low-gravity late-L dwarfs have significantly fainter, redder near-infrared photometry and $\approx200-300$~K cooler temperatures than old, high-gravity objects. Our sample now reveals such gravity dependence becomes weaker for T dwarfs, with young objects having comparable near-infrared photometry and $\approx100$~K cooler temperatures compared to old objects. Finally, we find that young objects have a larger amplitude $J$-band brightening than old objects, and also brighten at $H$~band as they cross the L/T transition.
\end{abstract}

\keywords{binaries: general -- brown dwarfs -- white dwarfs}

\section{Introduction}
\label{sec:introduction}

Over the past decade, direct imaging has invigorated the field of extrasolar planetary systems by revealing a population of giant planets and brown dwarf companions with orbital distances beyond $\sim 10$~au \citep[e.g.,][]{2016PASP..128j2001B}. Among this population, companions on ultrawide orbits ($\gtrsim 500$~au) are especially appealing as they can be very well characterized. Compared to closer companions, for which spectroscopic follow-up is complicated by the contaminating light of the primary star \citep[e.g., 51~Eri~b;][]{2017A&A...603A..57S}, wide-orbit companions can in principle have their physical properties robustly estimated via direct spectroscopic analysis, with the assistance of atmospheric models \citep[e.g.,][]{2012ApJ...756..172M, 2014IAUS..299..271A}. Compared to free-floating objects in the field, whose luminosities, ages, and masses are degenerate and thus cannot be determined without independent age or mass measurements \citep[e.g.,][]{2017ApJS..231...15D}, wide-orbit companions can have their physical properties established using the evolutionary models \citep[e.g.,][]{2008ApJ...689.1327S} thanks to ages from their primary stars.

Wide-orbit companions are therefore valuable benchmarks \citep[e.g.,][]{2006MNRAS.368.1281P, 2006ApJ...647..552S, 2007ApJ...660.1507L} to test and improve current models of ultracool atmospheres, which are limited by, e.g., incomplete molecular opacity line lists, assumptions of chemical and radiative-convective equilibria, and patchy and time-evolving clouds \citep[e.g.,][]{2015ARA&A..53..279M}. In addition, novel techniques have been developed for precisely characterizing stars through their interior structures and stellar activity, e.g., asteroseismology \citep[e.g.,][]{2013ARA&A..51..353C} and gyrochronology \citep[e.g.,][]{2007ApJ...669.1167B}. Binary systems composed of stellar primaries and ultracool companions can potentially extend these techniques from the relatively well-calibrated stellar regime down to the substellar regime. This would allow us to better understand the evolution of giant planets and brown dwarfs.

Nearly 40 wide-orbit ($\gtrsim 500$~au) ultracool companions have been thus far discovered, spanning M6--Y0 spectral types with ages from $\approx 1$~Myr to $\approx 10$~Gyr \citep[see summary in][]{2014ApJ...792..119D}. However, we lack knowledge of this sample's completeness, given that some were found as by-products from large-area searches for field brown dwarfs \citep[e.g., GJ~570D;][]{2000ApJ...531L..57B} and some were found from companion searches that focused only on young moving groups \citep[e.g., GU~Psc~b;][]{2014ApJ...787....5N}. A few dedicated large-scale searches for ultracool companions have been conducted \citep[e.g.,][]{2006MNRAS.368.1281P, 2014ApJ...792..119D}, but their primary star samples were heterogeneous (e.g., not volume-limited) and they relied on 2MASS \citep{2006AJ....131.1163S}, whose sensitivity has now been greatly surpassed by Pan-STARRS1 \citep[PS1;][]{2016arXiv161205560C}, the UKIDSS surveys including UKIDSS Hemisphere Survey \citep[UHS;][]{2018MNRAS.473.5113D}, Large Area Survey (LAS), Galactic Clusters Survey (GCS), and Galactic Plane Survey \citep[GPS;][]{2007MNRAS.379.1599L}, and AllWISE \citep[][]{2010AJ....140.1868W, 2014yCat.2328....0C}.

To create a larger and more complete catalog of wide-orbit ultracool benchmarks, we are carrying out the COol Companions ON Ultrawide orbiTS (COCONUTS) program.  We are targeting a volume-limited sample of primary stars ($\approx 3\times10^{5}$ objects) with reliable distances within $100$~pc, selected from {\it Gaia}~DR2 \citep{2016AandA...595A...1G, 2018AandA...616A...1G} and the extended {\it Hipparcos} catalog \citep{2012AstL...38..331A}. Using photometry and astrometry from {\it Gaia}~DR2, PS1, 2MASS, UKIDSS, and AllWISE, we are searching for ultracool companions ($\approx 5-70$~M$_{\rm Jup}$) with separations beyond $\approx 500$~au. Though the occurrence rate of wide-orbit substellar companions (separation $>500$~au and mass $5-70$~M$_{\rm Jup}$) could be as low as $\approx$~$0.1\%$ (as extrapolated from the \citealt{2014ApJ...794..159B} companion distribution spanning orbital separations of $10-100$~au and masses of $5-70$~M$_{\rm Jup}$), our target star sample should be large enough to yield many benchmarks. Also, as done for companions on closer orbits \citep[e.g., $10-100$~au;][]{2014ApJ...794..159B, 2019arXiv190405358N}, a well-defined sample of wide-orbit companions from our survey will establish the companion distributions beyond 500~au, which can shed light on the outer architecture of extrasolar planetary systems.

Here we present the first discovery from COCONUTS: a high-gravity mid-T companion to a white dwarf. We describe the identification of the system in Section~\ref{sec:system} and the spectroscopic follow-up in Section~\ref{sec:observations}. We then analyze the physical properties of each system component in Sections~\ref{sec:WD_analysis} and \ref{sec:T_analysis}, with a summary in Section~\ref{sec:summary}.

\section{Identifying the Co-Moving White Dwarf + T dwarf Pair}
\label{sec:system}

We queried the the PS1 Processing Version 3 database using the Desktop Virtual Observatory \citep[DVO;][]{2004PASP..116..449M, 2016arXiv161205242M} for objects within a projected separation of $10^{4}$~au around each primary star in the COCONUTS sample, and then cross-matched with {\it Gaia}~DR2, 2MASS, UKIDSS, and AllWISE to obtain the objects' multi-wavelength photometry and multi-epoch astrometry. We then apply color and magnitude cuts based on the photometry of ultracool dwarfs from \cite{2018ApJS..234....1B} and identify co-moving candidates if their proper motions are consistent with their primary stars. Finally, we remove previously known objects by cross-matching our candidate list with SIMBAD\footnote{http://simbad.u-strasbg.fr/simbad/} and existing catalogs of ultracool dwarfs \citep[e.g.,][]{2014ApJ...792..119D, 2018ApJS..234....1B}, and remove extended sources and detector artifacts by visually checking PS1 images. 

\begin{figure*}[h]
\begin{center}
\includegraphics[height=6in]{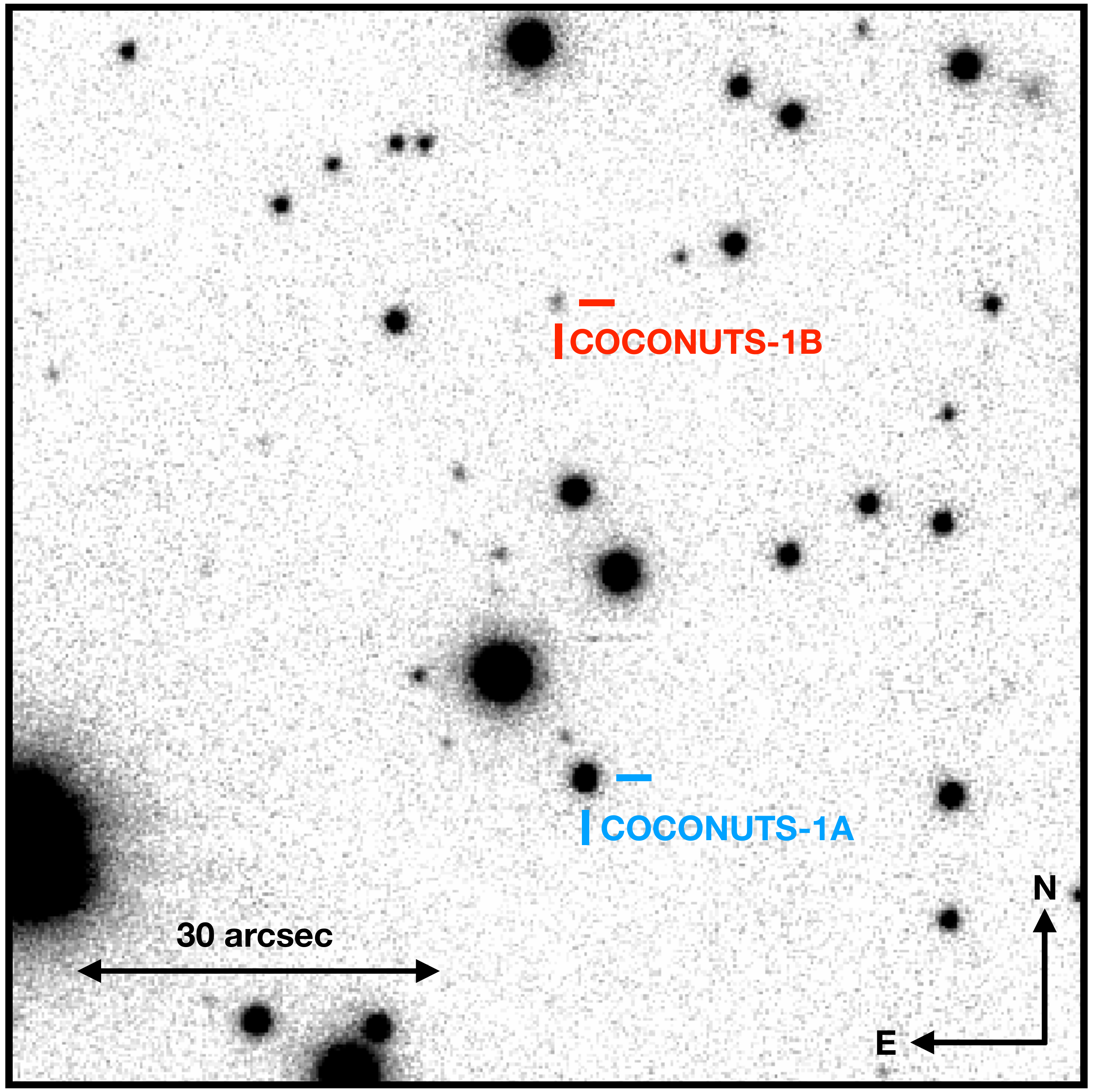}
\caption{The white dwarf primary COCONUTS-1A and the T dwarf companion COCONUTS-1B in the Pan-STARRS1 $y_{\rm P1}$-band image (size of 90~arcsec on each side). The two object are separated by $40.61 \pm 0.04$~arcsec, which corresponds to $1280 \pm 4$~au at the primary's distance.}
\label{fig:finder}
\end{center}
\end{figure*}

The above process yielded our first discovery COCONUTS-1, a T dwarf companion PSO~J058.9869+45.4296 to a white dwarf PSO~J058.9855+45.4184. We hereafter note the primary star as COCONUTS-1A and the companion as COCONUTS-1B. The {\it Gaia}~DR2 astrometric solutions of the primary are reliable given that they are derived from 14 independent {\it Gaia} observations (i.e., visibility\_periods\_used$ = 14$), and are consistent with the {\it Gaia}'s single-star model given that its renormalized unit weight error (RUWE $= 1.05$) is smaller than 1.4 \citep[as suggested by][]{Lindegren2018}. The primary has a distance of $31.51 \pm 0.09$~pc, derived from its {\it Gaia}~DR2 astrometry in a Bayesian fashion by \cite{2018AJ....156...58B}. In addition, we note the primary is a white dwarf, primarily based on its {\it Gaia}~DR2 photometry ($G$-band absolute magnitude of $14.681 \pm 0.007$~mag and $G_{\rm BP} - G_{\rm RP} = 1.085 \pm 0.010$~mag) and its optical and near-infrared spectra (Section~\ref{subsec:WD_obs}). The A and B components have an angular separation of $40.61 \pm 0.04$~arcsec from their PS1 coordinates, which is then converted into a projected physical separation of $= 1280 \pm 4$~au given the primary star's distance. We show the system in Figure~\ref{fig:finder}, and summarize its astrophysical properties in Table~\ref{tab:info}. 

\begin{figure*}[h]
\begin{center}
\includegraphics[height=6.5in]{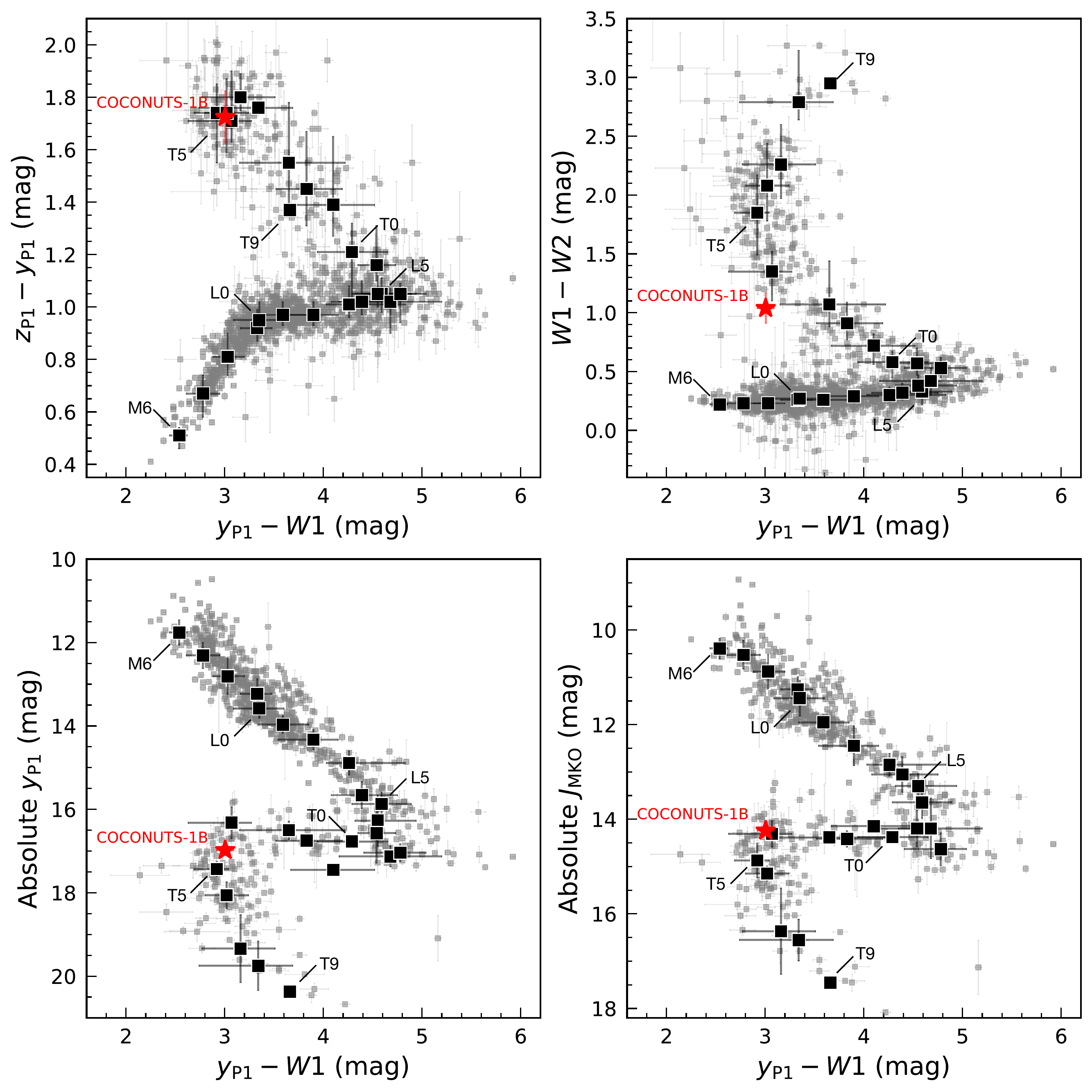}
\caption{Color-color/magnitude diagrams for the companion COCONUTS-1B (red). We use grey squares to show known field M6--T9 dwarfs from \cite{2018ApJS..234....1B} and Best et al. (submitted), which have infrared absolute magnitudes with S/N$>5$ and are not young, binaries, or subdwarfs. We use black squares and black error bars to show the typical photometry and $1\sigma$ confidence limits of these field dwarfs in each spectral type bin. The $y_{\rm P1}$ and $J_{\rm MKO}$ absolute magnitudes of COCONUTS-1B are computed by assuming its primary star's {\it Gaia} DR2 parallax. The companion's colors and magnitudes both suggest a mid-T dwarf located at the distance of the white dwarf primary.}
\label{fig:T_CCDCMD}
\end{center}
\end{figure*}

\begin{figure*}[t]
\begin{center}
\includegraphics[height=3.5in]{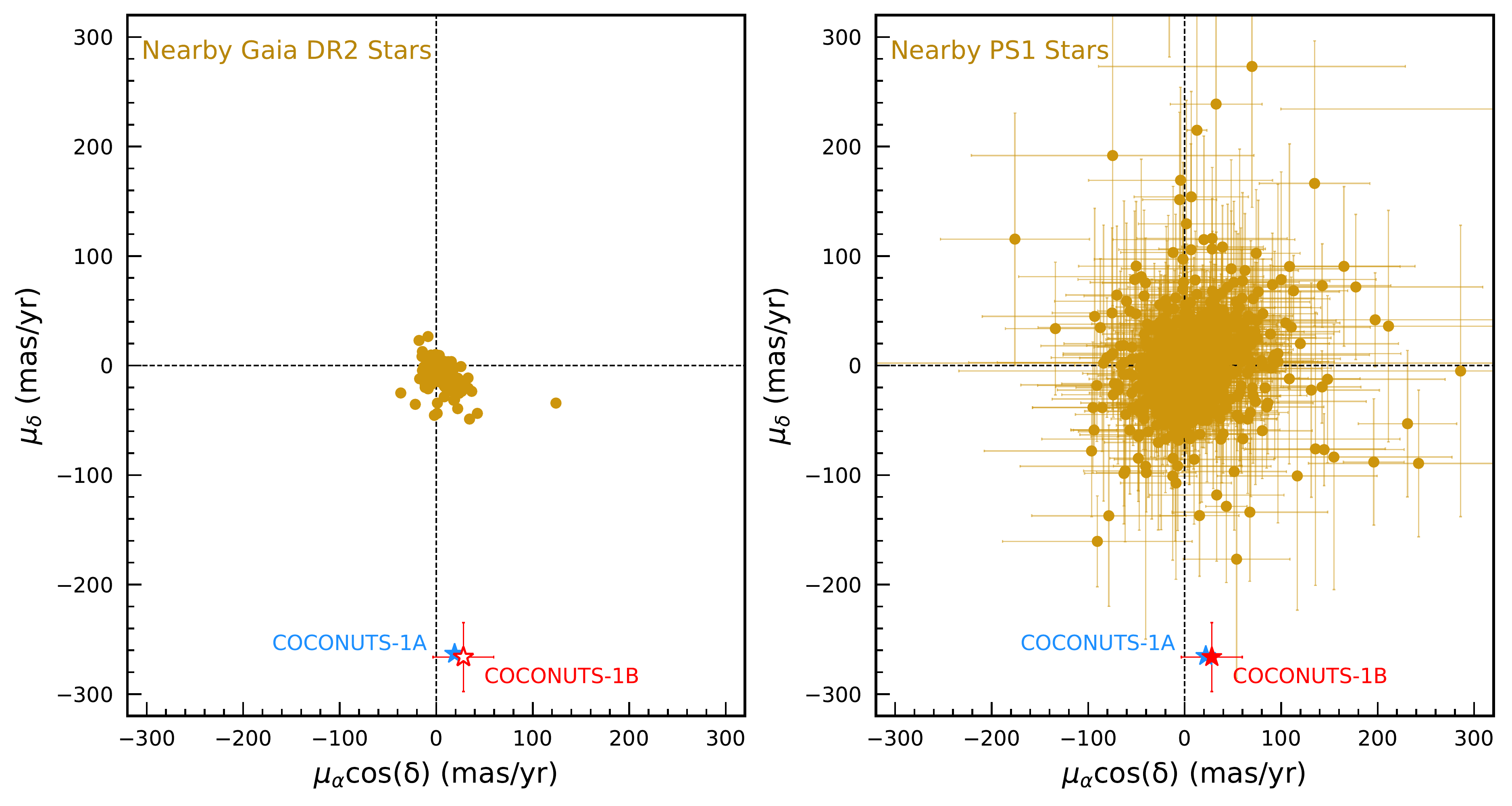}
\caption{{\it Left:} {\it Gaia}~DR2 proper motions of the primary star (blue) and stars within a radius of $10$~arcmin (brown), compared with the PS1 proper motion of the companion (since it has no {\it Gaia}~DR2 detection) shown as a red open symbol. The nearby {\it Gaia}~DR2 stars are all background with parallaxes smaller than the primary by at least $14$~mas. {\it Right:} PS1 proper motions of the primary, the companion, and stars within a radius of $10$~arcmin. These two diagrams validate the association between the primary and the companion, as they both have significant southward motion as opposed to the other stars in the field.}
\label{fig:common_pm}
\end{center}
\end{figure*}

The companion's colors and absolute magnitudes (assuming the white dwarf's {\it Gaia}~DR2 distance) are consistent with a mid-T dwarf located at the distance of the white dwarf (Figure~\ref{fig:T_CCDCMD}). In addition, the PS1 proper motion\footnote{\cite{2016arXiv161205242M} computed the positions, parallaxes, and proper motions of PS1 objects using iteratively re-weighted least squares fitting with outlier clipping, and then tied the objects' astrometry to {\it Gaia} DR1.} of the companion, $(\mu_{\alpha} {\rm cos}\delta, \mu_{\delta}) = (28.2 \pm 31.5, -266.2 \pm 31.4)$~mas/yr, is consistent with the primary, which also shows a significant southward motion of $(21.9 \pm 2.5, -265.2 \pm 1.2)$~mas/yr based on PS1 and $(19.1 \pm 0.2, -263.2 \pm 0.1)$~mas/yr based on {\it Gaia}~DR2. Their common proper motions are vastly different from the motions of the other stars in the neighborhood (Figure~\ref{fig:common_pm}), validating the association between the primary and the companion. 

\section{Observations}
\label{sec:observations}

\subsection{Spectroscopy of the Primary: UH~2.2m/SNIFS and IRTF/SpeX}
\label{subsec:WD_obs}

We obtained an optical spectrum of the white dwarf primary COCONUTS-1A using the SuperNova Integral Field Spectrograph \citep[SNIFS;][]{2002SPIE.4836...61A, 2004SPIE.5249..146L} mounted on the University of Hawaii 2.2~m telescope on Maunakea. SNIFS is a $6'' \times 6''$ Integral Field Unit (IFU) with a moderate spectral resolution of $R \approx 1200$, providing simultaneous coverage from $3300$~\AA\ to $9700$~\AA. Our observations were conducted on two consecutive nights (2018 December 11--12 UT), with total exposure times of $2700$~seconds and $1800$~seconds, respectively. The SNIFS data reduction followed the pipeline as described in \cite{2001MNRAS.326...23B}, which extracted the object's one-dimensional (1D) spectrum incorporated with dark, bias, and flat-field corrections, wavelength calibration, and sky subtraction. The dispersion in wavelength calibration is $\approx 0.38$~\AA\ for wavelength $<4700$~\AA\ and $\approx 2.11$~\AA\ for wavelength $>5300$~\AA. We flux-calibrated the 1D spectrum from each night using spectrophotometric standard stars observed in the same night (GD~71 and GD~153 for December 11; Feige~34 and HD~93521 for December 12). We combined both nights with a weighted average. The reduced SNIFS spectrum is in air wavelength and has signal-to-noise ratio (S/N) $\approx 20$ per pixel at $6000$~\AA.

We also obtained a near-infrared spectrum of COCONUTS-1A using the NASA Infrared Telescope Facility (IRTF) on 2018 October 23 UT with clear skies and $\approx 0.9''$ seeing. We used the facility spectrograph SpeX \citep{2003PASP..115..362R} in prism mode with the $0.8'' \times 15''$ slit ($R \approx 50-160$) and the wavelength coverage of $0.7-2.52$~$\mu$m and took 10 exposures with $60$~seconds each in a standard ABBA pattern. We observed the A0V standard star HD~23452 within $20$~minutes and $0.05$~airmass of the primary star for telluric correction. We reduced the data using version 4.1 of the Spextool software package \citep{2004PASP..116..362C}. The reduced SpeX spectrum is in vacuum wavelength and has S/N $\approx 40$ per pixel in $J$ band.

\subsection{Spectroscopy of the Companion: IRTF/SpeX}
\label{subsec:T_obs}
The companion COCONUTS-1B was observed using IRTF/SpeX in prism mode on 2018 October 22 (UT), with clear skies and $\approx 0.8''$ seeing. We used the $0.8'' \times 15''$ slit and took 18 exposures with $120$~seconds each in a standard ABBA pattern. The A0~V standard star HD~21038 was observed within $1$~hour and $0.1$~airmass of the companion. The dispersion in wavelength calibration of the prism data is $5.9$~\AA\ (M. Cushing, private communication). The reduced SpeX spectrum is in vacuum wavelength and has S/N $\approx 40$ per pixel in $J$ band.

\section{The White Dwarf Primary}
\label{sec:WD_analysis}

\begin{figure*}[t]
\begin{center}
\includegraphics[height=4.5in]{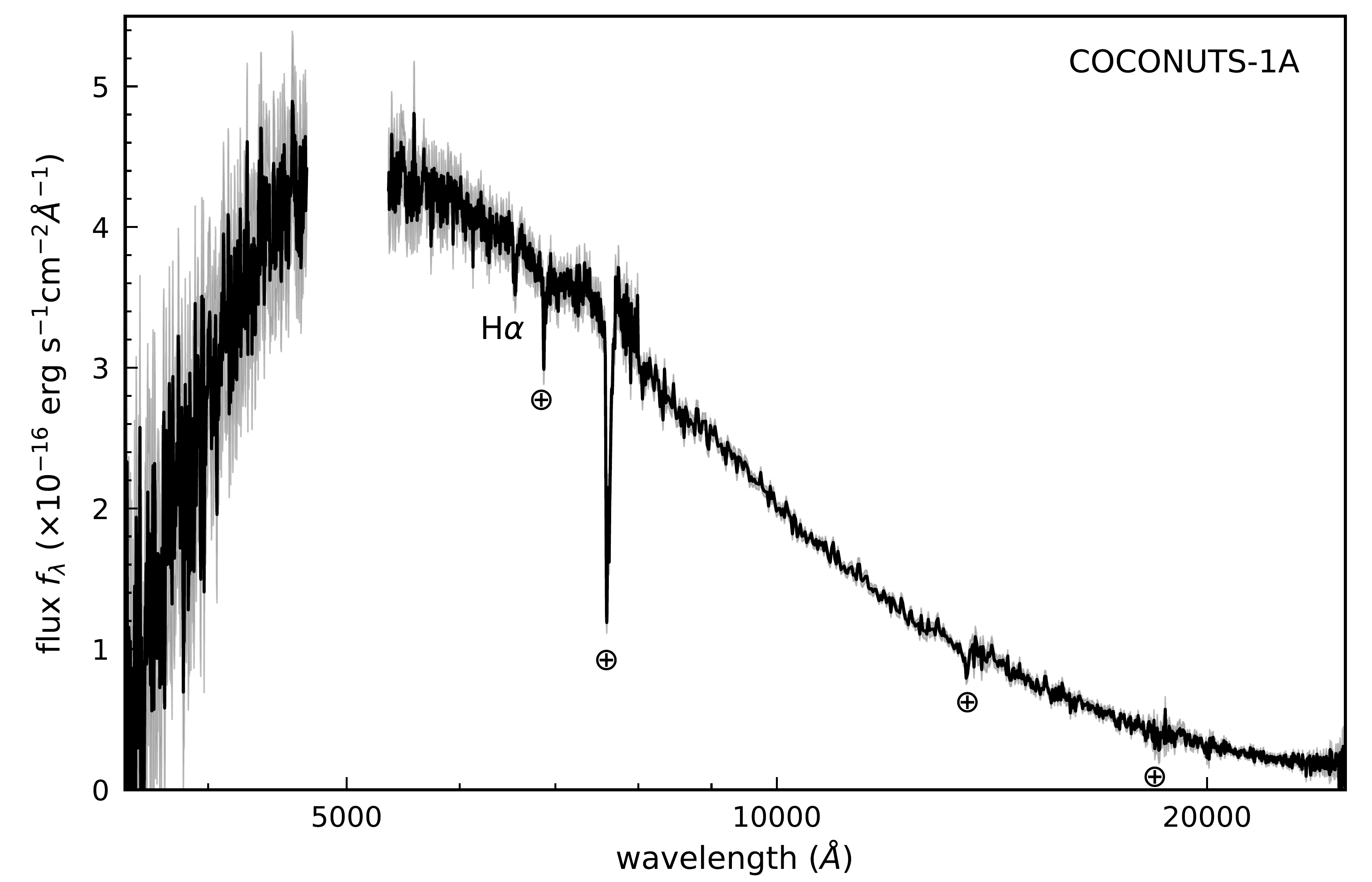}
\caption{The flux-calibrated UH~2.2m/SNIFS and IRTF/SpeX spectra of the white dwarf primary COCONUTS-1A. Flux uncertainties are shown as the grey shadow and prominent telluric absorption features are noted by ``$\oplus$''. Here we convert the wavelength of the white dwarf's SpeX spectrum from vacuum to air based on \cite{1996ApOpt..35.1566C} to match the wavelength of the SNIFS spectrum, as well as the \cite{2011ApJ...730..128T} model spectrum shown in Figure~\ref{fig:WD_bestfit}. The white dwarf is mostly featureless except for the H$\alpha$ line, suggesting COCONUTS-1A is hydrogen-dominated (DA). }
\label{fig:WD_spec}
\end{center}
\end{figure*}

We present the spectrum of the white dwarf primary in Figure~\ref{fig:WD_spec}. The white dwarf is mostly featureless except for the H${\alpha}$ line, suggesting COCONUTS-1A is a hydrogen-dominated (DA) white dwarf. The UH~2.2m/SNIFS spectrum has a relatively low S/N of $\approx 10$ at $\leqslant 4800$~\AA, lacks strong Balmer lines, and has a wavelength gap at $4800-5300$~\AA. We therefore do not derive physical properties from a spectroscopic analysis \citep[e.g.,][]{1992ApJ...394..228B, 1995ApJ...449..258B, 2005ApJS..156...47L}. As a robust alternative \citep[e.g.,][]{2017ApJ...848...11B, 2018MNRAS.480.3942H}, we conduct a photometric analysis by fitting its spectral energy distribution (SED) from optical to near-infrared wavelengths \citep[e.g.,][]{1997ApJS..108..339B, 2012ApJS..199...29G}. We have only used the observed spectrum to confirm our SED-based model fits.

We follow \cite{1997ApJS..108..339B} to perform our SED analysis \citep[also see][]{2019MNRAS.482.4570G}. We first construct the SED of the white dwarf using photometry from {\em Gaia}~DR2, PS1, and the UKIDSS Hemisphere Survey (Table~\ref{tab:info}), and convert magnitudes into fluxes using zero-point fluxes from \cite{2018AandA...616A...4E}, \cite{2012ApJ...750...99T}, and \cite{2007MNRAS.379.1599L}, respectively. We do not include 2MASS photometry, whose uncertainties are too large to constrain models. We assume reddening is negligible given the system's distance ($31.51$~pc). We then fit the white dwarf's SED using hydrogen-dominated atmospheric models from \cite{2011ApJ...730..128T}, which span $1.5 \times 10^{3} - 1.4 \times 10^{5}$~K in effective temperature $T_{\rm eff}$ and $6.0 - 10.0$~dex in logarithmic surface gravity $\log\ g$. For a given choice of $\{T_{\rm eff}, \log\ g\}$, we obtain a model spectrum by linearly interpolating the \cite{2011ApJ...730..128T} models, and then synthesize photometry by convolving this model spectrum with the corresponding filter response curves\footnote{The filter response in $J_{\rm MKO}$ band from the UKIDSS Hemisphere Survey is provided by \cite{2006MNRAS.367..454H}, and the filters of the other bands are from the same aforementioned references that provide the zero-point fluxes.}, scaled by the {\em Gaia}~DR2 distance and white dwarf radius \citep[i.e., Equation~2--3 of][]{1997ApJS..108..339B}. The radius is not a free parameter but rather derived from the $\{T_{\rm eff}, \log\ g\}$ values using the \cite{2001PASP..113..409F} evolutionary sequences, which assume C/O-cores (where carbon and oxygen are mixed uniformly with equal mass) and thick hydrogen layers (where the hydrogen envelope constitutes $10^{-4}$ of the total white dwarf mass).

\begin{figure*}[t]
\begin{center}
\includegraphics[height=5.5in]{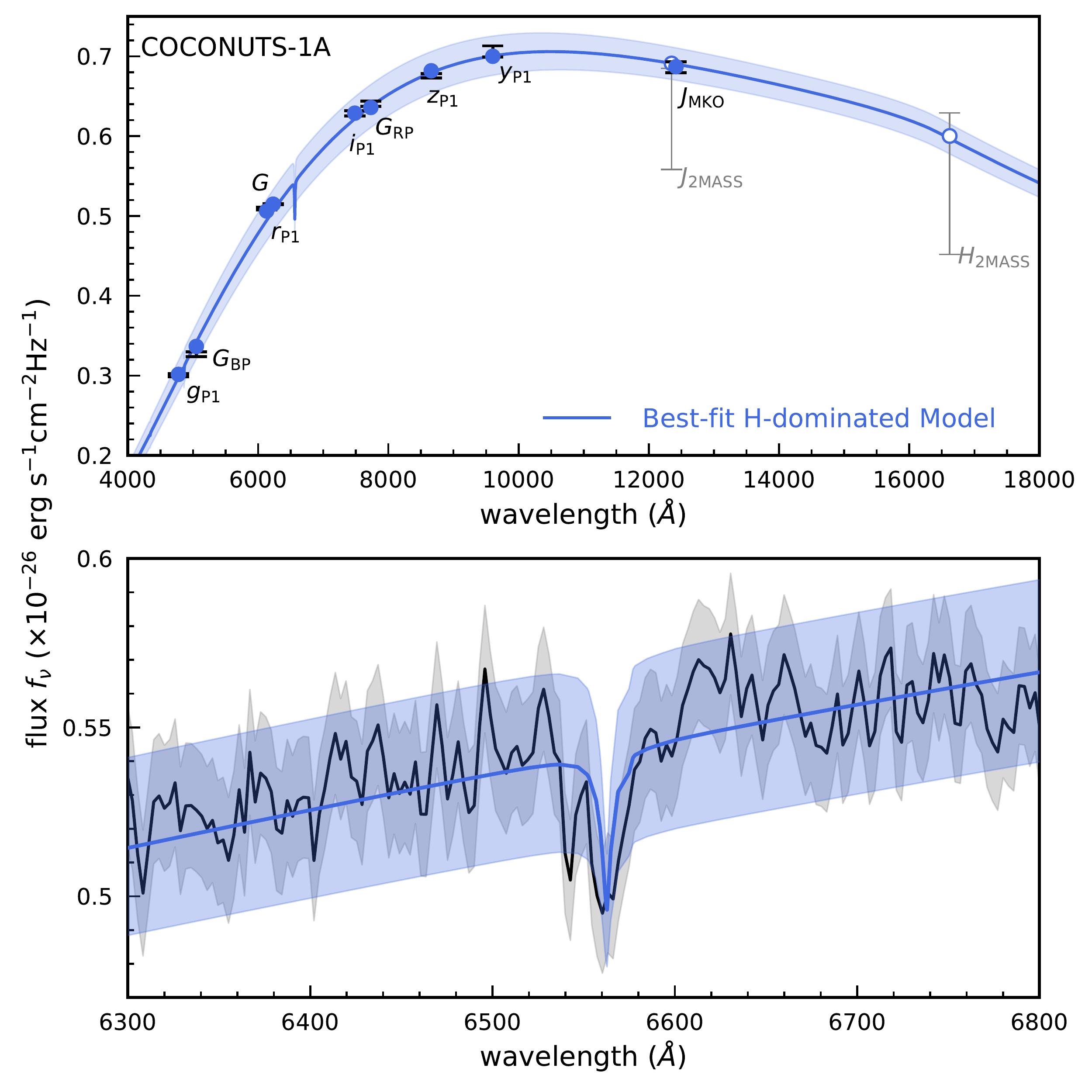}
\caption{The upper panel presents the best-fit hydrogen-dominated model spectrum (blue line) with $1\sigma$ model uncertainties (computed from the parameter errors; blue shadow) for the SED of the white dwarf primary COCONUTS-1A. The observed fluxes and uncertainties from {\it Gaia}~DR2, PS1, UHS, and 2MASS are shown as error bars, and the 2MASS photometry are plotted in grey since they are not included in our SED analysis (Section~\ref{sec:WD_analysis}). The synthesized broadband fluxes using the best-fit model spectrum are consistent with observations and are shown as blue circles (solid circles for {\it Gaia}~DR2, PS1, and UHS, and open circles for 2MASS). The lower panel compares the best-fit model spectrum (blue) with the flux-calibrated UH~2.2m/SNIFS spectrum (black) and flux uncertainties (grey shadow) near the H$\alpha$ line. The model spectrum is convolved at the SNIFS resolution. We note a possible spectral feature around $6540$~\AA, but a higher S/N is needed to assess the reality of this feature. The agreement between the SED-derived model and the observed H$\alpha$ line confirms the hydrogen-dominated atmosphere of COCONUTS-1A.}
\label{fig:WD_bestfit}
\end{center}
\end{figure*}

We derive $\{T_{\rm eff}, \log\ g\}$ and uncertainties using the nonlinear least-squares method \citep{1992nrfa.book.....P}, where the $\chi^{2}$ values are computed by summing the difference between observed and synthetic fluxes in all bands, weighted by errors that incorporate the parallax and magnitude uncertainties. Finally, we obtain $T_{\rm eff} = 5115 \pm 61$~K and $\log\ g = 7.945 \pm 0.038$. Here we adopt external uncertainties of 1.2\% on effective temperature and 0.038~dex on surface gravity, as derived from the spectroscopic analysis by \cite{2005ApJS..156...47L}, to account for the underestimated error budget of physical parameters. We present our best-fit model spectra in Figure~\ref{fig:WD_bestfit}. The agreement between the model and the observed H$\alpha$ line confirms the almost pure-hydrogen atmosphere interpretation.

Using the SED-derived $\{T_{\rm eff}, \log\ g\}$ and uncertainties, we interpolate the \cite{2001PASP..113..409F} isochrones and obtain a white dwarf mass as $0.548 \pm 0.023$~M$_{\odot}$. These models suggest the star has been cooling as a white dwarf for $4.6 \pm 0.6$~Gyr. While we do not attempt a correction, we note that cool white dwarfs ($T_{\rm eff} \lesssim 6000$~K) appear over-luminous compared to astrophysical expectations (cooling at constant mass), leading to an underestimated mass by $\approx 5\%$ \citep[e.g.,][]{2018MNRAS.480.3942H, 2019ApJ...878...63B, 2019ApJ...876...67B}.

We derive the mass of the zero-age-main-sequence white dwarf progenitor to be $1.54\pm0.20$~M$_{\odot}$ using a mean of three cluster-calibrated initial-to-final mass relations (IFMRs) \citep{2008ApJ...676..594K, 2009MNRAS.395.1795C, 2009ApJ...693..355W}. Using the MIST isochrones \citep{2016ApJ...823..102C}, we can convert this progenitor mass into a main-sequence lifetime of $2.7^{+1.4}_{-0.8}$~Gyr. Unfortunately, IFMRs are poorly calibrated for low-mass ($<0.6$~M$_{\odot}$) white dwarfs. However, theoretical IFMRs also do not accurately match higher-mass stars, which could be related to the adopted efficiency of convective overshoot \citep[e.g.,][]{2009ApJ...692.1013S, 2016ApJ...823...46F}. Without detailed calibrations for the theoretical IFMRs, we adopt our progenitor mass from the empirical, cluster-calibrated IFMRs.

Combining the cooling age of the white dwarf and the main-sequence lifetime of its progenitor, we derive a total age of the white dwarf to be $7.3^{+2.8}_{-1.6}$~Gyr. Here the age uncertainties are doubled from those of the white dwarf progenitor's main-sequence lifetime in order to account for errors from the white dwarf's cooling age and the IFMR systematics. We assume the T dwarf companion and white dwarf are coeval, given that the A and B components of the system are co-moving and associated (Section~\ref{sec:system}).

\section{The T Dwarf Companion}
\label{sec:T_analysis}

\subsection{Spectral Type}
\label{subsec:T_spt}

We calculate the spectral type for the companion using the \cite{2007ApJ...659..655B} polynomial fits for the five spectral indices defined by \cite{2006ApJ...637.1067B}, which trace CH$_{4}$ and H$_{2}$O absorption in ultracool atmospheres. We obtain spectral types of T3.9 from its H$_{2}$O-$J$ index of $0.377$, T1.3 from its CH$_{4}$-$J$ index of $0.604$, T4.0 from its H$_{2}$O-$H$ index of $0.417$, T4.3 from its CH$_{4}$-$H$ index of $0.534$, and T4.7 from its CH$_{4}$-$K$ index of $0.248$. The final index-based spectral type is T3.6$\pm$1.4, computed as the average and the standard deviation over all the five indices.

\begin{figure*}[t]
\begin{center}
\includegraphics[height=7.5in]{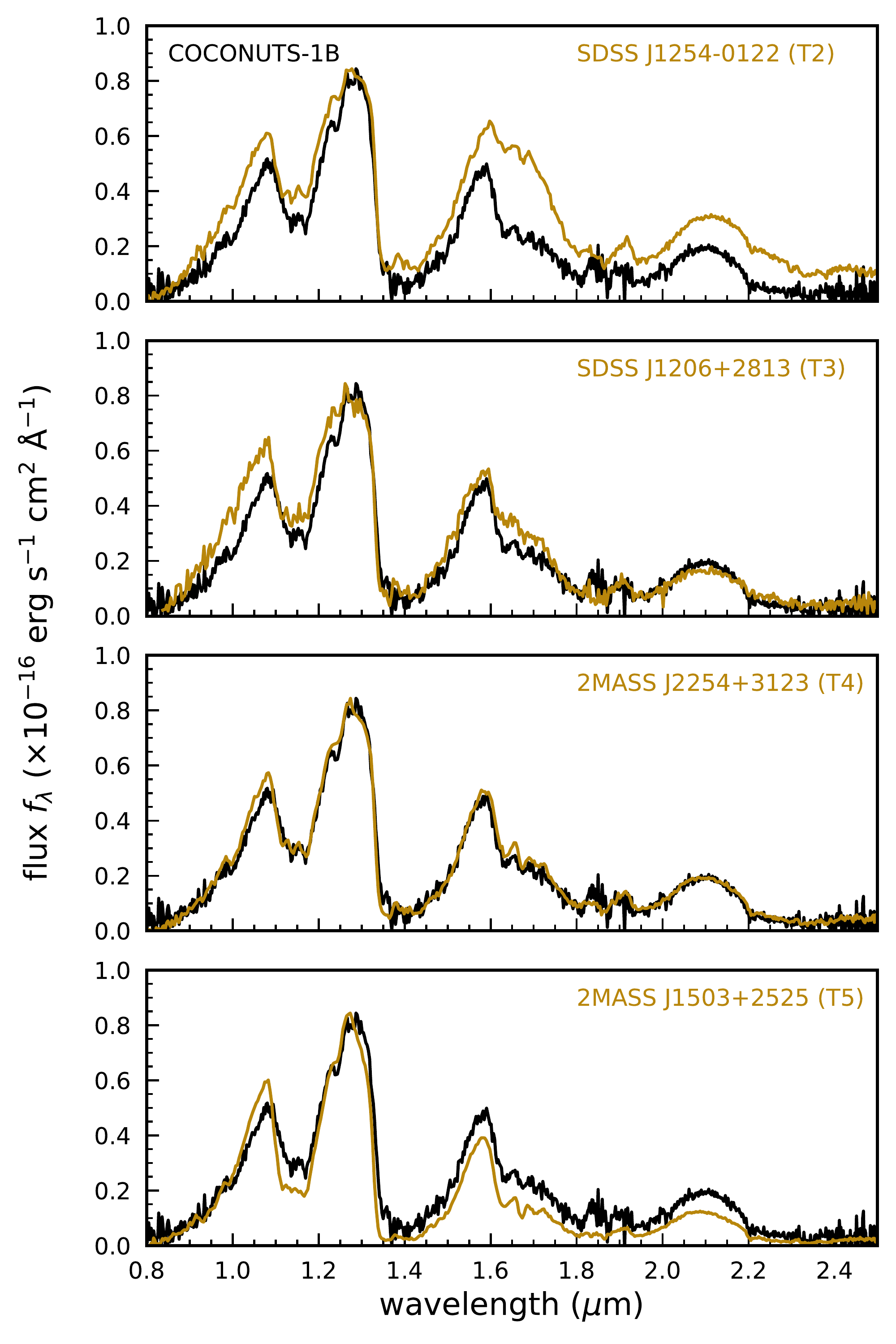}
\caption{The IRTF/SpeX near-infrared spectra of the companion COCONUTS-1B (black), as well as the T2--T5 spectral standards \citep[brown;][]{2006ApJ...637.1067B, 2010ApJ...722..311L}. The spectrum of COCONUTS-1B is flux calibrated using its $J_{\rm MKO}$ magnitude from UHS (Section~\ref{subsubsec:atm_fm}), and the other spectral standards are normalized by their peak fluxes. We thereby derive a visual type of T4$\pm 0.5$ for COCONUTS-1B, although its $Y$-band emission is relatively suppressed compared to the T4 standard.}
\label{fig:T_spec}
\end{center}
\end{figure*}

We also compare the companion's spectrum with T-type spectral standards \citep{2006ApJ...637.1067B} (Figure~\ref{fig:T_spec}). As suggested by \cite{2010ApJ...722..311L}, we use SDSS~J1206+2813 as the T3 spectral standard, given that both the original \cite{2006ApJ...637.1067B} standard 2MASS~J1209−1004 \citep[T2.0+T7.5;][]{2010ApJ...722..311L} and the alternative \cite{2006ApJ...637.1067B} standard SDSS~J1021−0304AB \citep[T1+T5;][]{2006ApJS..166..585B} are binaries. We derive a visual spectral type of T4 with an uncertainty of 0.5 subtype. We notice that the companion's spectrum around the $Y$-band peak is relatively suppressed compared to the T4 spectral standard. 

Both the index-based and visual spectral types are consistent with the companion's colors and absolute magnitudes (Figure~\ref{fig:T_CCDCMD}) and we adopt the visual type of T4$\pm 0.5$ as the final spectral type. In addition, we find the companion is unlikely an unresolved binary based on criteria suggested by \cite{2010ApJ...710.1142B}.

\subsection{Companionship Assessment}
\label{subsec:companion}
We use the observed space density of ultracool dwarfs to estimate the probability that our T dwarf discovery is a field interloper not bound to the white dwarf. Precise estimates of the space densities of L/T transition ($\approx$L6$-$T6) objects has been difficult because of their short-lived nature in this evolutionary stage \citep[e.g.,][]{2013MNRAS.430.1171D, 2015MNRAS.449.3651M}. Recently, \cite{2018PhDT.......159B} have compiled a volume-limited sample of L0-T8 field dwarfs within 25~pc and derived precise space densities. We adopt their results for T4 dwarfs of $(4.5\pm1.1) \times 10^{-4}$~pc$^{-3}$.

We then derive the spectrophotometric distance of COCONUTS-1B by comparing its observed photometry in $z_{\rm P1}$, $y_{\rm P1}$, $J_{\rm MKO}$, W1, and W2 bands, to the absolute magnitudes of T3$-$T5 field dwarfs (excluding binaries, subdwarfs, and young objects) from the \cite{2018ApJS..234....1B} catalog. We obtain typical absolute magnitudes of $M_{z_{\rm P1}} = 18.7 \pm 0.3$~mag, $M_{y_{\rm P1}} = 17.0 \pm 0.4$~mag, $M_{J_{\rm MKO}} = 14.4 \pm 0.4$~mag, $M_{W1} = 14.0 \pm 0.6$~mag, and $M_{W2} = 12.5 \pm 0.3$~mag using 13, 14, 23, 21, and 21 field dwarfs, respectively. We thereby derive spectrophotometric distances of $d_{z_{\rm P1}} = 32 \pm 5$~pc, $d_{y_{\rm P1}} = 31 \pm 6$~pc, $d_{J_{\rm MKO}} = 29 \pm 5$~pc, $d_{W1} = 31 \pm 8$~pc, and $d_{W2} = 39 \pm 6$~pc. We note spectrophotometric distances from all bands except for W2\footnote{The relatively farther photometric distance from W2 may be intriguing. Using the white dwarf's distance, the T dwarf's W2-band absolute magnitude is $12.93 \pm 0.11$~mag, which, compared with the typical $M_{W2}$ value, suggests its emergent flux around $\approx 4.6$~$\mu$m is fainter than other T3$-$T5 dwarfs by a factor of $1.5 \pm 0.4$ (or equivalently fainter by $0.4 \pm 0.3$~mag).} are consistent with the {\it Gaia}~DR2 distance $31.51 \pm 0.09$~pc of the white dwarf, again supporting the common distance between the T dwarf and the white dwarf. We adopt a spectrophotometric distance of $31 \pm 3$~pc by averaging all the bands except W2 with the uncertainties propagated.

We use the aforementioned space density to compute the expected number of the T4-type field interloper in a volume approximated as a circular region with a generous radius of $10^4$~au (Section~\ref{sec:system}) and a depth of $6$~pc (i.e., 2 times the uncertainty of the companion's photometric distance), leading to $(2.6 \pm 0.6) \times 10^{-6}$~objects. Considering that our T dwarf discovery also has a common proper motion with the white dwarf (Figure~\ref{fig:common_pm}), the contamination probability will be even lower. Thus, we conclude the white dwarf and the T dwarf form a physical pair.

\subsection{Atmospheric Model Analysis}
\label{subsec:atm}

\subsubsection{Model Atmospheres}
We model the near-infrared spectrum of COCONUTS-1B using the cloudless Sonora grids (\citealt{2017AAS...23031507M}; Marley et al. 2019 in prep). The models are generated over $0.4-50$~$\mu$m in wavelength assuming chemical equilibrium, a helium abundance of $Y=0.28$, and a solar carbon-to-oxygen ratio. The model grids span $200 - 2400$~K in effective temperature ($T_{\rm eff}$; with spacing of $25$~K in $200-600$~K, $50$~K in $600-1000$~K, and $100$~K in $1000-2400$~K) and $3.25 - 5.5$~dex in logarithmic surface gravity (${\rm log}\ g$; cgs units; with spacing of 0.25~dex). The grids include three metallicities ($Z$): sub-solar ($-0.5$~dex), solar ($0$~dex), and super-solar ($+0.5$~dex). 

Are cloudless models suitable for our companion? While clouds play a crucial role shaping the emergent spectra of ultracool dwarfs, they become less significant as brown dwarfs evolve through the L/T transition to cooler temperatures \citep[$T_{\rm eff} \lesssim 1400$~K; e.g.,][]{2005ARA&A..43..195K, 2008ApJ...689.1327S, 2010ApJ...723L.117M}. This process results in clearer atmospheres at the near-infrared wavelengths. Therefore the early- to mid-T dwarfs, such as our T4 companion discovery, have relatively brighter $J$-band emission and bluer $J-K$ colors than earlier-type L/T transition objects \citep[e.g.,][also see Figures~\ref{fig:T_CCDCMD} and \ref{fig:CMD_LT}]{2012ApJS..201...19D}. We can test the influence of clouds by comparing our available set of models to COCONUTS-1B. If its observed spectrum significantly deviates from the model atmospheres or implies unphysical properties for an object given the age of its white dwarf primary, then we can gain insight about cloud formation in ultracool atmospheres during the mid-T evolutionary stage.

\subsubsection{Forward Modeling}
\label{subsubsec:atm_fm}
We fit our SpeX spectrum using the Sonora model grids over $0.9-2.3$~$\mu$m in wavelength, where the companion's spectrum has S/N $\gtrsim 5$~per pixel. We convolve each model spectrum with the instrumental profile corresponding to the SpeX prism with the $0.8''$ slit \citep{2003PASP..115..362R}, including the wavelength-dependent spectral resolution in the convolution process \citep[using scripts of the Starfish package by][]{2015ApJ...812..128C}. The convolved model spectra therefore match the spectral resolution of the observed data. We also incorporate the radial velocity $v_{r}$, the projected rotational velocity $v\ \sin i$, and the solid angle $\Omega = (R/d)^{2}$, where $R$ is the object's radius and $d$ is distance, into the model spectra when comparing with data.

We then flux-calibrate the observed spectrum by using the companion's $J$-band magnitude from the UKIDSS Hemisphere Survey, which provides the only observed near-infrared photometry for this object. We use the WFCAM $J$-band filter from \cite{2006MNRAS.367..454H} and obtain the zero-point flux from \cite{2007MNRAS.379.1599L}.

We then use the Markov chain Monte Carlo (MCMC) algorithm {\it emcee} \citep{2013PASP..125..306F} with 24 walkers to determine the posteriors for the six physical parameters $\{T_{\rm eff}, {\rm log}\ g, Z, v_{r}, v\ {\rm sin}\ i, \log\ \Omega\}$. We use linear interpolation to construct model spectra with $\{T_{\rm eff}, \log\ g, Z\}$ between the Sonora grid points, with the interpolation conducted in logarithmic units for $T_{\rm eff}$. To evaluate model parameters, we construct the covariance matrix by placing the squared measurement uncertainties along the diagonal axis. 

We assume uniform priors for $T_{\rm eff}$ between $200$~K and $2400$~K, ${\rm log}\ g$ between $3.25$ and $5.5$, $Z$ between $-0.5$ and $+0.5$, and $v\ \sin i$ between $0$ and $v_{\rm max}$. The maximum rotational velocity $v_{\rm max}$ is determined based on the object's oblateness, expressed as $f = 2Cv_{\rm rot}^{2} / 3gd\sqrt{\Omega}$ \citep[][]{2003ApJ...588..545B, 2011MNRAS.417.2874M}, where $v_{\rm rot}$ is the equatorial rotational velocity, $d$ is the {\it Gaia}~DR2 distance of the white dwarf primary, and $C=0.9669$ \citep{1939isss.book.....C} corresponds to the $n=1.5$ polytropic index which well approximates fully convective brown dwarfs \citep[e.g.,][]{1993RvMP...65..301B}. Requiring the object's oblateness to be below the stability limit for the $n=1.5$ polytope, i.e., $f \leqslant 0.385 \equiv f_{\rm crit}$ \citep[][]{1964ApJ...140..552J}, we therefore derive the following constraints on $v_{\rm rot}$ and $v\ \sin i$:
\begin{equation} \label{eq:v_max}
0 \leqslant v\ \sin i \leqslant v_{\rm rot} \leqslant \Omega^{1/4} \left(\frac{3f_{\rm crit}gd}{2C}\right)^{1/2} \equiv v_{\rm max}
\end{equation}
For each step of each {\it emcee} walker, we thus compute $v_{\rm max}$ and assume a uniform prior between $0$ and $v_{\rm max}$ for the projected rotational velocity $v\ \sin i$. In addition, we assume uniform priors for both $v_{r}$ and $\log\ \Omega$ between $-\infty$ and $+\infty$.

\subsubsection{Results}
\label{subsubsec:atm_res}
\begin{figure*}[t]
\begin{center}
\includegraphics[height=6.5in]{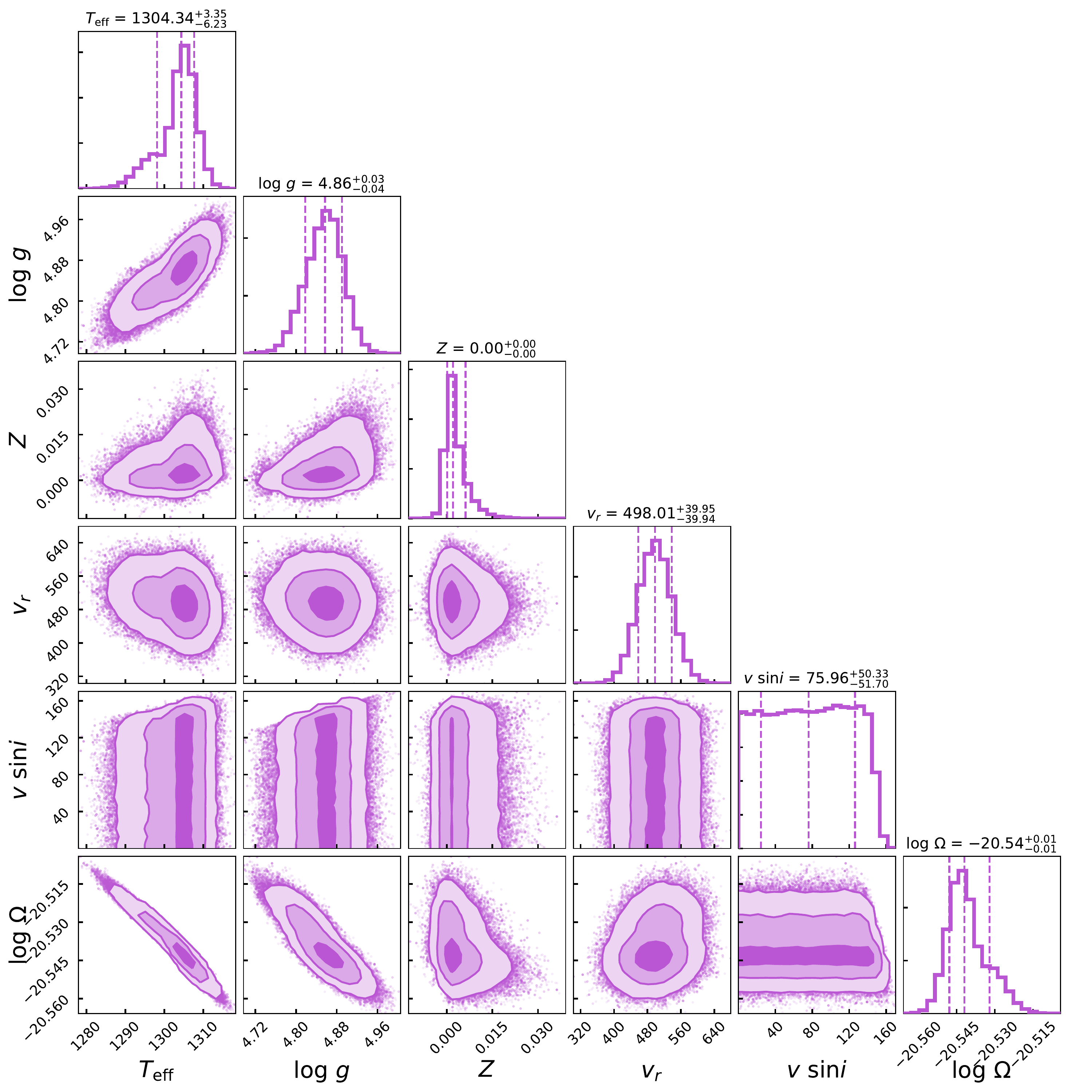}
\caption{Posterior distributions of COCONUTS-1B properties derived using the Sonora atmospheric models. The physical parameters plotted are effective temperature ($T_{\rm eff}$; in units of K), logarithmic surface gravity (${\rm log}\ g$), metallicity ($Z$), radial velocity ($v_{r}$; in units of km/s), projected rotational velocity ($v\ \sin i$; in units of km/s), and the logarithmic solid angle ${\rm log}\ \Omega$. }
\label{fig:atm_posteriors}
\end{center}
\end{figure*}

We present the posteriors of the six physical parameters in Figure~\ref{fig:atm_posteriors}, with results summarized in Table~\ref{tab:atmevo}. Using the {\it Gaia}~DR2 distance $31.51\pm0.09$~pc of the white dwarf primary, we convert the best-fit ${\rm log}\ \Omega = -20.542^{+0.010}_{-0.006}$~dex into the companion's radius of $R = 0.730^{+0.009}_{-0.006}$~$R_{\rm Jup}$. We further use $\log\ g$ and $R$ and compute the companion's mass as $M = 15.4^{+0.9}_{-0.8}$~M$_{\rm Jup}$. Neither the radius nor the mass are plausible, which we discuss in Section~\ref{subsec:benchmarking}.

\begin{figure*}[t]
\begin{center}
\includegraphics[height=7.5in]{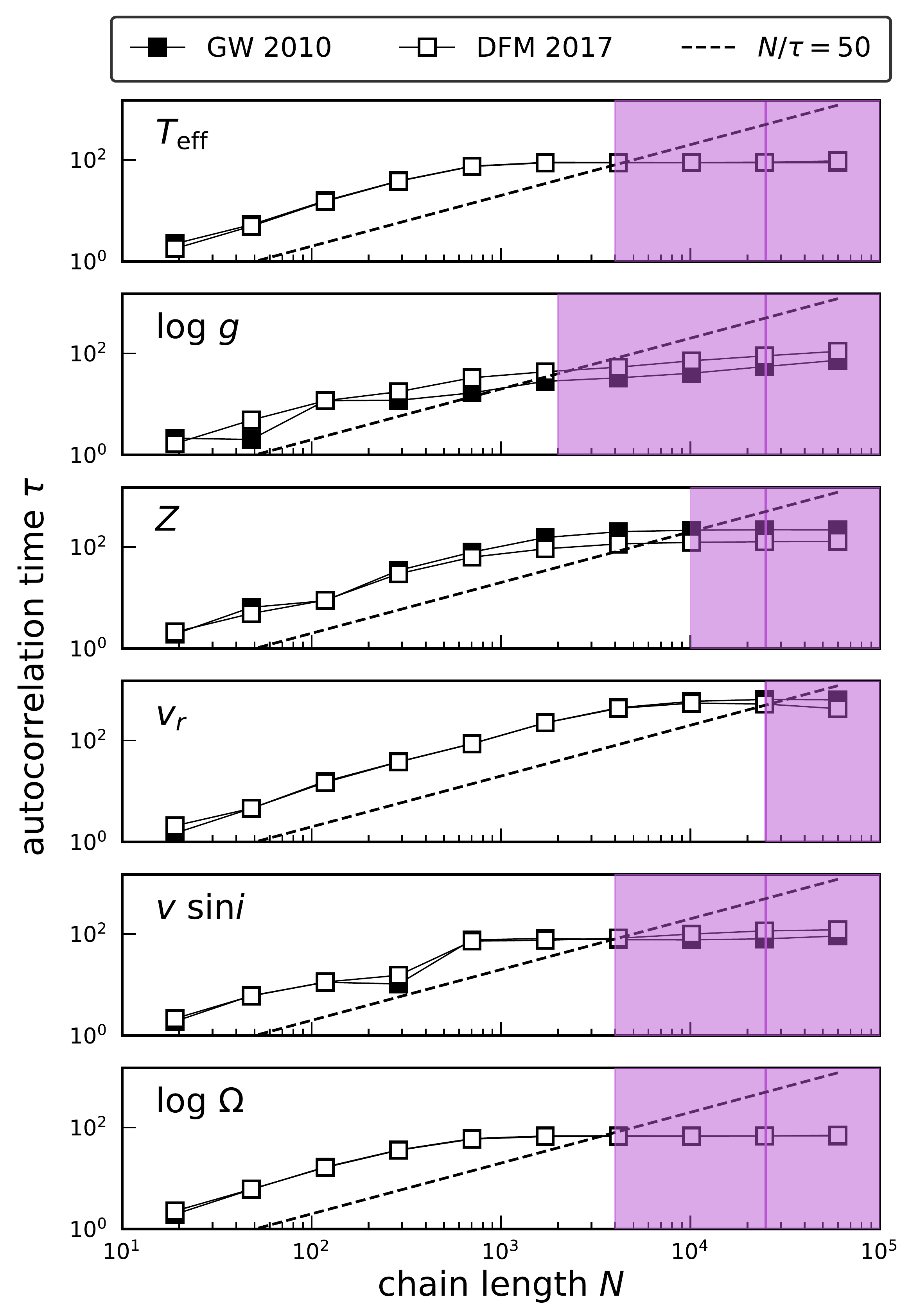}
\caption{The integrated autocorrelation time for the MCMC chains of each physical parameter from our atmospheric model analysis, computed with different chain lengths based on the \cite{2010CAMCS...5...65G} method (GW~2010; solid squares) and the revised version suggested by Foreman-Mackey (DFM~2017; open squares), both of which produce consistent estimates. As suggested by the {\it emcee} documentation, the chains are supposed to converge once their lengths exceeds 50 times the estimated autocorrelation time from both (dashed line). While the chains for different parameters require different chain lengths (purple shadows), all of our chains converge after $2.5\times10^{4}$ samples (purple vertical lines). }
\label{fig:atm_convergence}
\end{center}
\end{figure*}

We test the convergence of the resulting MCMC chains based on their integrated autocorrelation time, following the {\it emcee} documentation\footnote{\url{https://emcee.readthedocs.io/en/latest/tutorials/autocorr/}}. As we increase the number of iterations, we estimate the average autocorrelation time for each physical parameter based on the \cite{2010CAMCS...5...65G} method and the revised version suggested by Daniel Foreman-Mackey\footnote{\url{https://github.com/dfm/emcee/issues/209}}. The chains are supposed to converge once their lengths exceed 50 times the average autocorrelation time. As shown in Figure~\ref{fig:atm_convergence}, all of our chains have converged after $2.5\times 10^{4}$ iterations. We run the fitting process with $6 \times 10^{4}$ iterations and use the second half of chains to produce parameter posteriors.

\begin{figure*}[t]
\begin{center}
\includegraphics[height=6.in]{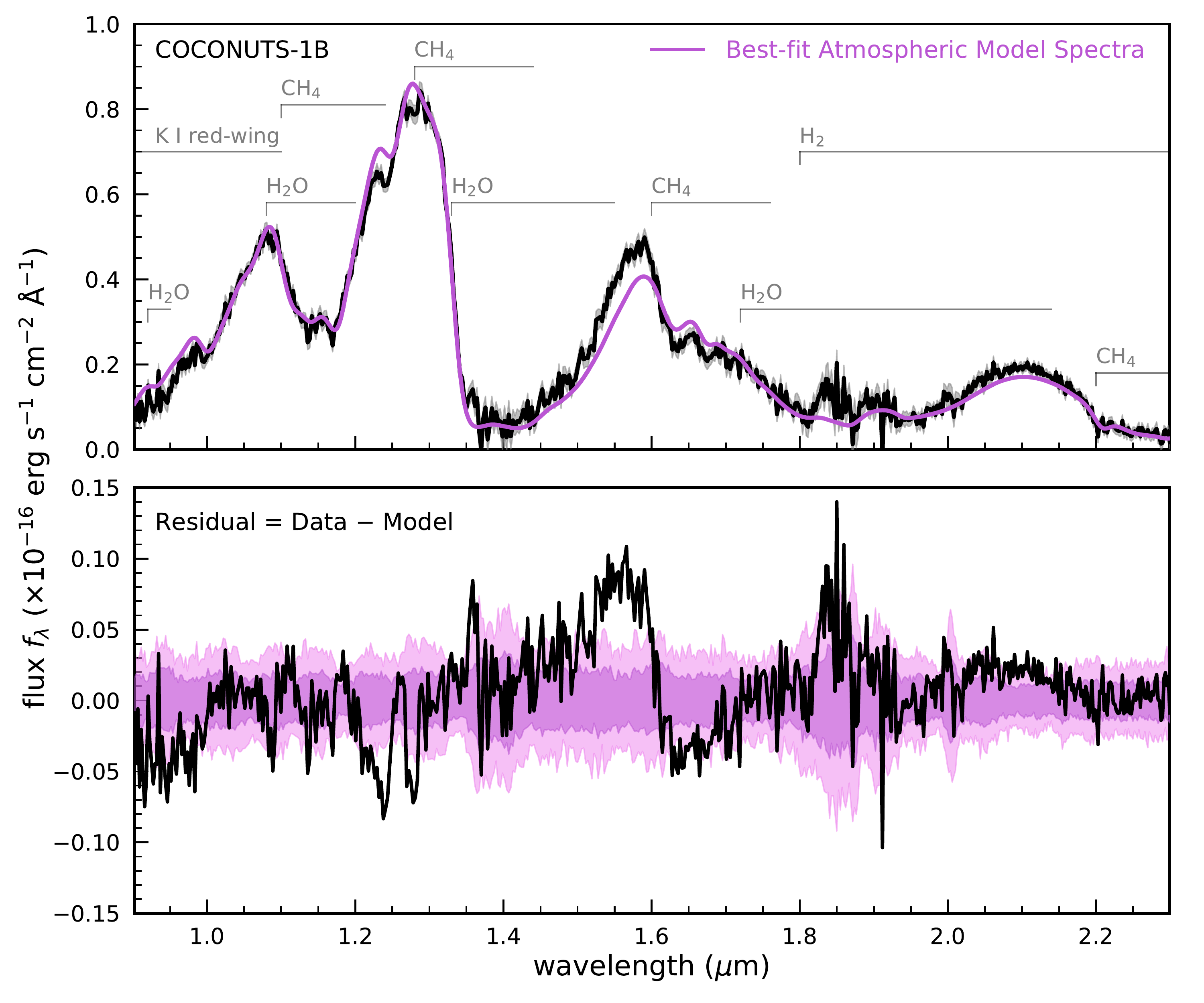}
\caption{The upper panel presents the observed spectra of COCONUTS-1B (black) with $1\sigma$ measurement uncertainties (grey shadow) and the Sonora atmospheric model spectra interpolated at the parameters drawn from the MCMC chains (purple). The lower panel shows the fitting residual (data $-$ model; black), and the purple shadows are $1\sigma$ and $2\sigma$ dispersions of 50,000 random draws from the covariance matrix, composed of measurement uncertainties along the diagonal axis.}
\label{fig:atm_spec}
\end{center}
\end{figure*}

Figure~\ref{fig:atm_spec} compares the observed data with Sonora model spectra interpolated at physical parameters drawn from the MCMC chains. While the model spectra match the overall observed spectral morphology, the residuals in several wavelength ranges are much larger than the measurement uncertainties. At the blue wing of the $Y$ band ($<1$~$\mu$m), the models over-predict the emergent flux. The flux in this region is a sensitive function of both the potassium abundance and the pressure-broadened red wing of the potassium doublet at $0.77$~$\mu$m \citep[e.g.,][]{1999ApJ...520L.119T, 2003ApJ...583..985B, 2007A&A...474L..21A}. In addition, the models over-predict the emergent flux around the $J$-band peak while under-predicting the $H$-band peak. The excess $J$-band flux from models likely arises from the lack of a deep cloud deck in these models, which would attenuate the flux emerging from the deep atmosphere. The under-predicted $H$-band flux from models could well be a consequence of the fitting procedure responding to the overestimate of the $J$-band flux by choosing models that under-predict the $H$-band flux. 

Our derived physical parameters all have very small uncertainties, likely because model systematics and uncertainties from the linear interpolation of the model grid are not accounted for. Also, the posteriors of $T_{\rm eff}$ and $Z$ are close to the model grid point of $T_{\rm eff} = 1300$~K and $Z = 0$~dex, respectively (Figure~\ref{fig:atm_posteriors}). This is likely due to the sparsely sampled atmospheric models (the $T_{\rm eff}$ spacing is $100$~K and the $Z$ spacing is $0.5$~dex), where linear interpolation may bias the posteriors toward model grid points \citep[e.g.,][]{2014ApJ...794..125C, 2015ApJ...812..128C}.  A more sophisticated analysis is beyond the scope of this paper. Given the artificially small parameter errors, we inflate the uncertainties of $\{T_{\rm eff}, \log\ g, Z\}$ by adopting halves of the local grid spacing, namely $50$~K, $0.13$~dex, and $0.25$~dex, respectively. These then lead to inflated errors of $0.056$~R$_{\rm Jup}$ in radius and $5.2$~M$_{\rm Jup}$ in mass (Table~\ref{tab:atmevo}). 

As a result of the fitting process, we note that the companion's radial velocity $v_{r}$ and projected rotational velocity $v\ \sin i$ might behave like nuisance parameters aiming for better fitting results rather than preserving their physical meaning, given that the main differences between data and models result from model systematics. We adjust the error of $v_{r}$ by incorporating the dispersion in wavelength calibration of the prism data ($5.9$~\AA; Section~\ref{subsec:T_obs}), which corresponds to a velocity of $200 - 80$~km/s in the $0.9-2.3$~$\mu$m wavelength. We thus adopt a $v_{r}$ uncertainty of $200$~km/s. We keep the uncertainty of $v\ \sin i$, which has a significance of only 1.5$\sigma$ from our forward modeling analysis ($76 ^{+52}_{-50}$~km/s; Table~\ref{tab:atmevo}).

\subsection{Evolutionary Model Analysis}
\label{subsec:evo}
We also use the hot-start cloudless Sonora evolutionary models with solar metallicity (\citealt{2017AAS...23031507M}; Marley et al. 2019 in prep) to derive the companion's $T_{\rm eff}$, ${\rm log}\ g$, and $R$. These evolutionary-derived parameters can be directly compared to the atmospheric modeling results, as the best-fit metallicity from our atmospheric models is $Z = 0$ (Figure~\ref{fig:atm_posteriors}). We adapt the method of \cite{2006ApJ...647..552S} for the evolutionary model analysis. The \cite{2006ApJ...647..552S} method aims to determine the $\{T_{\rm eff}, \log\ g\}$ values, where the bolometric luminosity $L_{\rm bol}$ is the same between the one derived using the object's near-infrared spectrum and the atmospheric model-based bolometric correction, and the one derived using evolutionary models. Here we adapt this method in a Bayesian framework.

We first calculate the companion's near-infrared luminosity $L_{0.9-2.3\, \mu\rm m}$ by integrating its flux-calibrated IRTF/SpeX spectrum in $0.9-2.3$~$\mu$m and using the {\it Gaia}~DR2 distance of the primary, with uncertainties in flux and distance incorporated. We then initiate an MCMC process with free parameters being $T_{\rm eff}$ and $\log\ g$. For a given choice of $\{T_{\rm eff}, \log\ g\}$, we construct a model spectrum by linearly interpolating the Sonora atmospheric models. We then use the interpolated model spectrum to compute a bolometric correction, defined as the ratio between its integrated fluxes between $0.9-2.3$~$\mu$m and between $0.4-50$~$\mu$m. We find this ratio ranges from $0.6-0.72$ for the cloudless Sonora models of $T_{\rm eff} > 1200$~K, with changes of $\lesssim 0.008$ across halves of the grid spacing. We thus assign an uncertainty of 0.01 to the computed flux ratio to account for variations in the model-based bolometric correction. We then apply the ratio to the companion's measured $L_{0.9-2.3\, \mu\rm m}$ to derive its atmospheric-based bolometric luminosity $L_{\rm bol, atm}$, with uncertainties propagated from $L_{0.9-2.3\, \mu\rm m}$ and the bolometric correction.

\begin{figure*}[t]
\begin{center}
\includegraphics[height=6.5in]{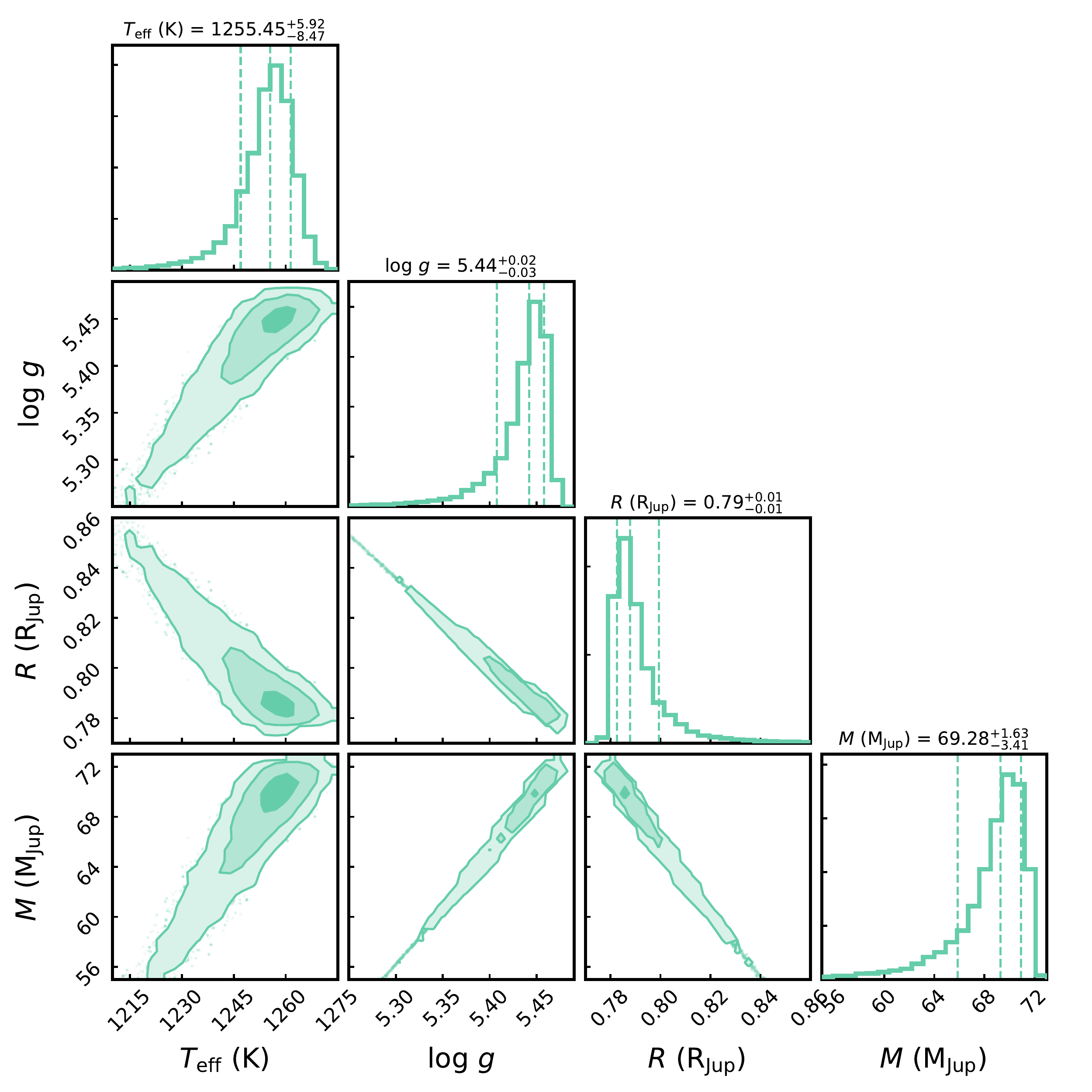}
\caption{Posterior distributions of COCONUTS-1B properties derived from evolutionary models and the age of the white dwarf primary. The parameters plotted are effective temperature ($T_{\rm eff}$), logarithmic surface gravity (${\rm log}\ g$), radius ($R$), and mass ($M$). }
\label{fig:evo_posteriors}
\end{center}
\end{figure*}

In addition, with the same $\{T_{\rm eff}, \log\ g\}$ values, we interpolate the Sonora evolutionary models, in logarithmic units for $T_{\rm eff}$ and age, to derive the companion's predicted bolometric luminosity $L_{\rm bol, evo}$ and age $t_{\rm evo}$. We evaluate $\{T_{\rm eff}, \log\ g\}$ using the following likelihood function
\begin{equation}
\mathcal{L} = p(L_{\rm bol, atm}\ |\ L_{\rm bol, evo}) \times p(t_{\rm WD}\ |\ t_{\rm evo}) \times p(T_{\rm eff}) \times p(\log\ g)
\end{equation}
where $t_{\rm WD} = 7.3_{-1.6}^{+2.8}$~Gyr is the age of the white dwarf primary. To compute $p(L_{\rm bol, atm}\ |\ L_{\rm bol, evo})$, we assume the bolometric luminosity follows a normal distribution with mean and standard deviation corresponding to the value and uncertainty of $L_{\rm bol, atm}$, respectively. To compute $p(t_{\rm WD}\ |\ t_{\rm evo})$, we assume the age follows a distribution composed of two half-Gaussians joined at $7.3$~Gyr and extending from $0$~Gyr to $12$~Gyr. We assign standard deviations of $1.6$~Gyr and $2.8$~Gyr to the Gaussians younger and older than $7.3$~Gyr, respectively, to account for asymmetric error bars. Priors of $T_{\rm eff}$ and $\log\ g$ are assumed to be uniform distributions within the parameter space of the Sonora models, as done in Section~\ref{subsubsec:atm_fm}. We use {\it emcee} to execute the above process and derive posteriors of $\{T_{\rm eff}, \log\ g\}$ as $T_{\rm eff} = 1255^{+6}_{-8}$~K and $\log\ g = 5.44^{+0.02}_{-0.03}$~dex. We then use the resulting chains of $\{T_{\rm eff}, \log\ g\}$ with the interpolated Sonora evolutionary models to obtain $\log\ (L_{\rm bol} / L_{\odot}) = -4.832 \pm 0.007$, $R = 0.789^{+0.011}_{-0.005}$~R$_{\rm Jup}$, and $M = 69.3^{+1.6}_{-3.4}$~M$_{\rm Jup}$. We present posteriors for the evolutionary model parameters in Figure~\ref{fig:evo_posteriors}, with the best-fit values summarized in Table~\ref{tab:atmevo}.

We also try alternative methods to determine the companion's properties, which first calculate the companion's bolometric luminosity $L_{\rm bol}$ from its observed spectrum and then combine it with the white dwarf's age to use the interpolated Sonora evolutionary models. We use two approaches to compute $L_{\rm bol}$. Our first approach directly integrates the best-fit atmospheric model spectra over $0.4-50$~$\mu$m and uses the {\it Gaia}~DR2 distance of the primary, with parameter uncertainties incorporated. We derive $\log\ (L_{\rm bol} / L_{\odot}) = -4.84 \pm 0.03$, which is consistent with our method adapted from \cite{2006ApJ...647..552S}. The relatively larger uncertainty from this approach results from the inflated uncertainties of atmospheric-derived parameters (Section~\ref{subsubsec:atm_res}).  

Our second approach follows \cite{2004AJ....127.3516G}, which combines the $0.9-2.3$~$\mu$m IRTF/SpeX spectrum with the observed broadband fluxes converted from $z_{\rm P1}$, $W1$, and $W2$ magnitudes, using the filter responses and zero-point fluxes from \cite{2012ApJ...750...99T} and \cite{2011ApJ...735..112J}. We linearly interpolate fluxes between the spectrum and broadband fluxes. At shorter wavelength, we linearly extrapolate fluxes to zero, and at longer wavelength, we assume a Rayleigh-Jeans tail longword of $W2$. We integrate the spectrum constructed as above and derive the $L_{\rm bol}$ as $\log\ (L_{\rm bol} / L_{\odot}) = -4.901 \pm 0.003$. Compared to our adopted method based on \cite{2006ApJ...647..552S}, the $L_{\rm bol}$ estimated from this approach is a factor of $\approx 1.17$ fainter, which is close to the correction factor of 1.2 suggested by \cite{2004AJ....127.3516G}. Given that the aforementioned extrapolation process is inadequate to account for CH$_{4}$ and CO absorption of mid-to-late T dwarfs in the $\approx 3-5$~$\mu$m range, \cite{2004AJ....127.3516G} determined such a correction based on the observed $L$- and $M$-band spectra of Gl~229B (T7). In conclusion, both of our alternative approaches can produce consistent $L_{\rm bol}$, thereby evolutionary model parameters, with our adopted method.

\subsection{Benchmarking}
\label{subsec:benchmarking}

\begin{figure*}[t]
\begin{center}
\includegraphics[height=6.5in]{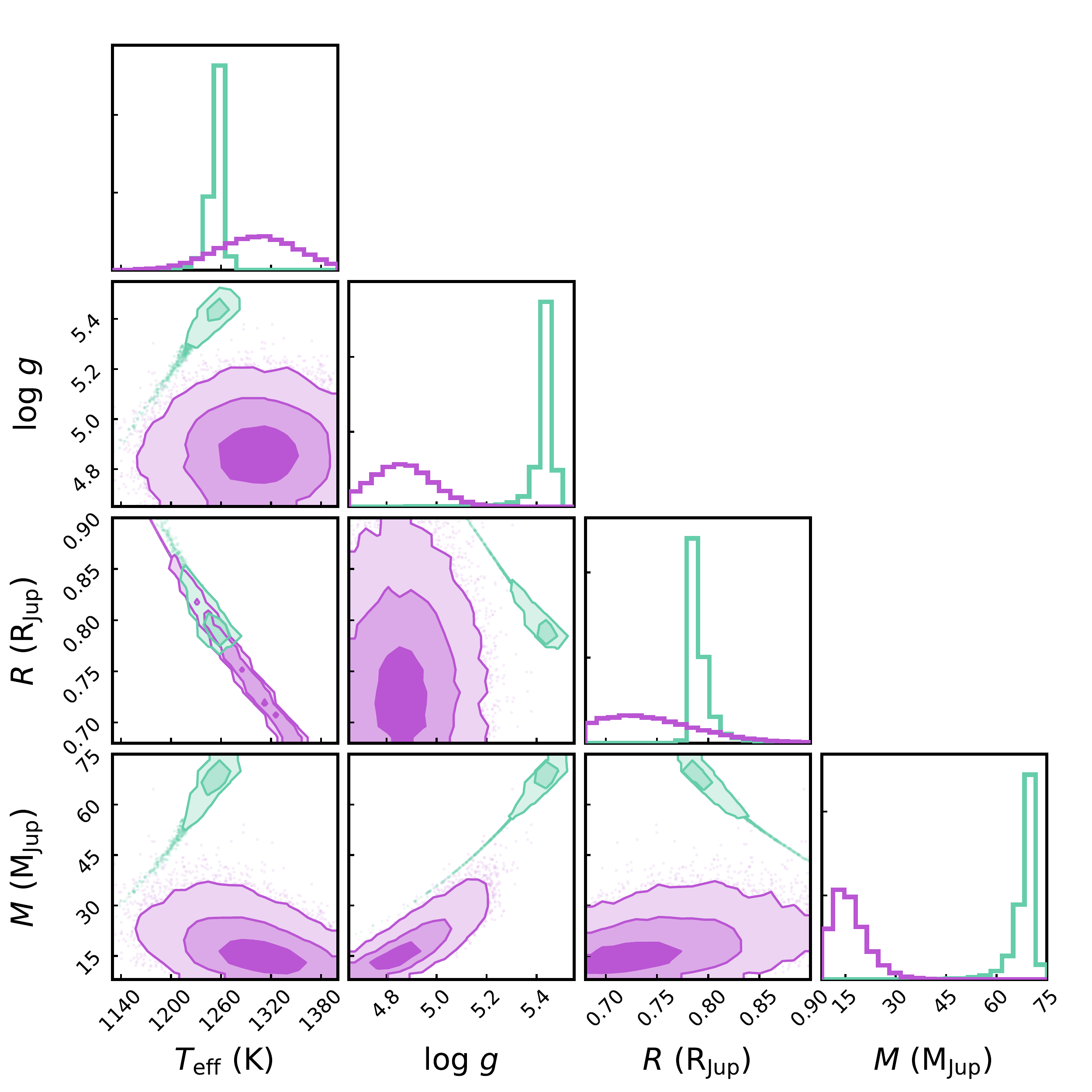}
\caption{Posterior distributions of the COCONUTS-1B physical properties derived from both atmospheric models (purple) and evolutionary models (green). The physical parameters plotted are effective temperature ($T_{\rm eff}$), logarithmic surface gravity (${\rm log}\ g$), radius ($R$), and mass ($M$). The posteriors of atmospheric-derived parameters are not from the MCMC chains of our forward modeling shown in Figure~\ref{fig:atm_posteriors}, but rather re-generated using adjusted uncertainties as described in Section~\ref{subsec:benchmarking}. We note that $T_{\rm eff}$ and $R$ derived from the two models are consistent within $50$~K and $0.06$~R$_{\rm Jup}$, but the atmospheric models produce a much lower $\log\ g$ by $\approx 0.6$~dex, thus a $\approx 5\times$ lower $M$ estimate. }
\label{fig:atm_evo_posteriors}
\end{center}
\end{figure*}

Having derived the companion's properties using both atmospheric models (Section~\ref{subsec:atm}) and evolutionary models (Section~\ref{subsec:evo}), we now compare the physical parameters derived from these two model sets, an approach often referred to as ``benchmarking'' \citep[e.g.,][]{2006MNRAS.368.1281P, 2008ApJ...689..436L}. Figure~\ref{fig:atm_evo_posteriors} compares the posteriors of effective temperature, surface gravity, radius, and mass derived from atmospheric models and evolutionary models. The posteriors of the atmospheric-derived parameters are not directly from the MCMC chains of our forward modeling analysis, but are rather those derived from inflating the uncertainties to halves of the model grid spacing (Section~\ref{subsec:atm}). Specifically, we generate Gaussian posteriors for $T_{\rm eff}$ and $\log\ g$, assuming the mean are their best-fit values from our atmospheric model analysis, and the standard deviation is 50 K and 0.13 dex, respectively. We then generate the radius $R$ posterior using the new $T_{\rm eff}$ posterior and the $R^{2} T_{\rm eff}^{4}$ chain values from the original atmospheric model analysis (i.e., the Stephen-Boltzmann law). We generate the mass $M$ posterior using the $\log\ g$ and $R$ posteriors. We note that the atmospheric and evolutionary models predict consistent $T_{\rm eff}$ and $R$ values, with insignificant differences of $49 \pm 51$~K and $0.06 \pm 0.06$~R$_{\rm Jup}$, respectively. 

However, the ${\rm log}\ g$ values are vastly different from the atmospheric and evolutionary models, with the former predicting a lower value by $0.58 \pm 0.13$~dex, leading to a $\approx 5 \times$ lower mass estimate. Also, based on the Sonora evolutionary models, the $T_{\rm eff}$ and $\log\ g$ of COCONUTS-1B derived from the atmospheric models correspond to an unphysically young age of $\approx 380 \pm 230$~Myr, which is in direct contradiction to our age estimate of $7.3^{+2.8}_{-1.6}$~Gyr for the white dwarf\footnote{Age discrepancies have been previously found in binary systems composed of white dwarfs and main-sequence stars with projected separations $\leqslant 60$~au \citep[e.g.,][]{2006A&A...459..955L, 2013A&A...554A..21Z, 2014ApJ...783L..25M}. In these systems, the age of the stellar component, derived from gyrochronology and/or chromospheric activities, is notably younger than that of the white dwarf, derived from its cooling age and the progenitor's main-sequence lifetime. One likely explanation is that the angular momentum lost during white dwarf formation is transferred to the stellar component, causing the star to spin up and thereby appear younger when using rotation-based age-dating techniques \citep[e.g.,][]{1996MNRAS.279..180J, 1997ApJ...482L.175K, 2007A&A...462..345D}. However, these systems are not analogs of COCONUTS-1, where the white dwarf and the brown dwarf are widely separated ($\approx 1280$~au; Section~\ref{sec:system}) and therefore likely have evolved in isolation.} (Section~\ref{sec:WD_analysis}). We conclude that for COCONUTS-1, the old white dwarf age is more reliable than the much younger age of the brown dwarf companion derived from Sonora atmospheric models, especially since our white dwarf analysis is based on high-S/N photometry and spectroscopy spanning from optical to near-infrared, with systematics in white dwarf model atmospheres and IFMRs carefully incorporated (Section~\ref{sec:WD_analysis}). The unphysically young age of the brown dwarf companion based on atmospheric models is likely due to shortcomings of these cloudless models. Therefore, we adopt the Sonora evolutionary model parameters as the characteristics of the brown dwarf companion. We also note that the companion's $\log\ g = 5.44^{+0.02}_{-0.03}$~dex is among the highest surface gravity that brown dwarfs can reach over the cosmic time.

\begin{figure*}[t]
\begin{center}
\includegraphics[height=4.5in]{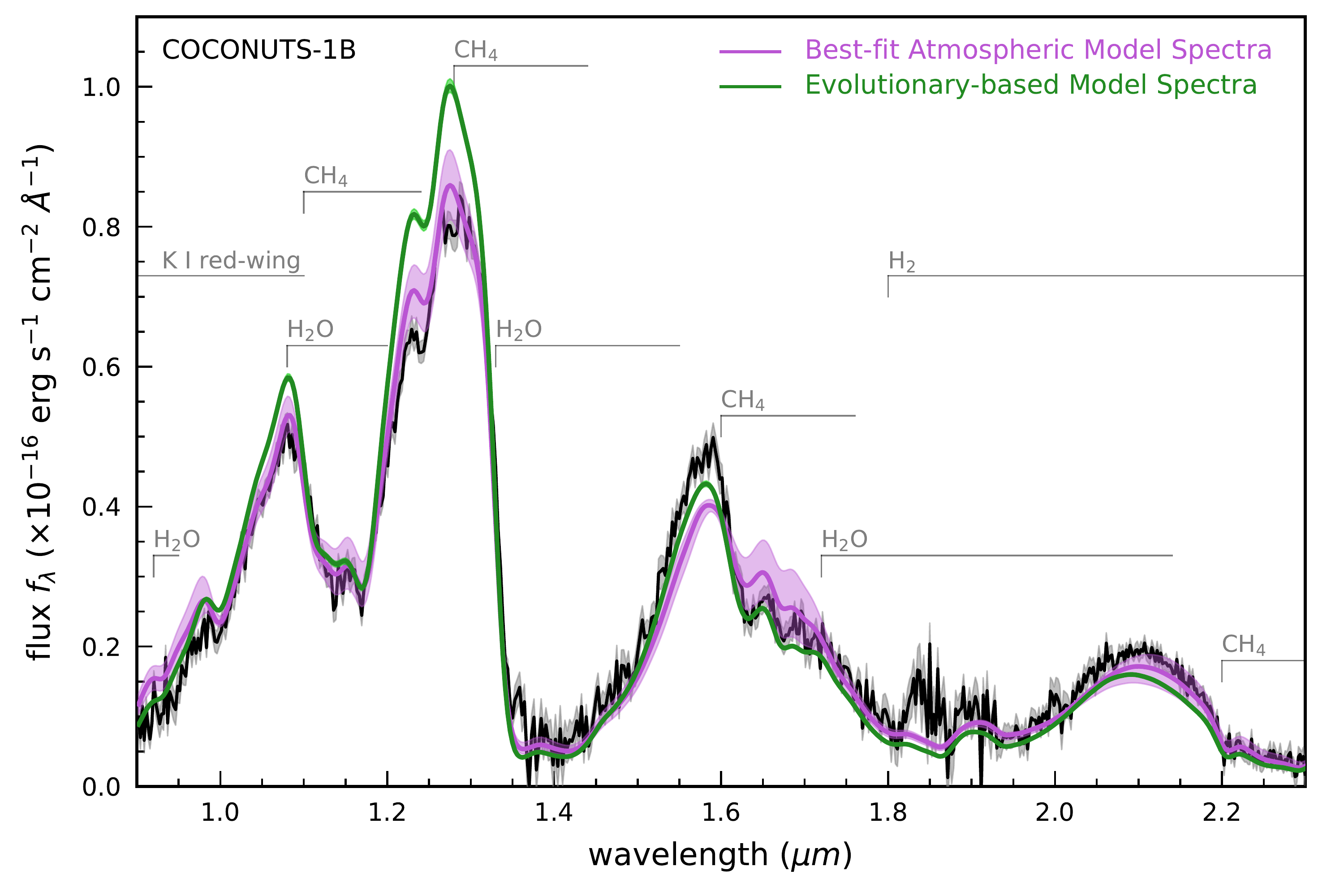}
\caption{The atmospheric model spectra interpolated at the parameters derived from the evolutionary model analysis of COCONUTS-1B (green) with $1\sigma$ uncertainties (green shadows). Overlaid are the observed spectra of COCONUTS-1B (black) and the Sonora atmospheric model spectra interpolated from the posteriors shown in Figure~\ref{fig:atm_evo_posteriors}, using the same format as in Figure~\ref{fig:atm_spec}.  }
\label{fig:atm_evo_NIRspec}
\end{center}
\end{figure*}

Here we explore the possible shortcomings of the cloudless Sonora model atmospheres. We interpolate and scale the Sonora atmospheric model spectra using the $\{T_{\rm eff}, \log\ g, R\}$ values inferred from the evolutionary model analysis and the white dwarf's distance. Then we compare these evolutionary-based model spectra to the best-fit atmospheric model spectra, as well as the observed data. As shown in Figure~\ref{fig:atm_evo_NIRspec}, the evolutionary-based model spectra do not match the observed spectrum, as the former are much bluer and have higher flux in $Y$ and $J$ band. In these cloudless models, the flux at $Y$- and $J$-band peaks emerges from $\approx 30$~bar where the atmospheric temperature is about $1900$~K (with the $J$ band probing a few bars deeper and $\approx 50$~K warmer than the $Y$ band). Most of the large peak flux differences arises because of the great wavelength sensitivity of the Planck function on the Wien tail.

Two different opacity sources could be responsible for the mismatch between the evolutionary-based model spectra and the data. First, silicate clouds condense at somewhat lower temperatures \citep[e.g., Figure 3 of][]{2013cctp.book..367M} and thus can be expected to limit the depth from which flux emerges, conceptually shaving off the peaks of $Y$ and $J$ bands \citep[e.g.,][]{2012ApJ...756..172M}, which might result in the evolutionary-based model spectra that better match with data. Such a comparison suggests that the silicate clouds still influence the emitted flux at $1300$~K, and cloudy models should provide a more accurate interpretation of the observed spectrum.

The second opacity which could play a role in the mismatch is the highly pressure broadened wings of the K resonance lines at $0.77\,\rm \mu m$ \citep{2000ApJ...531..438B}. The Sonora grid uses a line shape theory by \citet{2007A&A...465.1085A} which is valid to molecular hydrogen densities up to $10^{20}\,\rm cm^{-3}$.  These authors have recently developed a newer theory \citep[][]{2016A&A...589A..21A}, superseding their previous work, valid to higher densities. At 30~bar and 1900~K, the molecular hydrogen densities in the atmosphere of our companion are just above the upper limit of the previous \cite{2007A&A...465.1085A} theory and the strength of the broadened wing is uncertain, particularly in $Y$ band.

\begin{figure*}[t]
\begin{center}
\includegraphics[height=4.5in]{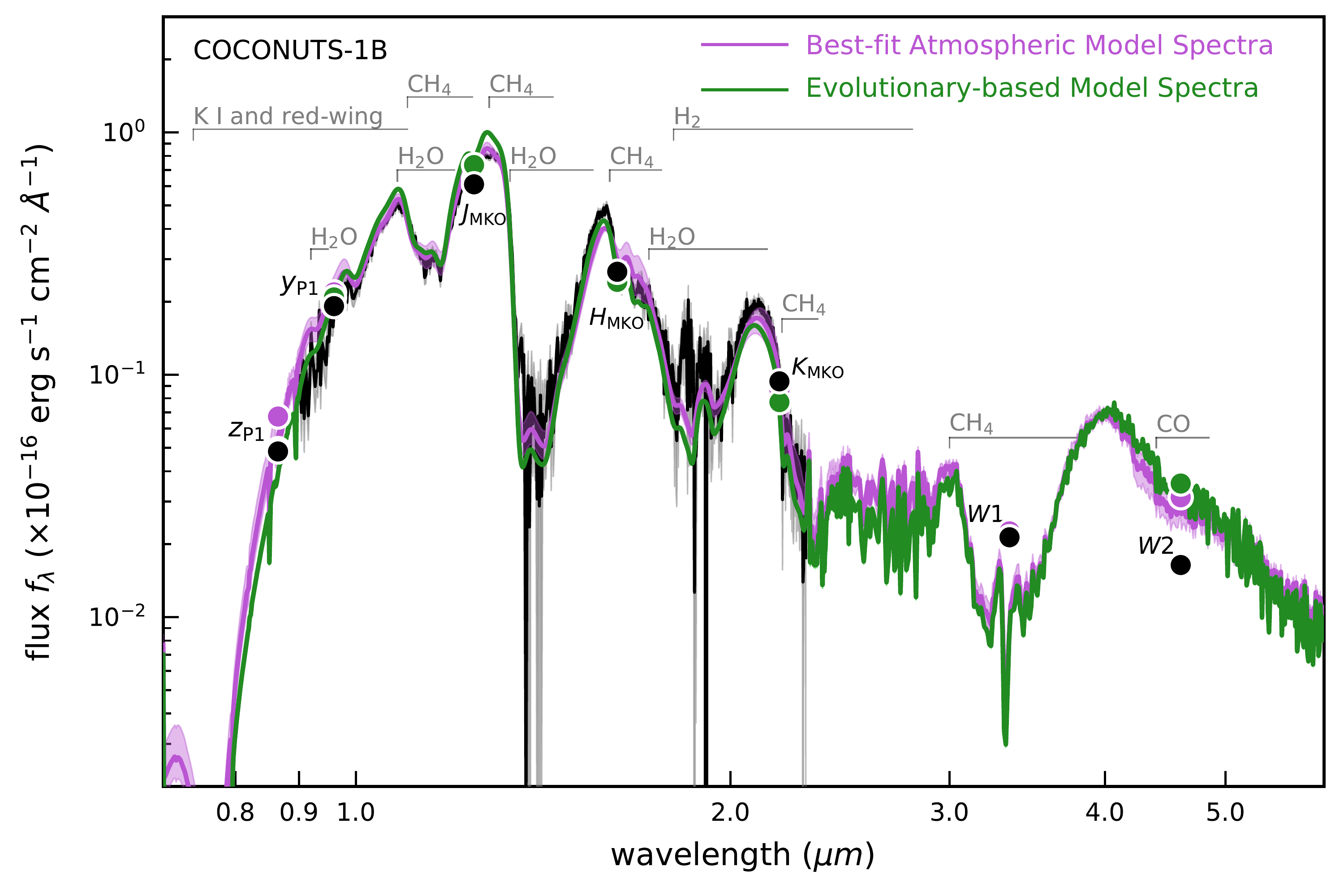}
\caption{The evolutionary-based model spectra (green), the best-fit atmospheric model spectra (purple), and the observed spectra (black), using the same format as in Figure~\ref{fig:atm_evo_NIRspec}. Circles mark the fluxes derived from the observed $z_{\rm P1}$, $y_{\rm P1}$, $J_{\rm MKO}$, $W1$, and $W2$ photometry (black), synthesized $H_{\rm MKO}$ and $K_{\rm MKO}$ photometry from the observed spectrum (black), and those synthesized from model spectra (purple and green). }
\label{fig:atm_evo_fullwlspec}
\end{center}
\end{figure*}

\begin{figure*}[t]
\begin{center}
\includegraphics[height=7.in]{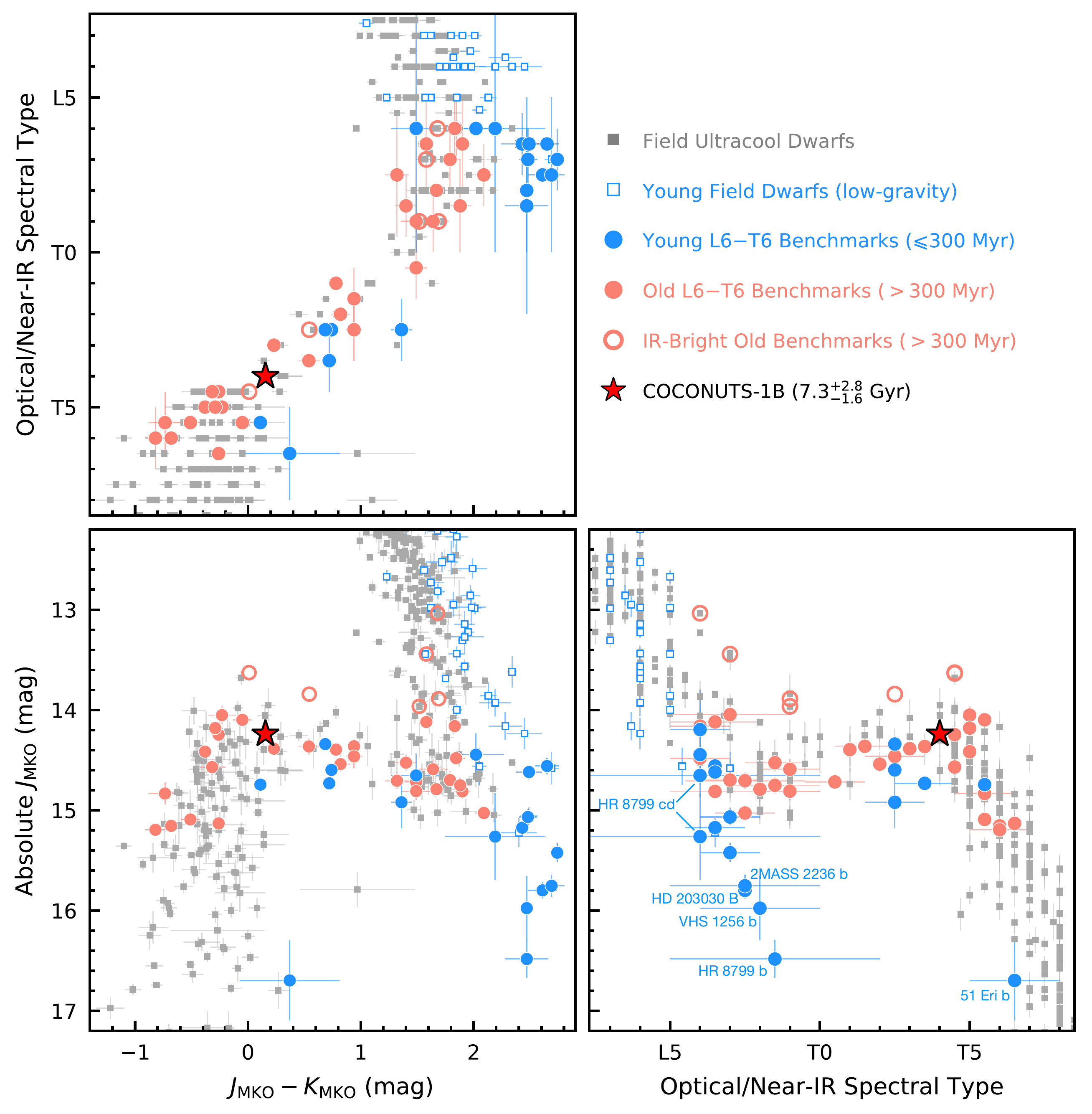}
\caption{Infrared properties of L/T transition benchmarks, including our newly discovered high-gravity T4 companion COCONUTS-1B (red), young L6$-$T6 objects with ages $\leqslant 300$~Myr (blue solid circles), seven infrared-bright old benchmarks (orange open circles; Section~\ref{subsec:suspicious}), and the normal old benchmarks ($>300$~Myr; orange solid circles). We use grey squares to show known field dwarfs (Best et al. submitted) which have infrared absolute magnitudes with S/N $> 5$ and are not young, binaries, or subdwarfs. We use blue open squares to show young moving group members or low-gravity field dwarfs. The upper left and lower right panel show the objects' absolute $J_{\rm MKO}$ magnitudes and $J_{\rm MKO}-K_{\rm MKO}$ colors as a function of spectral type, and the lower left panel is the corresponding color-magnitude diagram. We do not show error bars for spectral types if they are smaller than 1 subtype. Young L6$-$T6 objects have fainter infrared absolute magnitudes and redder colors than their older counterparts, although such differences between the two populations appear greater for late-L dwarfs than early-/mid-T dwarfs. In addition, young objects become brighter in $J$ band by $\approx 1.5$~mag as they evolve from late-L to T types, significantly larger than the $\approx 0.5$~mag $J$-band brightening seen in field L/T objects \citep[][]{2012ApJS..201...19D}. }
\label{fig:CMD_LT}
\end{center}
\end{figure*}

In Figure~\ref{fig:atm_evo_fullwlspec}, we plot both the atmospheric and evolutionary-based model spectra to a wider wavelength range of $0.7-6.0$~$\mu$m and compare with the companion's broadband photometry. We synthesize broadband fluxes from the models and compute fluxes from the companion's observed $z_{\rm P1}$, $y_{\rm P1}$, $J_{\rm MKO}$, $W1$, and $W2$ magnitudes, using the filter responses and zero-point fluxes from \cite{2012ApJ...750...99T} for Pan-STARRS1, \cite{2006MNRAS.367..454H} and \cite{2007MNRAS.379.1599L} for UKIDSS Hemisphere Survey, and \cite{2011ApJ...735..112J} for WISE. We also synthesize the $H_{\rm MKO}$ and $K_{\rm MKO}$ photometry of COCONUTS-1B using $J_{\rm MKO}$ from the observed near-infrared spectrum and compare to the synthesized values from models. We note that broadband photometry are consistent with the atmospheric model spectra and/or the evolutionary-based model spectra, except for W2, where the observed magnitude is much fainter than the model spectra. Such a mismatch around $4.6$~$\mu$m can be caused by the absorption from CO when it is dredged up from deeper, warmer layers of atmospheres into the photosphere through non-equilibrium mixing \citep[e.g.,][]{2002ApJ...568..335M, 2009ApJ...702..154S, 2014ApJ...797...41Z}. Therefore, non-equilibrium models are probably needed to better understand the atmosphere processes of COCONUTS-1B.

\section{The Surface-Gravity Dependence of the L/T Transition}
\label{sec:LT}

Spectrophotometric observations of benchmark planets and brown dwarfs have suggested a surface-gravity dependence of the L/T transition. Younger, lower-gravity L/T dwarfs tend to have fainter, redder near-infrared photometry \cite[e.g.,][]{2008Sci...322.1348M, 2011ApJ...733...65B, 2016ApJS..225...10F, 2016ApJ...833...96L}, higher variability amplitudes \citep[e.g.,][]{2015ApJ...813L..23B, 2017ApJ...841L...1G, 2019MNRAS.483..480V}, and cooler effective temperatures \citep[e.g.,][]{2006ApJ...651.1166M, 2007ApJ...654..570L, 2009ApJ...699..168D, 2013ApJ...774...55B, 2013ApJ...777L..20L, 2018ApJ...854L..27G} compared to their older, higher-gravity counterparts. Such gravity dependence is significant especially for the transition L~dwarfs (L6$-$L9). Less is known for the transition T~dwarfs (T0$-$T5) due to the small number of such objects with low surface gravities discovered thus far, but it appears that the gravity dependence is less pronounced than for L~dwarfs \citep[e.g.,][]{2015ApJ...810..158F, 2015ApJ...808L..20G, 2016ApJS..225...10F, 2016ApJ...833...96L}. 

To better understand the role that gravity plays in the L/T transition, a large sample of benchmarks spanning a wide range in surface gravities (or equivalently ages) is essential. Since the most recent large photometric analyses \citep[e.g.,][]{2016ApJS..225...10F, 2016ApJ...833...96L}, the census of planetary and substellar benchmarks has been expanded, as (1) new young moving group members and companions to stars or white dwarfs have been discovered, whose ages can be determined from their host associations or primary stars \citep[e.g., COCONUTS-1 in this work;][]{2017AJ....153...18B, 2017MNRAS.467.1126D, 2017ApJ...841L...1G, 2018ApJ...854L..27G, 2019MNRAS.487.1149G}, and (2) more substellar binaries and companions have measured dynamical masses thanks to ongoing astrometric monitoring programs \citep[e.g.,][]{2017ApJS..231...15D, 2019AJ....158..174D}. In addition, the current census of benchmarks can now have very precise absolute magnitudes and physical properties (e.g., bolometric luminosities and effective temperatures), as a result of the high-precision parallaxes from {\it Gaia}~DR2 \citep[][]{2016AandA...595A...1G, 2018AandA...616A...1G}. 

\subsection{A Sample of L/T Transition Benchmarks}
\label{subsec:LT_benchmarks}

We have combined COCONUTS-1B with all known L6$-$T6 planetary and substellar benchmarks (Table~\ref{tab:benchmarks_sptphotparage}). Our sample contains 9 free-floating members of stellar associations (nearby young moving groups, the Pleiades, and the Hyades), 38 single/binary companions to stars or white dwarfs, and 8 field substellar binaries, leading to a total of 60 objects in 50 systems. The ages of free-floating objects and companions to stars or white dwarfs in the literature have been determined from their host associations or primary stars (except for WISE~J$072003.20-084651.2$B [WISE~0720B]; \citealt{2015AJ....150..180B, 2019AJ....158..174D}). The ages of field substellar binaries and WISE~0720B in the literature have been determined from evolutionary models using their measured dynamical masses and bolometric luminosities. In order to investigate the impact of surface gravity on L/T transition properties, we divide our sample into ``young'' (20 objects) and ``old'' (40 objects) subsets using a dividing line of $300$~Myr. After $\sim 300$~Myr, the radii of ultracool dwarfs contract notably slowly, weakening the correlation between surface gravities and ages \citep[e.g.,][]{2001RvMP...73..719B, 2008ApJ...689.1295K, 2013ApJ...772...79A}.

\subsubsection{Bolometric Luminosities, Effective Temperatures, Surface Gravities, and Masses}
\label{subsec:LbolTeff}

Table~\ref{tab:benchmarks_lbolteff} collects bolometric luminosities and effective temperatures for our sample from the literature. Bolometric luminosities of objects were computed mainly from three approaches:\footnote{The only object in our sample whose $L_{\rm bol}$ was not derived from these three approaches is CFBDS~J$111807-064016$, for which \cite{2014AandA...561A..66R} interpolated the \cite{2003AandA...402..701B} evolutionary models using the age of the primary star and the object's $T_{\rm eff}$ derived from atmospheric models \citep[][]{2012RSPTA.370.2765A, 2013MSAIS..24..128A} to compute the $L_{\rm bol}$.} (1) integrating spectral energy distributions (SEDs), (2) applying empirical relations between bolometric luminosities and infrared absolute magnitudes \citep[e.g.,][]{2017ApJS..231...15D}, or (3) applying bolometric corrections based on the objects' spectral types to infrared absolute magnitudes \citep[e.g.,][]{2004AJ....127.3516G, 2010ApJ...722..311L, 2015ApJ...810..158F}. These methods usually produce consistent results, with the first method having the best precision. Results from the third method can have large uncertainties because spectral types of some L/T dwarfs (e.g., 2MASS~J$22362452+4751425$~b [2MASS~$2236$~b] in \citealt{2017AJ....153...18B}; the HR~8799 planets in \citealt{2018AJ....155..226G}) are not well-determined.
 
For objects whose literature $L_{\rm bol}$ values are either lacking or derived from bolometric corrections (method~3), we have (re-)computed their $L_{\rm bol}$ using the empirical relations between $L_{\rm bol}$ and $H$- or $K$-band absolute magnitudes from \cite{2017ApJS..231...15D}. These relations are based on field dwarfs with ages $\gtrsim 0.5$~Gyr from \cite{2015ApJ...810..158F}, who derived the objects' bolometric luminosities from a uniform SED analysis. We use the objects' $H$-band absolute magnitudes to compute their $L_{\rm bol}$ and switch to $K$ band when their $H$-band absolute magnitudes are fainter than $13.3$~mag, as suggested by \cite{2017ApJS..231...15D}. 

HIP~65426~b \citep[L$6\pm1$;][]{2017AandA...605L...9C}, VHS~J$125601.92-125723.9$~b \citep[VHS~$1256$~b; L$8\pm2$;][]{2015ApJ...804...96G}, LTT~7251B \citep[L$7\pm2$;][]{2018MNRAS.474.1826S}, and HIP~70849B \citep[T$4.5\pm0.5$;][]{2014AandA...569A.120L} are four exceptions to our re-calculation. HIP~65426~b has a too young age \citep[$14 \pm 4$~Myr;][]{2017AandA...605L...9C} for the \cite{2017ApJS..231...15D} relations to be applied ($\geqslant 0.5$~Gyr). \cite{2017AandA...605L...9C} derived $L_{\rm bol}$ of HIP~65426~b by using the {\it Gaia}~DR1 parallax ($8.98 \pm 0.30$~mas) of the primary and a bolometric correction from four dusty L5--L7.5 dwarfs \citep{2015ApJ...810..158F}. Here we update the $L_{\rm bol}$ of HIP~65426~b by using the {\it Gaia}~DR2 parallax ($9.16 \pm 0.06$~mas) of the primary star, the companion's estimated $J_{\rm 2MASS}$ magnitude \citep[$19.50 \pm 0.40$~mag; computed from the primary star's $J_{\rm 2MASS}$ and the photometric contrast in][]{2017AandA...605L...9C}, and the $J$-band bolometric correction for young L$6$ dwarfs from \cite{2015ApJ...810..158F}. 

VHS~$1256$~b also has a very young age \citep[$0.15-0.3$~Gyr;][]{2015ApJ...804...96G} that prevents the application of the \cite{2017ApJS..231...15D} relations. \cite{2015ApJ...804...96G} derived $L_{\rm bol}$ of VHS~$1256$~b using their own preliminary parallax of $78.8 \pm 6.4$~mas ($12.7 \pm 1.0$~pc) and the bolometric correction for PSO~J$318.5338-22.8603$ \citep[][]{2013ApJ...777L..20L}. After the discovery of VHS~$1256$~b, \cite{2016ApJ...818L..12S} found the primary is a $0.1''$-wide, equal-brightness binary, with a spectrophotometric distance of $17.1 \pm 2.5$~pc. Such a distance is in tension with the \cite{2015ApJ...804...96G} parallactic distance, but is more consistent with the parallax measured by the Hawaii Infrared Parallax Program (T. Dupuy, private communication). We thus adopt a parallax of $58.5 \pm 8.6$~mas based on the \cite{2016ApJ...818L..12S} spectrophotometric distance. We then use the object's measured $K_{\rm 2MASS}$ magnitude \citep[$14.57 \pm 0.12$~mag][]{2003yCat.2246....0C} and the $K$-band bolometric correction for young L8 dwarfs from \cite{2015ApJ...810..158F} to update the $L_{\rm bol}$ of VHS~$1256$~b. 

LTT~7251B is a L7 companion to a G8 dwarf star found by \cite{2018MNRAS.474.1826S} using their infrared astrometric catalog of the VISTA Variables in the Via Lactea (VVV) survey \citep[][]{2010NewA...15..433M, 2015A&A...575A..25S}. Its $H$-band absolute magnitude is too faint to use the \cite{2017ApJS..231...15D} relations and has a $K$-band magnitude from VVV, which is in neither the 2MASS nor MKO photometric systems \citep[e.g.,][]{2018MNRAS.474.5459G}. We therefore compute the companion's $L_{\rm bol}$ from its estimated $J_{\rm 2MASS}$ magnitude ($17.02 \pm 0.13$; converted from its $J_{\rm MKO}$ based on known L5-L9 dwarfs in Best et al. submitted) and the $J$-band bolometric correction for field L7 dwarfs from \cite{2015ApJ...810..158F}.

HIP~70849B is a T4.5 companion to a K7 dwarf star found by \cite{2014AandA...569A.120L}. The primary star also hosts a close-in ($P = 5-90$~years) giant planet ($M\ \sin\ i = 3-15$~M$_{\rm Jup}$) detected via radial velocity \citep[][]{2011AandA...535A..54S}. Similar to LTT~7251B, HIP~70849B has a too faint $H$-band absolute magnitude and a $K$-band magnitude from the VISTA Hemisphere Survey \citep[VHS;][]{2013Msngr.154...35M}. We therefore compute the object's $L_{\rm bol}$ by using its measured $J_{\rm 2MASS}$ magnitude \citep[$15.89 \pm 0.07$;][]{2003yCat.2246....0C} and the $J$-band bolometric correction for field T4.5 dwarfs from \cite{2015ApJ...810..158F}. 

Effective temperatures of objects have been computed in the literature mainly from two approaches:\footnote{Objects in our sample whose $T_{\rm eff}$ were not derived from these two approaches are PHL~$5038$B \citep[][]{2009AandA...500.1207S}, LSPM~J$1459+0857$B \citep[][]{2011MNRAS.410..705D}, WISE~J$004701.06+680352.1$ \citep[][]{2012AJ....144...94G, 2015ApJ...799..203G}, and Luhman~16AB \citep[][]{2013ApJ...767L...1L}. $T_{\rm eff}$ of the first two objects were converted from spectral types based on temperature scales \citep{2004AJ....127.3516G, 2004AJ....127.2948V}, and the $T_{\rm eff}$ of the latter two were estimated from the measured $L_{\rm bol}$ by assuming a radii from evolutionary models \citep[e.g.,][]{1997ApJ...491..856B, 1998AandA...337..403B, 2000ApJ...542..464C}. } (1) fitting the objects' near-infrared spectra using atmospheric models, or (2) interpolating hot-start evolutionary models using the objects' $L_{\rm bol}$ and independent ages or dynamical masses. Ideally, the two methods should produce consistent results, although the evolutionary model parameters are usually more robust and less vulnerable to well-noted shortcomings of atmospheric models (Section~\ref{subsec:benchmarking}). In addition, three groups of evolutionary models have been used in literature: (1) the \cite{2008ApJ...689.1327S} models and Sonora models (\citealt{2017AAS...23031507M}; Marley et al. 2019 in prep), (2) the \cite{1997ApJ...491..856B} models, and (3) the Lyon group's models \citep[][]{2003AandA...402..701B, 2015AandA...577A..42B}. These three sets of models have made different assumptions about initial interior composition, atmosphere boundary conditions, and electron conduction, but are generally consistent \citep[][]{2008ApJ...689.1327S}.

Similar to $L_{\rm bol}$, we have (re-)computed $T_{\rm eff}$ for most of our sample. We have also computed the objects' surface gravities and masses. We adopt the existing effective temperatures, surface gravities, and dynamical masses for the substellar binary components and WISE~0720B from \cite{2017ApJS..231...15D} and \cite{2019AJ....158..174D}, respectively. The $T_{\rm eff}$ and $\log\ g$ values of these objects were determined from interpolating the \cite{2008ApJ...689.1327S} hybrid evolutionary models using their measured $L_{\rm bol}$ and dynamical masses. For the remaining objects, we compute $T_{\rm eff}$, $\log\ g$, and $M$ by interpolating the same \cite{2008ApJ...689.1327S} hybrid models using the objects' $L_{\rm bol}$ and ages. We assume the objects' bolometric luminosities follow a normal distribution and assume their ages follow a uniform distribution (if the object's age is given as a range in the literature) or a Gaussian distribution (if the age is given with an error bar in the literature)\footnote{For ages with asymmetric errors, we assume a distribution composed of two half-Gaussians, with upper and lower uncertainties corresponding to standard deviations of the two Gaussians, similar to our Section~\ref{subsec:evo}.} truncated within $0-10$~Gyr. The evolutionary model parameters for COCONUTS-1B derived from the \cite{2008ApJ...689.1327S} hybrid models are consistent with our analysis using the cloudless Sonora models (Section~\ref{subsec:evo}), so we keep the Sonora-based results. For 51~Eri~b, we also compute physical parameters using the \cite{2008ApJ...683.1104F} cold-start evolutionary models \citep[also see][]{2017AJ....154...10R}, which assume objects are formed from core accretion with low initial entropy, whereas all the afore-mentioned hot-start models assume objects are formed with high initial entropy without any subsequent accretion. Finally, we note the young objects ($\leqslant 300$~Myr) in our sample have $\log\ g = 3.5-4.6$~dex and $M = 2-20$~M$_{\rm Jup}$, and the old ones ($>300$~Myr) have higher gravities and masses of $\log\ g = 4.8-5.5$~dex and $M=25-80$~M$_{\rm Jup}$.

\subsubsection{Infrared-Bright Old Benchmarks}
\label{subsec:suspicious}

Seven old companions ($>300$~Myr) in our sample have notably brighter ($\approx 0.3-1.0$~mag) infrared absolute magnitudes than field dwarfs with similar spectral types. We checked that these companions' proper motions are consistent with the {\it Gaia}~DR2 proper motions of their primary stars, so their companionship remain secure. Based on the literature, the spectra of these companions are mostly normal when compared to spectral standards, except for 2MASS~J$00150206+2959323$ \citep[2MASS~J$0015+2959$;][]{2014ApJ...792..119D}. This object has a peculiar near-infrared spectrum \citep[L7.5~pec;][]{2010ApJS..190..100K} with a blue $J_{\rm MKO}-K_{\rm MKO} = 1.58 \pm 0.07$~mag, about $0.4$~mag bluer than typical field L7 dwarfs (e.g., Best et al. submitted). However, this object's $JHK$-band absolute magnitudes are all significantly brighter than other field dwarfs, suggesting its blue color is not the only peculiarity .

Among these 7 infrared-bright objects, SDSS~J$213154.43-011939.3$ (SDSS~J$2131-0119$) is a L9 companion to a $0.64''$ M3$+$M6 binary NLTT~51469AB at a projected separation of $82''$ recently found by \cite{2019MNRAS.487.1149G}. They noted the {\it Gaia}~DR2 parallactic distance ($46.6 \pm 1.3$~pc) and the spectrophotometric distance ($34^{+10}_{-13}$~pc) of the primary stellar binary are only marginally consistent. One explanation is that the {\it Gaia}~DR2 astrometry of the primary might be affected by the orbital motion of the binary as it is unresolved by {\it Gaia} \citep[also see][]{2018AJ....156...57D}. \citeauthor{2019MNRAS.487.1149G} also noted the {\it Gaia}~DR2 astrometry of NLTT~51469AB possess an astrometric excess noise of $2.74$~mas with a very high significance ($2202~\sigma$) and was determined from only 8 independent groups of {\it Gaia} observations (i.e., visibility\_periods\_used = 8), whereas a larger number (e.g., $\geqslant 10$) would result in more reliable astrometry \citep{2018A&A...616A...2L}. We find that NLTT~51469~AB has a very large RUWE of $12$, indicating that the {\it Gaia}~DR2 astrometry of the primary can not be well-explained by a single-star model. In addition, the spectrophotometric distance of the L9 companion SDSS~J$2131-0119$ is $\approx 40 \pm 11$~pc, consistent with both parallactic and spectrophotometric distance of the primary star, making the true distance of this co-moving system unclear. If we adopt the spectrophotometric distance of the primary, then the companion's near-infrared absolute magnitudes are consistent with field L9 dwarfs. 

We find that the primary stars of the remaining 6 infrared-bright companions all have consistent {\it Gaia}~DR2 parallactic distances and spectrophotometric distances, both of which are farther than the companions' spectrophotometric distances, suggesting that these companions are either unresolved binaries/multiples associated with their primary stars or foreground interlopers. High-precision parallaxes for both companions and primary stars in all 7 systems are therefore needed to verify if they are located at the same distance and thus physically associated. Also, high spatial-resolution imaging would be helpful to examine the binarity of these L/T dwarfs. In our subsequent analysis, we simply adopt the primary stars' {\it Gaia}~DR2 distances for these 7 objects.

\subsection{Discussion}
\label{subsec:LT_discussion}

\begin{figure*}[t]
\begin{center}
\includegraphics[height=7.in]{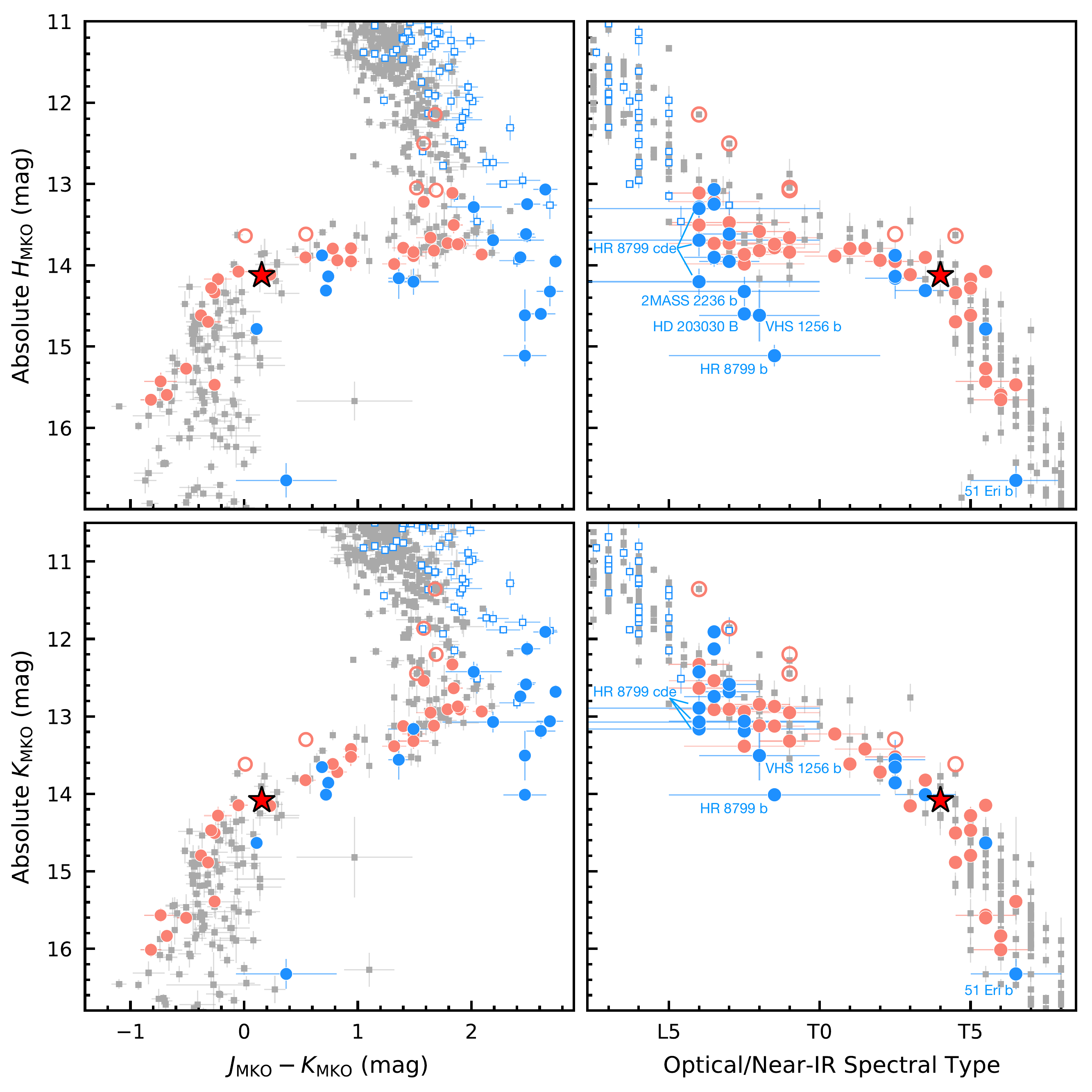}
\caption{Same format as Figure~\ref{fig:CMD_LT}, but with absolute $H_{\rm MKO}$ and $K_{\rm MKO}$ magnitudes. Combined with Figure~\ref{fig:CMD_LT}, these figures show that the difference in infrared absolute magnitudes between young and old populations is a function of wavelength, being most pronounced in $J$ band and decreasing for $H$ and $K$ bands. In addition, young objects show the $\approx 0.6$~mag $H$-band brightening as they evolve from late-L to T types, and the evolution in $K$-band photometry between young and old L/T dwarfs are nearly identical as a function of spectral type. }
\label{fig:CMD_LT_HK}
\end{center}
\end{figure*}

\begin{figure*}[t]
\begin{center}
\includegraphics[height=3.5in]{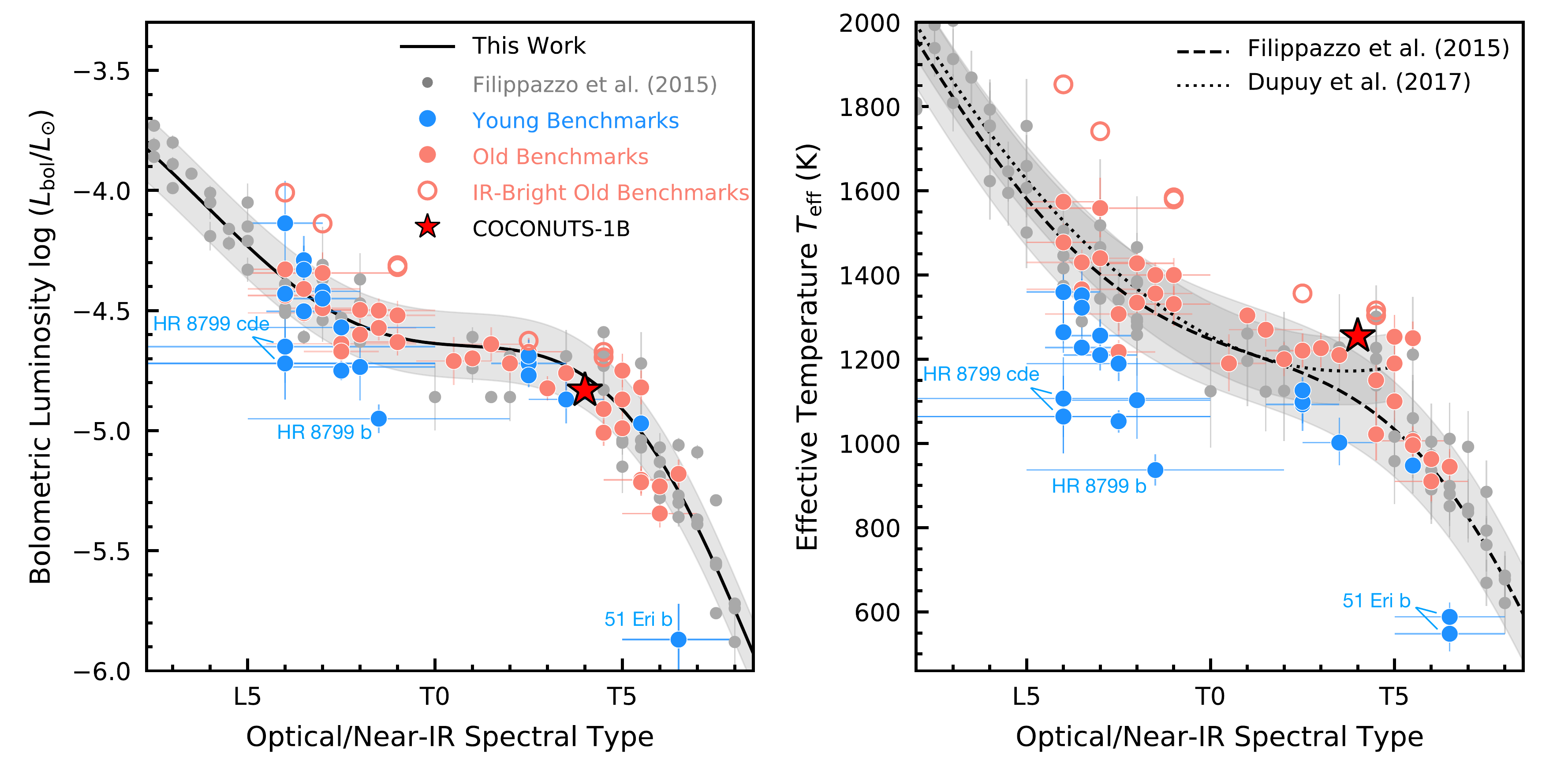}
\caption{Bolometric luminosities and effective temperatures of our sample, overlaid with field dwarfs (grey circles) from \cite{2015ApJ...810..158F}. In the left panel, we also overlay the polynomial of the $L_{\rm bol}$ vs. SpT relation derived in this work, using the \cite{2015ApJ...810..158F} sample (see our footnote~\ref{footnote:F15} for details), and in the right panel, we show the polynomial relation from both \cite{2015ApJ...810..158F} and \cite{2017ApJS..231...15D}. While luminosities of young objects are consistent with or only slightly fainter than the old objects at same spectral types, the former has cooler temperatures across the entire L/T transition, reinforcing that the L/T transition is gravity dependent. The young benchmark sample is very sparse near T0, primarily because of their rarity \citep[e.g.,][]{2013MNRAS.430.1171D, 2015MNRAS.449.3651M}. }
\label{fig:LbolTeff_LT}
\end{center}
\end{figure*}

We investigate the photometric and physical properties of L/T benchmarks with different ages (surface gravities) in Figures~\ref{fig:CMD_LT}, \ref{fig:CMD_LT_HK}, and \ref{fig:LbolTeff_LT}. As is well-known, going from L6 to L9 spectral types, young benchmark objects ($\leqslant 300$~Myr) become gradually fainter in $JHK$ absolute magnitudes and redder in $J_{\rm MKO}-K_{\rm MKO}$ (by $0.5-1.0$~mag) than old objects ($> 300$~Myr), constituting a natural photometric extension of earlier-type young moving group members and low-gravity field dwarfs \citep[e.g.,][]{2016ApJS..225...10F, 2016ApJ...833...96L}. The difference in near-infrared absolute magnitudes of the four young L7.5$-$L9 dwarfs (2MASS~$2236$~b, HD~203030~B, VHS~$1256$~b, and HR~8799~b) compared to the old benchmarks of same spectral types (excluding infrared-bright old benchmarks discussed in Section~\ref{subsec:suspicious}) is a function of wavelengths, as has been already noted \citep[e.g.,][]{2013AJ....145....2F, 2016ApJS..225...10F, 2016ApJ...833...96L}. Compared to field objects, the absolute magnitudes of young L7.5$-$L9 dwarfs are fainter by $0.8-2.0$~mag in $J_{\rm MKO}$, and such magnitude difference between young and old populations decreases to $0.3-1.5$~mag in $H_{\rm MKO}$, and finally become indistinguishable or only slight of $\lesssim 1.0$~mag in $K_{\rm MKO}$, suggesting that condensate clouds alter the spectral energy distributions of young, low-gravity late-L dwarfs more significantly than old, high-gravity objects \citep[e.g.,][]{2012ApJ...754..135M, 2013AJ....146..161M, 2016ApJ...830...96H}. 

Excluding the four HR~8799 planets, we find bolometric luminosities of young L6$-$L9 dwarfs (Figure~\ref{fig:LbolTeff_LT}) are consistent with or only slightly fainter (by $<0.1$~dex) than old objects \citep[e.g.,][]{2015ApJ...810..158F, 2016ApJS..225...10F}. While HR~8799~bcde have fainter $L_{\rm bol}$ by $\approx 0.3-0.5$~dex than their older counterparts, their spectral types are only loosely determined \cite[e.g.,][]{2010ApJ...723..850B, 2018AJ....155..226G} and thus their positions in Figure~\ref{fig:LbolTeff_LT} might be shifted horizontally to be more or less consistent with the old population. In addition, young L6$-$L9 dwarfs span a wide range in effective temperatures (mostly $1100-1300$~K, with the HR~8799~b of $\approx 950$~K), significantly cooler than old objects ($1300-1600$~K) with similar spectral types, and comparable to the temperatures of old T0$-$T6 dwarfs. This reinforces that the L/T transition occurs at much cooler temperatures at low surface gravities \citep[e.g.,][]{2006ApJ...651.1166M, 2007ApJ...654..570L, 2009ApJ...699..168D, 2011ApJ...733...65B, 2012ApJ...754..135M, 2013ApJ...774...55B, 2013ApJ...777L..20L, 2018ApJ...854L..27G}.

Young T0$-$T5 dwarfs still possess redder $J_{\rm MKO}-K_{\rm MKO}$ colors (by $0.2-0.5$~mag) than old objects, but their near-infrared absolute magnitudes are more similar to old objects than is the case for the late L types. Compared to field objects, the 5 young T2$-$T5 dwarfs (2MASS~J$13243553+6358281$, HN~Peg~B, SIMP~J$013656.5+093347.3$, GU~Psc~b, and SDSS~J$111010.01+011613.1$) have fainter absolute magnitudes by $\lesssim 0.5$~mag in $J_{\rm MKO}$ and $\lesssim 0.2$~mag in $H_{\rm MKO}$, and similar magnitudes in $K_{\rm MKO}$, indicating that the redder $J-K$ colors of young T dwarfs are largely due to their fainter $J$-band absolute magnitudes. Therefore, our sample suggests that infrared absolute magnitudes of low-gravity dwarfs are only slightly fainter than (if not consistent with) high-gravity dwarfs for T0$-$T5 spectral types (or equivalently with $J_{\rm MKO} - K_{\rm MKO}$ from $1.5$~mag to $-0.5$~mag). This behavior is the opposite of the \cite{2008ApJ...689.1327S} hybrid evolutionary models, which predict young early-T dwarfs to be brighter than old objects. The \cite{2012ApJ...754..135M} evolutionary models added a gravity dependence in the L/T transition to their group's models when analyzing the HR~8799 planets and suggested the $K$-band absolute magnitudes become fainter toward lower surface gravities. However, our sample finds such a $K$-band magnitude difference is very subtle between young and old objects. In addition, we note the color-magnitude diagram locus of young and old T dwarfs are mostly similar or only slightly different (as suggested by \citealt{2016ApJ...833...96L} with a smaller sample of objects), again in contrast to the \cite{2012ApJ...754..135M} models.

Unlike late-L dwarfs, bolometric luminosities of young T0$-$T5 benchmarks are more similar to the old benchmarks and field population\footnote{In Figure~\ref{fig:LbolTeff_LT}, we compare bolometric luminosities of our L/T benchmarks with the field-age dwarfs in \cite{2015ApJ...810..158F}, who also provided a polynomial describing the $L_{\rm bol}$~vs.~SpT relation for field dwarfs. We note the precision of the polynomial coefficients reported in their Table~10 produce an offset between their sample and the constructed polynomial. Such offset is insignificant in L types, but the polynomial is systematically brighter compared to the data by $0.05$~dex at T0, $0.15$~dex at T5, and $0.33$~dex at T9. We have therefore used the same data from \cite{2015ApJ...810..158F} and performed a $6$th order polynomial fit for M6$-$T9, namely $L_{\rm bol} = \sum_{i=0}^{n=6}\ c_{i}\ ({\rm SpT})^{i}$, where SpT $= 6$ for M6, $=10$ for L0, etc. Our coefficients are $c_{0} = 1.355808e+01$, $c_{1} = -6.988722e+00$, $c_{2} = 1.173746e+00$, $c_{3} = -1.013430e-01$, $c_{4} = 4.653312e-03$, $c_{5} = -1.077500e-04$, $c_{6} = 9.858614e-07$, with a rms of $0.137$~dex. The round-off error from our coefficients is $0.001$~dex. \label{footnote:F15}} (Figure~\ref{fig:LbolTeff_LT}). As a consequence, effective temperatures of young objects are cooler by $\approx 100$~K than old objects, given that their radii are still in the process of contraction \citep[e.g.,][]{2001RvMP...73..719B}. In comparison, COCONUTS-1B has contracted to a very small radii ($0.79$~R$_{\rm Jup}$; Table~\ref{tab:atmevo}) given its very old age of $7.3^{+2.8}_{-1.6}$~Gyr, thereby resulting in a relatively high temperature for its spectral type. 

In addition, field dwarfs are known to exhibit the $J$-band brightening phenomenon, as their $J$-band absolute magnitudes become brighter by $\approx$0.5~mag when they evolve from late-L to mid-T types, and then resume with fainter absolute magnitudes for later spectral types \citep[e.g.,][]{2002AJ....124.1170D, 2003AJ....126..975T, 2004AJ....127.2948V, 2006ApJ...647.1393L, 2012ApJS..201...19D}. No such brightening is clearly seen at $H$~and $K$~bands for field objects. The phenomenon is likely a hallmark of cloud evolution in the L/T transition, with the dominant opacity source in 1.0--1.3~$\mu$m region being condensate opacity rather than gas opacity \citep[e.g.,][]{2001ApJ...556..872A, 2006ApJ...640.1063B}. Our benchmark sample shows that for young objects, the $J$-band brightening has a significantly larger amplitude ($\approx$1.5~mag) than the $\approx$0.5~mag seen in the old population. Moreover, young objects appear to also undergo a $\approx$1~mag brightening in $H$~band as they evolve through the L/T transition. This longer wavelength brightening is broadly consistent with the notion that the atmospheres of young L dwarfs have a lower sedimentation efficiency (e.g., \citealt{2012ApJ...754..135M}; see also Figure~22 of \citealt{2016ApJ...833...96L}), leading to a more vertically extended cloud (which apparently can influence the flux in both the $J$- and $H$-band molecular opacity windows) as well as smaller particle sizes (which leads to an increase in condensate opacity for a given mass of condensates). Systematic modeling of cloud opacity evolution will be required to tease out such effects.

As an alternative modeling approach of ultracool atmospheres, \cite{2016ApJ...817L..19T} have proposed the thermal-chemical instability to explain the spectrophotometric evolution of brown dwarfs across the L/T transition \citep[also the T/Y transition;][]{2015ApJ...804L..17T} without invoking condensate clouds. For late-L dwarfs, such instability triggers local compositional convection that drives the abundance of CO and CH$_{4}$ out of chemical equilibrium and reduces the temperature gradient in the atmosphere, leading to the objects' observed fainter and redder near-infrared photometry. For early-T dwarfs, such CO/CH$_{4}$-related instability dissipates, warming up the deep layers of the atmosphere and leading to the observed $J$-band brightening phenomenon. The \citeauthor{2016ApJ...817L..19T} models have been tested to reproduce properties of several L/T dwarfs of different temperatures and gravities \citep[e.g.,][]{2016ApJ...817L..19T, 2017ApJ...850...46T}. These models should be further tested to account for the gravity-dependent $J$-band brightening and the $H$-band brightening of young objects as seen from our sample of L6--T6 benchmarks.

In summary, we find that the $K$-band absolute magnitudes and $L_{\rm bol}$ among different ages are nearly identical as a function of spectral type, but the effective temperatures of young L/T dwarfs are notably cooler than old dwarfs. The difference in $J$- and $H$-band absolute magnitudes between young ($\leqslant 300$~Myr) and old ($> 300$~Myr) populations changes from being substantial in late-L dwarfs to being nearly negligible in early-T dwarfs, indicating that the magnitude of the gravity dependence becomes weaker with later spectral types across the L/T transition. Finally, the brightening of near-infrared magnitudes across the transition is larger in both amplitude and wavelength range for young objects as compared to old ones.

\section{Summary}
\label{sec:summary}

We have reported the first discovery from our COol Companions ON Ultrawide orbiTS (COCONUTS) program, a large-scale survey for wide-orbit planetary and substellar companions within 100~pc. We have discovered the co-moving system COCONUTS-1, located at $31.51$~pc and composed of a DA white dwarf primary and a T4 companion with a projected separation of $41''$ ($1280$~au). Our photometric analysis of the white dwarf suggest it has a cool effective temperature ($5115$~K) and low mass ($0.548$~M$_{\odot}$). Combining the white dwarf cooling age with its progenitor's main-sequence lifetime, we estimate the system's age as $7.3_{-1.6}^{+2.8}$~Gyr. 

The multi-wavelength photometry and near-infrared spectrum of COCONUTS-1B both support its common distance with the white dwarf. We estimate an extremely low contamination probability by field interlopers, further supporting the physical association between the T dwarf and the white dwarf. 

We fit the cloudless Sonora atmospheric models to the near-infrared spectrum of COCONUTS-1B to study its physical properties. The best-fit Sonora model spectra generally match the observed spectrum but mismatch occurs at the blue wing of the $Y$ band, which is sensitive to the potassium abundance and the pressure-broadened red wing of the potassium doublet. We note mismatches near the peaks of $J$ and $H$ bands suggest a deep cloud deck in the atmospheres, which is not included in our set of model atmospheres.

We also use the cloudless Sonora evolutionary models to estimate the companion's physical properties, based on the companion's bolometric luminosity and the system's age from the white dwarf. Both the atmospheric and evolutionary models predict consistent effective temperatures and radii, but the atmospheric models imply a much smaller surface gravity, and thus a very young age of $\approx 380$~Myr, in contradiction to the age of the white dwarf. The unphysically young age of the brown dwarf companion reflects shortcomings of the cloudless model atmospheres. We therefore adopt the evolutionary model parameters as the characteristics of the companion. The companion's $\log\ g$ from the evolutionary models ($5.44^{+0.02}_{-0.03}$~dex) is among the highest surface gravity that brown dwarfs can reach over the cosmic time.

In order to understand the shortcomings of the cloudless atmospheric models, we interpolate the model spectra at the $\{T_{\rm eff}, \log\ g\}$ values derived from the evolutionary models. The resulting model atmospheres have a relatively bluer near-infrared color and more emergent flux in $Y$ and $J$ bands as compared to the observed spectrum. Adding silicate clouds or adopting different potassium line profiles into the model atmospheres may help to relieve the discrepancies. We also note the observed W2-band flux of COCONUTS-1B is fainter than the model spectra, which might be explained by the non-equilibrium abundances of CO.

For planetary and substellar benchmarks, white dwarf primaries can provide among the most precise ages for companions. Thus far, only a handful binaries composed of white dwarfs and resolved substellar companions have been found, and COCONUTS-1 is among the oldest in this sample (Table~\ref{tab:known_WD}), thereby probing the high surface gravity regime of the L/T transition. 

In order to better understand the gravity dependence in the L/T transition, we have compiled all 60 known L6$-$T6 benchmarks, including members of nearby associations, companions to stars or white dwarfs, and substellar binary components. Many of these benchmarks now have precise parallaxes thanks to {\it Gaia}~DR2. We have also (re-)computed bolometric luminosities, effective temperatures, surface gravities, and masses for most of these L/T benchmarks for a more uniform comparison.

We find infrared photometry of young ($\leqslant 300$~Myr) late-L dwarfs are significantly fainter (by $0.8-2.0$~mag in $J$-band absolute magnitude) and redder (by $0.5-1.0$~mag in $J-K$) than their older ($>300$~Myr) counterparts, as has been already well-noted \citep[e.g.,][]{2016ApJS..225...10F, 2016ApJ...833...96L}. Also, young late-L's have consistent or only slightly fainter bolometric luminosities than old objects, with their effective temperatures $\approx 200-300$~K cooler than the latter at same spectral types.

Our large benchmark sample shows the gravity dependence appears weaker for T dwarfs. While young T0$-$T5 benchmarks still possess redder $J-K$ colors (by $0.2-0.5$~mag) than old objects, their near-infrared absolute magnitudes are more similar to field objects, especially in $H$ and $K$ bands, than is the case for the late-L dwarfs. The redder $J-K$ colors of young T dwarfs are largely due to their fainter $J$-band absolute magnitudes. The observed behavior is the opposite of the \cite{2008ApJ...689.1327S} hybrid evolutionary models, which predict young early-T dwarfs to be brighter than old objects. The \cite{2012ApJ...754..135M} evolutionary models with a gravity-dependent L/T transition suggest $K$-band absolute magnitudes become notably fainter toward lower surface gravities, but our sample finds such $K$-band magnitude difference is subtle between young and old objects. In addition, bolometric luminosities of young T dwarfs agree well with the old population, resulting in cooler effective temperatures (by $\approx 100$~K) given they are still in process of contraction. 

In summary, our sample shows that ultracool dwarfs with different ages have vastly different photometric properties in late L but similar near-infrared absolute magnitudes and bolometric luminosities as they evolve to early T. Notably, our sample reveals that young L/T objects exhibit the $J$-band brightening phenomenon with a significantly larger amplitude ($\approx 1.5$~mag) as compared to the $\approx 0.5$~mag brightening seen in field L/T objects \citep[e.g.,][]{2012ApJS..201...19D}, and also extending to $H$ band in the young objects. 

Finally, we note more discoveries of young T0$-$T1 benchmarks would bridge the observed properties of late-L and T dwarfs, establishing a more comprehensive understanding of the L/T evolution. These objects are very rare so deep imaging surveys like UHS and VHS exploring a larger volume in solar neighborhood would be very helpful. Also, a uniform spectroscopic analysis of planetary-mass and substellar benchmarks will help us to better understand the wavelength ranges where atmospheric models cannot explain the data. Such work will provide useful suggestions to improve contemporary model atmospheres.

\acknowledgments
We thank the anonymous referee for helpful comments. We thank Didier Saumon, Caroline Morley, Paul Molli\`{e}re, William Best, and Jennifer van Saders for insightful discussions and comments. We thank Ian Czekala for discussions about the Starfish package, and thank Michael Gully-Santiago for implementing Starfish for IRTF/SpeX prism data. We thank Michael Cushing for discussions about the IRTF/SpeX wavelength calibration and thank Trent Dupuy for discussions about the Hawaii Infrared Parallax Program. This work benefited from the Exoplanet Summer Program in the Other Worlds Laboratory (OWL) at the University of California, Santa Cruz, a program funded by the Heising-Simons Foundation. M.C.L. acknowledges National Science Foundation (NSF) grant AST-1518339. M.A.T. acknowledges support from the DOE CSGF through grant DE-SC0019323. The advanced computing resources from the University of Hawaii Information Technology Services -- Cyberinfrastructure and the technical support from Curt Dodds are gratefully acknowledged. This work presents results from the European Space Agency (ESA) space mission {\it Gaia}. {\it Gaia} data are being processed by the {\it Gaia} Data Processing and Analysis Consortium (DPAC). Funding for the DPAC is provided by national institutions, in particular the institutions participating in the {\it Gaia} MultiLateral Agreement (MLA). The {\it Gaia} mission website is https://www.cosmos.esa.int/gaia. The {\it Gaia} archive website is https://archives.esac.esa.int/gaia. The Pan-STARRS1 Surveys (PS1) have been made possible through contributions of the Institute for Astronomy, the University of Hawaii, the Pan-STARRS Project Office, the Max-Planck Society and its participating institutes, the Max Planck Institute for Astronomy, Heidelberg and the Max Planck Institute for Extraterrestrial Physics, Garching, The Johns Hopkins University, Durham University, the University of Edinburgh, Queen's University Belfast, the Harvard-Smithsonian Center for Astrophysics, the Las Cumbres Observatory Global Telescope Network Incorporated, the National Central University of Taiwan, the Space Telescope Science Institute, the National Aeronautics and Space Administration under Grant No. NNX08AR22G issued through the Planetary Science Division of the NASA Science Mission Directorate, the National Science Foundation under Grant No. AST-1238877, the University of Maryland, and Eotvos Lorand University (ELTE). This publication makes use of data products from the Two Micron All Sky Survey, which is a joint project of the University of Massachusetts and the Infrared Processing and Analysis Center/California Institute of Technology, funded by the National Aeronautics and Space Administration and the National Science Foundation. This work is based in part on data obtained as part of the UKIRT Infrared Deep Sky Survey. The UHS is a partnership between the UK STFC, The University of Hawaii, The University of Arizona, Lockheed Martin and NASA. This publication makes use of data products from the Wide-field Infrared Survey Explorer, which is a joint project of the University of California, Los Angeles, and the Jet Propulsion Laboratory/California Institute of Technology, and NEOWISE, which is a project of the Jet Propulsion Laboratory/California Institute of Technology. WISE and NEOWISE are funded by the National Aeronautics and Space Administration. This research has made use of the SIMBAD database and the VizieR catalog access tool developed and operated at CDS, Strasbourg, France. This work was greatly facilitated by the TOPCAT software written by Mark Taylor (http://www.starlink.ac.uk/topcat/). Finally, the authors wish to recognize and acknowledge the very significant cultural role and reverence that the summit of Maunakea has always had within the indigenous Hawaiian community.  We are most fortunate to have the opportunity to conduct observations from this mountain.

\facilities{UH~2.2m (SNIFS), IRTF (SpeX), Pan-STARRS, 2MASS, UKIRT, WISE}

\software{emcee \citep{2013PASP..125..306F}, Starfish \citep{2015ApJ...812..128C}, TOPCAT \citep{2005ASPC..347...29T}, Spextool \citep[v4.1;][]{2004PASP..116..362C}, Astropy \citep{2013A&A...558A..33A, 2018AJ....156..123A}, IPython \citep{PER-GRA:2007}, Numpy \citep{numpy},  Scipy \citep{scipy}, Matplotlib \citep{Hunter:2007}.}

\end{CJK*}

\clearpage
\bibliographystyle{aasjournal}
\bibliography{ms}

\begin{thebibliography}{}
\expandafter\ifx\csname natexlab\endcsname\relax\def\natexlab#1{#1}\fi
\providecommand{\url}[1]{\href{#1}{#1}}
\providecommand{\dodoi}[1]{doi:~\href{http://doi.org/#1}{\nolinkurl{#1}}}
\providecommand{\doeprint}[1]{\href{http://ascl.net/#1}{\nolinkurl{http://ascl.net/#1}}}
\providecommand{\doarXiv}[1]{\href{https://arxiv.org/abs/#1}{\nolinkurl{https://arxiv.org/abs/#1}}}

\bibitem[{{Abt}(2009)}]{2009ApJS..180..117A}
{Abt}, H.~A. 2009, \apjs, 180, 117, \dodoi{10.1088/0067-0049/180/1/117}

\bibitem[{{Abt} \& {Morrell}(1995)}]{1995ApJS...99..135A}
{Abt}, H.~A., \& {Morrell}, N.~I. 1995, \apjs, 99, 135, \dodoi{10.1086/192182}

\bibitem[{{Ackerman} \& {Marley}(2001)}]{2001ApJ...556..872A}
{Ackerman}, A.~S., \& {Marley}, M.~S. 2001, \apj, 556, 872,
  \dodoi{10.1086/321540}

\bibitem[{{Albert} {et~al.}(2011){Albert}, {Artigau}, {Delorme}, {Reyl{\'e}},
  {Forveille}, {Delfosse}, \& {Willott}}]{2011AJ....141..203A}
{Albert}, L., {Artigau}, {\'E}., {Delorme}, P., {et~al.} 2011, \aj, 141, 203,
  \dodoi{10.1088/0004-6256/141/6/203}

\bibitem[{{Aldering} {et~al.}(2002){Aldering}, {Adam}, {Antilogus}, {Astier},
  {Bacon}, {Bongard}, {Bonnaud}, {Copin}, {Hardin}, \&
  {Henault}}]{2002SPIE.4836...61A}
{Aldering}, G., {Adam}, G., {Antilogus}, P., {et~al.} 2002, in Society of
  Photo-Optical Instrumentation Engineers (SPIE) Conference Series, Vol. 4836,
  Survey and Other Telescope Technologies and Discoveries, ed. J.~A. {Tyson} \&
  S.~{Wolff}, 61--72

\bibitem[{{Allard}(2014)}]{2014IAUS..299..271A}
{Allard}, F. 2014, in IAU Symposium, Vol. 299, Exploring the Formation and
  Evolution of Planetary Systems, ed. M.~{Booth}, B.~C. {Matthews}, \& J.~R.
  {Graham}, 271--272

\bibitem[{{Allard} {et~al.}(2007{\natexlab{a}}){Allard}, {Allard}, {Homeier},
  {Kielkopf}, {McCaughrean}, \& {Spiegelman}}]{2007A&A...474L..21A}
{Allard}, F., {Allard}, N.~F., {Homeier}, D., {et~al.} 2007{\natexlab{a}},
  \aap, 474, L21, \dodoi{10.1051/0004-6361:20078362}

\bibitem[{{Allard} {et~al.}(2001){Allard}, {Hauschildt}, {Alexander},
  {Tamanai}, \& {Schweitzer}}]{2001ApJ...556..357A}
{Allard}, F., {Hauschildt}, P.~H., {Alexander}, D.~R., {Tamanai}, A., \&
  {Schweitzer}, A. 2001, \apj, 556, 357, \dodoi{10.1086/321547}

\bibitem[{{Allard} {et~al.}(2011){Allard}, {Homeier}, \&
  {Freytag}}]{2011ASPC..448...91A}
{Allard}, F., {Homeier}, D., \& {Freytag}, B. 2011, Astronomical Society of the
  Pacific Conference Series, Vol. 448, {Model Atmospheres From Very Low Mass
  Stars to Brown Dwarfs}, ed. C.~{Johns-Krull}, M.~K. {Browning}, \& A.~A.
  {West}, 91

\bibitem[{{Allard} {et~al.}(2012){Allard}, {Homeier}, \&
  {Freytag}}]{2012RSPTA.370.2765A}
---. 2012, Philosophical Transactions of the Royal Society of London Series A,
  370, 2765, \dodoi{10.1098/rsta.2011.0269}

\bibitem[{{Allard} {et~al.}(2013){Allard}, {Homeier}, {Freytag},
  {Schaffenberger}, {}, \& {Rajpurohit}}]{2013MSAIS..24..128A}
{Allard}, F., {Homeier}, D., {Freytag}, B., {et~al.} 2013, Memorie della
  Societa Astronomica Italiana Supplementi, 24, 128.
\newblock \doarXiv{1302.6559}

\bibitem[{{Allard} {et~al.}(2007{\natexlab{b}}){Allard}, {Spiegelman}, \&
  {Kielkopf}}]{2007A&A...465.1085A}
{Allard}, N.~F., {Spiegelman}, F., \& {Kielkopf}, J.~F. 2007{\natexlab{b}},
  \aap, 465, 1085, \dodoi{10.1051/0004-6361:20066616}

\bibitem[{{Allard} {et~al.}(2016){Allard}, {Spiegelman}, \&
  {Kielkopf}}]{2016A&A...589A..21A}
---. 2016, \aap, 589, A21, \dodoi{10.1051/0004-6361/201628270}

\bibitem[{{Allers} \& {Liu}(2013)}]{2013ApJ...772...79A}
{Allers}, K.~N., \& {Liu}, M.~C. 2013, \apj, 772, 79,
  \dodoi{10.1088/0004-637X/772/2/79}

\bibitem[{{Alonso-Floriano} {et~al.}(2015){Alonso-Floriano}, {Morales},
  {Caballero}, {Montes}, {Klutsch}, {Mundt}, {Cort{\'e}s-Contreras}, {Ribas},
  {Reiners}, {Amado}, {Quirrenbach}, \& {Jeffers}}]{2015AandA...577A.128A}
{Alonso-Floriano}, F.~J., {Morales}, J.~C., {Caballero}, J.~A., {et~al.} 2015,
  \aap, 577, A128, \dodoi{10.1051/0004-6361/201525803}

\bibitem[{{Anderson} \& {Francis}(2012)}]{2012AstL...38..331A}
{Anderson}, E., \& {Francis}, C. 2012, Astronomy Letters, 38, 331,
  \dodoi{10.1134/S1063773712050015}

\bibitem[{{Artigau} {et~al.}(2006){Artigau}, {Doyon}, {Lafreni{\`e}re},
  {Nadeau}, {Robert}, \& {Albert}}]{2006ApJ...651L..57A}
{Artigau}, {\'E}., {Doyon}, R., {Lafreni{\`e}re}, D., {et~al.} 2006, \apjl,
  651, L57, \dodoi{10.1086/509146}

\bibitem[{{Astropy Collaboration} {et~al.}(2013){Astropy Collaboration},
  {Robitaille}, {Tollerud}, {Greenfield}, {Droettboom}, {Bray}, {Aldcroft},
  {Davis}, {Ginsburg}, {Price-Whelan}, {Kerzendorf}, {Conley}, {Crighton},
  {Barbary}, {Muna}, {Ferguson}, {Grollier}, {Parikh}, {Nair}, {Unther},
  {Deil}, {Woillez}, {Conseil}, {Kramer}, {Turner}, {Singer}, {Fox}, {Weaver},
  {Zabalza}, {Edwards}, {Azalee Bostroem}, {Burke}, {Casey}, {Crawford},
  {Dencheva}, {Ely}, {Jenness}, {Labrie}, {Lim}, {Pierfederici}, {Pontzen},
  {Ptak}, {Refsdal}, {Servillat}, \& {Streicher}}]{2013A&A...558A..33A}
{Astropy Collaboration}, {Robitaille}, T.~P., {Tollerud}, E.~J., {et~al.} 2013,
  \aap, 558, A33, \dodoi{10.1051/0004-6361/201322068}

\bibitem[{{Astropy Collaboration} {et~al.}(2018){Astropy Collaboration},
  {Price-Whelan}, {Sip{\H o}cz}, {G{\"u}nther}, {Lim}, {Crawford}, {Conseil},
  {Shupe}, {Craig}, {Dencheva}, {Ginsburg}, {VanderPlas}, {Bradley},
  {P{\'e}rez-Su{\'a}rez}, {de Val-Borro}, {Aldcroft}, {Cruz}, {Robitaille},
  {Tollerud}, {Ardelean}, {Babej}, {Bach}, {Bachetti}, {Bakanov}, {Bamford},
  {Barentsen}, {Barmby}, {Baumbach}, {Berry}, {Biscani}, {Boquien}, {Bostroem},
  {Bouma}, {Brammer}, {Bray}, {Breytenbach}, {Buddelmeijer}, {Burke},
  {Calderone}, {Cano Rodr{\'{\i}}guez}, {Cara}, {Cardoso}, {Cheedella},
  {Copin}, {Corrales}, {Crichton}, {D'Avella}, {Deil}, {Depagne}, {Dietrich},
  {Donath}, {Droettboom}, {Earl}, {Erben}, {Fabbro}, {Ferreira}, {Finethy},
  {Fox}, {Garrison}, {Gibbons}, {Goldstein}, {Gommers}, {Greco}, {Greenfield},
  {Groener}, {Grollier}, {Hagen}, {Hirst}, {Homeier}, {Horton}, {Hosseinzadeh},
  {Hu}, {Hunkeler}, {Ivezi{\'c}}, {Jain}, {Jenness}, {Kanarek}, {Kendrew},
  {Kern}, {Kerzendorf}, {Khvalko}, {King}, {Kirkby}, {Kulkarni}, {Kumar},
  {Lee}, {Lenz}, {Littlefair}, {Ma}, {Macleod}, {Mastropietro}, {McCully},
  {Montagnac}, {Morris}, {Mueller}, {Mumford}, {Muna}, {Murphy}, {Nelson},
  {Nguyen}, {Ninan}, {N{\"o}the}, {Ogaz}, {Oh}, {Parejko}, {Parley}, {Pascual},
  {Patil}, {Patil}, {Plunkett}, {Prochaska}, {Rastogi}, {Reddy Janga},
  {Sabater}, {Sakurikar}, {Seifert}, {Sherbert}, {Sherwood-Taylor}, {Shih},
  {Sick}, {Silbiger}, {Singanamalla}, {Singer}, {Sladen}, {Sooley},
  {Sornarajah}, {Streicher}, {Teuben}, {Thomas}, {Tremblay}, {Turner},
  {Terr{\'o}n}, {van Kerkwijk}, {de la Vega}, {Watkins}, {Weaver}, {Whitmore},
  {Woillez}, {Zabalza}, \& {Astropy Contributors}}]{2018AJ....156..123A}
{Astropy Collaboration}, {Price-Whelan}, A.~M., {Sip{\H o}cz}, B.~M., {et~al.}
  2018, \aj, 156, 123, \dodoi{10.3847/1538-3881/aabc4f}

\bibitem[{{Bacon} {et~al.}(2001){Bacon}, {Copin}, {Monnet}, {Miller},
  {Allington-Smith}, {Bureau}, {Carollo}, {Davies}, {Emsellem}, \&
  {Kuntschner}}]{2001MNRAS.326...23B}
{Bacon}, R., {Copin}, Y., {Monnet}, G., {et~al.} 2001, \mnras, 326, 23,
  \dodoi{10.1046/j.1365-8711.2001.04612.x}

\bibitem[{{Bailer-Jones} {et~al.}(2018){Bailer-Jones}, {Rybizki}, {Fouesneau},
  {Mantelet}, \& {Andrae}}]{2018AJ....156...58B}
{Bailer-Jones}, C.~A.~L., {Rybizki}, J., {Fouesneau}, M., {Mantelet}, G., \&
  {Andrae}, R. 2018, \aj, 156, 58, \dodoi{10.3847/1538-3881/aacb21}

\bibitem[{{Baraffe} {et~al.}(1998){Baraffe}, {Chabrier}, {Allard}, \&
  {Hauschildt}}]{1998AandA...337..403B}
{Baraffe}, I., {Chabrier}, G., {Allard}, F., \& {Hauschildt}, P.~H. 1998, \aap,
  337, 403.
\newblock \doarXiv{astro-ph/9805009}

\bibitem[{{Baraffe} {et~al.}(2003){Baraffe}, {Chabrier}, {Barman}, {Allard}, \&
  {Hauschildt}}]{2003AandA...402..701B}
{Baraffe}, I., {Chabrier}, G., {Barman}, T.~S., {Allard}, F., \& {Hauschildt},
  P.~H. 2003, \aap, 402, 701, \dodoi{10.1051/0004-6361:20030252}

\bibitem[{{Baraffe} {et~al.}(2015){Baraffe}, {Homeier}, {Allard}, \&
  {Chabrier}}]{2015AandA...577A..42B}
{Baraffe}, I., {Homeier}, D., {Allard}, F., \& {Chabrier}, G. 2015, \aap, 577,
  A42, \dodoi{10.1051/0004-6361/201425481}

\bibitem[{{Barman} {et~al.}(2011){Barman}, {Macintosh}, {Konopacky}, \&
  {Marois}}]{2011ApJ...733...65B}
{Barman}, T.~S., {Macintosh}, B., {Konopacky}, Q.~M., \& {Marois}, C. 2011,
  \apj, 733, 65, \dodoi{10.1088/0004-637X/733/1/65}

\bibitem[{{Barnaby} {et~al.}(2000){Barnaby}, {Spillar}, {Christou}, \&
  {Drummond}}]{2000AJ....119..378B}
{Barnaby}, D., {Spillar}, E., {Christou}, J.~C., \& {Drummond}, J.~D. 2000,
  \aj, 119, 378, \dodoi{10.1086/301155}

\bibitem[{{Barnes} \& {Fortney}(2003)}]{2003ApJ...588..545B}
{Barnes}, J.~W., \& {Fortney}, J.~J. 2003, \apj, 588, 545,
  \dodoi{10.1086/373893}

\bibitem[{{Barnes}(2007)}]{2007ApJ...669.1167B}
{Barnes}, S.~A. 2007, \apj, 669, 1167, \dodoi{10.1086/519295}

\bibitem[{{Becklin} \& {Zuckerman}(1988)}]{1988Natur.336..656B}
{Becklin}, E.~E., \& {Zuckerman}, B. 1988, \nat, 336, 656,
  \dodoi{10.1038/336656a0}

\bibitem[{{B{\'e}dard} {et~al.}(2017){B{\'e}dard}, {Bergeron}, \&
  {Fontaine}}]{2017ApJ...848...11B}
{B{\'e}dard}, A., {Bergeron}, P., \& {Fontaine}, G. 2017, \apj, 848, 11,
  \dodoi{10.3847/1538-4357/aa8bb6}

\bibitem[{{Bell} {et~al.}(2015){Bell}, {Mamajek}, \&
  {Naylor}}]{2015MNRAS.454..593B}
{Bell}, C. P.~M., {Mamajek}, E.~E., \& {Naylor}, T. 2015, \mnras, 454, 593,
  \dodoi{10.1093/mnras/stv1981}

\bibitem[{{Bergeron} {et~al.}(2019){Bergeron}, {Dufour}, {Fontaine}, {Coutu},
  {Blouin}, {Genest-Beaulieu}, {B{\'e}dard}, \& {Rolland
  }}]{2019ApJ...876...67B}
{Bergeron}, P., {Dufour}, P., {Fontaine}, G., {et~al.} 2019, The Astrophysical
  Journal, 876, 67, \dodoi{10.3847/1538-4357/ab153a}

\bibitem[{{Bergeron} {et~al.}(1997){Bergeron}, {Ruiz}, \&
  {Leggett}}]{1997ApJS..108..339B}
{Bergeron}, P., {Ruiz}, M.~T., \& {Leggett}, S.~K. 1997, \apjs, 108, 339,
  \dodoi{10.1086/312955}

\bibitem[{{Bergeron} {et~al.}(1992){Bergeron}, {Saffer}, \&
  {Liebert}}]{1992ApJ...394..228B}
{Bergeron}, P., {Saffer}, R.~A., \& {Liebert}, J. 1992, \apj, 394, 228,
  \dodoi{10.1086/171575}

\bibitem[{{Bergeron} {et~al.}(1995){Bergeron}, {Wesemael}, {Lamontagne},
  {Fontaine}, {Saffer}, \& {Allard}}]{1995ApJ...449..258B}
{Bergeron}, P., {Wesemael}, F., {Lamontagne}, R., {et~al.} 1995, \apj, 449,
  258, \dodoi{10.1086/176053}

\bibitem[{{Best}(2018)}]{2018PhDT.......159B}
{Best}, W. M.~J. 2018, PhD thesis, University of Hawai'i at Manoa

\bibitem[{{Best} {et~al.}(2015){Best}, {Liu}, {Magnier}, {Deacon}, {Aller},
  {Redstone}, {Burgett}, {Chambers}, {Draper}, {Flewelling}, {Hodapp},
  {Kaiser}, {Metcalfe}, {Tonry}, {Wainscoat}, \&
  {Waters}}]{2015ApJ...814..118B}
{Best}, W. M.~J., {Liu}, M.~C., {Magnier}, E.~A., {et~al.} 2015, \apj, 814,
  118, \dodoi{10.1088/0004-637X/814/2/118}

\bibitem[{{Best} {et~al.}(2018){Best}, {Magnier}, {Liu}, {Aller}, {Zhang},
  {Burgett}, {Chambers}, {Draper}, {Flewelling}, {Kaiser}, {Kudritzki},
  {Metcalfe}, {Tonry}, {Wainscoat}, \& {Waters}}]{2018ApJS..234....1B}
{Best}, W.~M.~J., {Magnier}, E.~A., {Liu}, M.~C., {et~al.} 2018, \apjs, 234, 1,
  \dodoi{10.3847/1538-4365/aa9982}

\bibitem[{{Biller} {et~al.}(2015){Biller}, {Vos}, {Bonavita}, {Buenzli},
  {Baxter}, {Crossfield}, {Allers}, {Liu}, {Bonnefoy}, {Deacon}, {Brandner},
  {Schlieder}, {Dupuy}, {Kopytova}, {Manjavacas}, {Allard}, {Homeier}, \&
  {Henning}}]{2015ApJ...813L..23B}
{Biller}, B.~A., {Vos}, J., {Bonavita}, M., {et~al.} 2015, \apjl, 813, L23,
  \dodoi{10.1088/2041-8205/813/2/L23}

\bibitem[{{Blouin} {et~al.}(2019){Blouin}, {Dufour}, {Thibeault}, \&
  {Allard}}]{2019ApJ...878...63B}
{Blouin}, S., {Dufour}, P., {Thibeault}, C., \& {Allard}, N.~F. 2019, The
  Astrophysical Journal, 878, 63, \dodoi{10.3847/1538-4357/ab1f82}

\bibitem[{{Bouvier} {et~al.}(2008){Bouvier}, {Kendall}, {Meeus}, {Testi},
  {Moraux}, {Stauffer}, {James}, {Cuilland re}, {Irwin}, {McCaughrean},
  {Baraffe}, \& {Bertin}}]{2008AandA...481..661B}
{Bouvier}, J., {Kendall}, T., {Meeus}, G., {et~al.} 2008, \aap, 481, 661,
  \dodoi{10.1051/0004-6361:20079303}

\bibitem[{{Bowler}(2016)}]{2016PASP..128j2001B}
{Bowler}, B.~P. 2016, \pasp, 128, 102001,
  \dodoi{10.1088/1538-3873/128/968/102001}

\bibitem[{{Bowler} {et~al.}(2010){Bowler}, {Liu}, {Dupuy}, \&
  {Cushing}}]{2010ApJ...723..850B}
{Bowler}, B.~P., {Liu}, M.~C., {Dupuy}, T.~J., \& {Cushing}, M.~C. 2010, \apj,
  723, 850, \dodoi{10.1088/0004-637X/723/1/850}

\bibitem[{{Bowler} {et~al.}(2013){Bowler}, {Liu}, {Shkolnik}, \&
  {Dupuy}}]{2013ApJ...774...55B}
{Bowler}, B.~P., {Liu}, M.~C., {Shkolnik}, E.~L., \& {Dupuy}, T.~J. 2013, \apj,
  774, 55, \dodoi{10.1088/0004-637X/774/1/55}

\bibitem[{{Bowler} {et~al.}(2017){Bowler}, {Liu}, {Mawet}, {Ngo}, {Malo},
  {Mace}, {McLane}, {Lu}, {Tristan}, {Hinkley}, {Hillenbrand}, {Shkolnik},
  {Benneke}, \& {Best}}]{2017AJ....153...18B}
{Bowler}, B.~P., {Liu}, M.~C., {Mawet}, D., {et~al.} 2017, \aj, 153, 18,
  \dodoi{10.3847/1538-3881/153/1/18}

\bibitem[{{Brandt} \& {Huang}(2015)}]{2015ApJ...807...24B}
{Brandt}, T.~D., \& {Huang}, C.~X. 2015, \apj, 807, 24,
  \dodoi{10.1088/0004-637X/807/1/24}

\bibitem[{{Brandt} {et~al.}(2014){Brandt}, {McElwain}, {Turner}, {Mede},
  {Spiegel}, {Kuzuhara}, {Schlieder}, {Wisniewski}, {Abe}, {Biller},
  {Brandner}, {Carson}, {Currie}, {Egner}, {Feldt}, {Golota}, {Goto}, {Grady},
  {Guyon}, {Hashimoto}, {Hayano}, {Hayashi}, {Hayashi}, {Henning}, {Hodapp},
  {Inutsuka}, {Ishii}, {Iye}, {Janson}, {Kandori}, {Knapp}, {Kudo}, {Kusakabe},
  {Kwon}, {Matsuo}, {Miyama}, {Morino}, {Moro-Mart{\'{\i}}n}, {Nishimura},
  {Pyo}, {Serabyn}, {Suto}, {Suzuki}, {Takami}, {Takato}, {Terada}, {Thalmann},
  {Tomono}, {Watanabe}, {Yamada}, {Takami}, {Usuda}, \&
  {Tamura}}]{2014ApJ...794..159B}
{Brandt}, T.~D., {McElwain}, M.~W., {Turner}, E.~L., {et~al.} 2014, \apj, 794,
  159, \dodoi{10.1088/0004-637X/794/2/159}

\bibitem[{{Burgasser}(2007)}]{2007ApJ...659..655B}
{Burgasser}, A.~J. 2007, \apj, 659, 655, \dodoi{10.1086/511027}

\bibitem[{{Burgasser} {et~al.}(2010){Burgasser}, {Cruz}, {Cushing}, {Gelino},
  {Looper}, {Faherty}, {Kirkpatrick}, \& {Reid}}]{2010ApJ...710.1142B}
{Burgasser}, A.~J., {Cruz}, K.~L., {Cushing}, M., {et~al.} 2010, \apj, 710,
  1142, \dodoi{10.1088/0004-637X/710/2/1142}

\bibitem[{{Burgasser} {et~al.}(2006{\natexlab{a}}){Burgasser}, {Geballe},
  {Leggett}, {Kirkpatrick}, \& {Golimowski}}]{2006ApJ...637.1067B}
{Burgasser}, A.~J., {Geballe}, T.~R., {Leggett}, S.~K., {Kirkpatrick}, J.~D.,
  \& {Golimowski}, D.~A. 2006{\natexlab{a}}, \apj, 637, 1067,
  \dodoi{10.1086/498563}

\bibitem[{{Burgasser} {et~al.}(2006{\natexlab{b}}){Burgasser}, {Kirkpatrick},
  {Cruz}, {Reid}, {Leggett}, {Liebert}, {Burrows}, \&
  {Brown}}]{2006ApJS..166..585B}
{Burgasser}, A.~J., {Kirkpatrick}, J.~D., {Cruz}, K.~L., {et~al.}
  2006{\natexlab{b}}, \apjs, 166, 585, \dodoi{10.1086/506327}

\bibitem[{{Burgasser} {et~al.}(2005{\natexlab{a}}){Burgasser}, {Kirkpatrick},
  \& {Lowrance}}]{2005AJ....129.2849B}
{Burgasser}, A.~J., {Kirkpatrick}, J.~D., \& {Lowrance}, P.~J.
  2005{\natexlab{a}}, \aj, 129, 2849, \dodoi{10.1086/430218}

\bibitem[{{Burgasser} {et~al.}(2003){Burgasser}, {Kirkpatrick}, {Reid},
  {Brown}, {Miskey}, \& {Gizis}}]{2003ApJ...586..512B}
{Burgasser}, A.~J., {Kirkpatrick}, J.~D., {Reid}, I.~N., {et~al.} 2003, \apj,
  586, 512, \dodoi{10.1086/346263}

\bibitem[{{Burgasser} {et~al.}(2015){Burgasser}, {Melis}, {Todd}, {Gelino},
  {Hallinan}, \& {Bardalez Gagliuffi}}]{2015AJ....150..180B}
{Burgasser}, A.~J., {Melis}, C., {Todd}, J., {et~al.} 2015, \aj, 150, 180,
  \dodoi{10.1088/0004-6256/150/6/180}

\bibitem[{{Burgasser} {et~al.}(2005{\natexlab{b}}){Burgasser}, {Reid},
  {Leggett}, {Kirkpatrick}, {Liebert}, \& {Burrows}}]{2005ApJ...634L.177B}
{Burgasser}, A.~J., {Reid}, I.~N., {Leggett}, S.~K., {et~al.}
  2005{\natexlab{b}}, \apjl, 634, L177, \dodoi{10.1086/498866}

\bibitem[{{Burgasser} {et~al.}(2013){Burgasser}, {Sheppard}, \&
  {Luhman}}]{2013ApJ...772..129B}
{Burgasser}, A.~J., {Sheppard}, S.~S., \& {Luhman}, K.~L. 2013, \apj, 772, 129,
  \dodoi{10.1088/0004-637X/772/2/129}

\bibitem[{{Burgasser} {et~al.}(2000){Burgasser}, {Kirkpatrick}, {Cutri},
  {McCallon}, {Kopan}, {Gizis}, {Liebert}, {Reid}, {Brown}, {Monet}, {Dahn},
  {Beichman}, \& {Skrutskie}}]{2000ApJ...531L..57B}
{Burgasser}, A.~J., {Kirkpatrick}, J.~D., {Cutri}, R.~M., {et~al.} 2000, \apjl,
  531, L57, \dodoi{10.1086/312522}

\bibitem[{{Burgasser} {et~al.}(2002){Burgasser}, {Kirkpatrick}, {Brown},
  {Reid}, {Burrows}, {Liebert}, {Matthews}, {Gizis}, {Dahn}, {Monet}, {Cutri},
  \& {Skrutskie}}]{2002ApJ...564..421B}
{Burgasser}, A.~J., {Kirkpatrick}, J.~D., {Brown}, M.~E., {et~al.} 2002, \apj,
  564, 421, \dodoi{10.1086/324033}

\bibitem[{{Burningham} {et~al.}(2010){Burningham}, {Pinfield}, {Lucas},
  {Leggett}, {Deacon}, {Tamura}, {Tinney}, {Lodieu}, {Zhang}, {Huelamo},
  {Jones}, {Murray}, {Mortlock}, {Patel}, {Barrado Y Navascu{\'e}s}, {Zapatero
  Osorio}, {Ishii}, {Kuzuhara}, \& {Smart}}]{2010MNRAS.406.1885B}
{Burningham}, B., {Pinfield}, D.~J., {Lucas}, P.~W., {et~al.} 2010, \mnras,
  406, 1885, \dodoi{10.1111/j.1365-2966.2010.16800.x}

\bibitem[{{Burningham} {et~al.}(2013){Burningham}, {Cardoso}, {Smith},
  {Leggett}, {Smart}, {Mann}, {Dhital}, {Lucas}, {Tinney}, {Pinfield}, {Zhang},
  {Morley}, {Saumon}, {Aller}, {Littlefair}, {Homeier}, {Lodieu}, {Deacon},
  {Marley}, {van Spaandonk}, {Baker}, {Allard}, {Andrei}, {Canty}, {Clarke},
  {Day-Jones}, {Dupuy}, {Fortney}, {Gomes}, {Ishii}, {Jones}, {Liu},
  {Magazz{\'u}}, {Marocco}, {Murray}, {Rojas-Ayala}, \&
  {Tamura}}]{2013MNRAS.433..457B}
{Burningham}, B., {Cardoso}, C.~V., {Smith}, L., {et~al.} 2013, \mnras, 433,
  457, \dodoi{10.1093/mnras/stt740}

\bibitem[{{Burrows} {et~al.}(2001){Burrows}, {Hubbard}, {Lunine}, \&
  {Liebert}}]{2001RvMP...73..719B}
{Burrows}, A., {Hubbard}, W.~B., {Lunine}, J.~I., \& {Liebert}, J. 2001,
  Reviews of Modern Physics, 73, 719, \dodoi{10.1103/RevModPhys.73.719}

\bibitem[{{Burrows} \& {Liebert}(1993)}]{1993RvMP...65..301B}
{Burrows}, A., \& {Liebert}, J. 1993, Reviews of Modern Physics, 65, 301,
  \dodoi{10.1103/RevModPhys.65.301}

\bibitem[{{Burrows} {et~al.}(2000){Burrows}, {Marley}, \&
  {Sharp}}]{2000ApJ...531..438B}
{Burrows}, A., {Marley}, M.~S., \& {Sharp}, C.~M. 2000, \apj, 531, 438,
  \dodoi{10.1086/308462}

\bibitem[{{Burrows} {et~al.}(2006){Burrows}, {Sudarsky}, \&
  {Hubeny}}]{2006ApJ...640.1063B}
{Burrows}, A., {Sudarsky}, D., \& {Hubeny}, I. 2006, \apj, 640, 1063,
  \dodoi{10.1086/500293}

\bibitem[{{Burrows} \& {Volobuyev}(2003)}]{2003ApJ...583..985B}
{Burrows}, A., \& {Volobuyev}, M. 2003, \apj, 583, 985, \dodoi{10.1086/345412}

\bibitem[{{Burrows} {et~al.}(1997){Burrows}, {Marley}, {Hubbard}, {Lunine},
  {Guillot}, {Saumon}, {Freedman}, {Sudarsky}, \&
  {Sharp}}]{1997ApJ...491..856B}
{Burrows}, A., {Marley}, M., {Hubbard}, W.~B., {et~al.} 1997, \apj, 491, 856,
  \dodoi{10.1086/305002}

\bibitem[{{Casagrande} {et~al.}(2011){Casagrande}, {Sch{\"o}nrich}, {Asplund},
  {Cassisi}, {Ram{\'\i}rez}, {Mel{\'e}ndez}, {Bensby}, \&
  {Feltzing}}]{2011AandA...530A.138C}
{Casagrande}, L., {Sch{\"o}nrich}, R., {Asplund}, M., {et~al.} 2011, \aap, 530,
  A138, \dodoi{10.1051/0004-6361/201016276}

\bibitem[{{Casewell} {et~al.}(2009){Casewell}, {Dobbie}, {Napiwotzki},
  {Burleigh}, {Barstow}, \& {Jameson}}]{2009MNRAS.395.1795C}
{Casewell}, S.~L., {Dobbie}, P.~D., {Napiwotzki}, R., {et~al.} 2009, Monthly
  Notices of the Royal Astronomical Society, 395, 1795,
  \dodoi{10.1111/j.1365-2966.2009.14593.x}

\bibitem[{{Chabrier} {et~al.}(2000){Chabrier}, {Baraffe}, {Allard}, \&
  {Hauschildt}}]{2000ApJ...542..464C}
{Chabrier}, G., {Baraffe}, I., {Allard}, F., \& {Hauschildt}, P. 2000, \apj,
  542, 464, \dodoi{10.1086/309513}

\bibitem[{{Chambers} {et~al.}(2016){Chambers}, {Magnier}, {Metcalfe},
  {Flewelling}, {Huber}, {Waters}, {Denneau}, {Draper}, {Farrow}, {Finkbeiner},
  {Holmberg}, {Koppenhoefer}, {Price}, {Rest}, {Saglia}, {Schlafly}, {Smartt},
  {Sweeney}, {Wainscoat}, {Burgett}, {Chastel}, {Grav}, {Heasley}, {Hodapp},
  {Jedicke}, {Kaiser}, {Kudritzki}, {Luppino}, {Lupton}, {Monet}, {Morgan},
  {Onaka}, {Shiao}, {Stubbs}, {Tonry}, {White}, {Ba{\~n}ados}, {Bell},
  {Bender}, {Bernard}, {Boegner}, {Boffi}, {Botticella}, {Calamida},
  {Casertano}, {Chen}, {Chen}, {Cole}, {Deacon}, {Frenk}, {Fitzsimmons},
  {Gezari}, {Gibbs}, {Goessl}, {Goggia}, {Gourgue}, {Goldman}, {Grant},
  {Grebel}, {Hambly}, {Hasinger}, {Heavens}, {Heckman}, {Henderson}, {Henning},
  {Holman}, {Hopp}, {Ip}, {Isani}, {Jackson}, {Keyes}, {Koekemoer}, {Kotak},
  {Le}, {Liska}, {Long}, {Lucey}, {Liu}, {Martin}, {Masci}, {McLean}, {Mindel},
  {Misra}, {Morganson}, {Murphy}, {Obaika}, {Narayan}, {Nieto-Santisteban},
  {Norberg}, {Peacock}, {Pier}, {Postman}, {Primak}, {Rae}, {Rai}, {Riess},
  {Riffeser}, {Rix}, {R{\"o}ser}, {Russel}, {Rutz}, {Schilbach}, {Schultz},
  {Scolnic}, {Strolger}, {Szalay}, {Seitz}, {Small}, {Smith}, {Soderblom},
  {Taylor}, {Thomson}, {Taylor}, {Thakar}, {Thiel}, {Thilker}, {Unger},
  {Urata}, {Valenti}, {Wagner}, {Walder}, {Walter}, {Watters}, {Werner},
  {Wood-Vasey}, \& {Wyse}}]{2016arXiv161205560C}
{Chambers}, K.~C., {Magnier}, E.~A., {Metcalfe}, N., {et~al.} 2016, arXiv
  e-prints.
\newblock \doarXiv{1612.05560}

\bibitem[{{Chandrasekhar}(1939)}]{1939isss.book.....C}
{Chandrasekhar}, S. 1939, {An introduction to the study of stellar structure}

\bibitem[{{Chaplin} \& {Miglio}(2013)}]{2013ARA&A..51..353C}
{Chaplin}, W.~J., \& {Miglio}, A. 2013, \araa, 51, 353,
  \dodoi{10.1146/annurev-astro-082812-140938}

\bibitem[{{Chauvin} {et~al.}(2017){Chauvin}, {Desidera}, {Lagrange}, {Vigan},
  {Gratton}, {Langlois}, {Bonnefoy}, {Beuzit}, {Feldt}, {Mouillet}, {Meyer},
  {Cheetham}, {Biller}, {Boccaletti}, {D'Orazi}, {Galicher}, {Hagelberg},
  {Maire}, {Mesa}, {Olofsson}, {Samland}, {Schmidt}, {Sissa}, {Bonavita},
  {Charnay}, {Cudel}, {Daemgen}, {Delorme}, {Janin-Potiron}, {Janson},
  {Keppler}, {Le Coroller}, {Ligi}, {Marleau}, {Messina}, {Molli{\`e}re},
  {Mordasini}, {M{\"u}ller}, {Peretti}, {Perrot}, {Rodet}, {Rouan}, {Zurlo},
  {Dominik}, {Henning}, {Menard}, {Schmid}, {Turatto}, {Udry}, {Vakili}, {Abe},
  {Antichi}, {Baruffolo}, {Baudoz}, {Baudrand}, {Blanchard}, {Bazzon}, {Buey},
  {Carbillet}, {Carle}, {Charton}, {Cascone}, {Claudi}, {Costille}, {Deboulbe},
  {De Caprio}, {Dohlen}, {Fantinel}, {Feautrier}, {Fusco}, {Gigan}, {Giro},
  {Gisler}, {Gluck}, {Hubin}, {Hugot}, {Jaquet}, {Kasper}, {Madec}, {Magnard},
  {Martinez}, {Maurel}, {Le Mignant}, {M{\"o}ller-Nilsson}, {Llored}, {Moulin},
  {Orign{\'e}}, {Pavlov}, {Perret}, {Petit}, {Pragt}, {Puget}, {Rabou},
  {Ramos}, {Rigal}, {Rochat}, {Roelfsema}, {Rousset}, {Roux}, {Salasnich},
  {Sauvage}, {Sevin}, {Soenke}, {Stadler}, {Suarez}, {Weber}, {Wildi},
  {Antoniucci}, {Augereau}, {Baudino}, {Brandner}, {Engler}, {Girard}, {Gry},
  {Kral}, {Kopytova}, {Lagadec}, {Milli}, {Moutou}, {Schlieder},
  {Szul{\'a}gyi}, {Thalmann}, \& {Wahhaj}}]{2017AandA...605L...9C}
{Chauvin}, G., {Desidera}, S., {Lagrange}, A.~M., {et~al.} 2017, \aap, 605, L9,
  \dodoi{10.1051/0004-6361/201731152}

\bibitem[{{Chiu} {et~al.}(2006){Chiu}, {Fan}, {Leggett}, {Golimowski}, {Zheng},
  {Geballe}, {Schneider}, \& {Brinkmann}}]{2006AJ....131.2722C}
{Chiu}, K., {Fan}, X., {Leggett}, S.~K., {et~al.} 2006, \aj, 131, 2722,
  \dodoi{10.1086/501431}

\bibitem[{{Choi} {et~al.}(2016){Choi}, {Dotter}, {Conroy}, {Cantiello},
  {Paxton}, \& {Johnson}}]{2016ApJ...823..102C}
{Choi}, J., {Dotter}, A., {Conroy}, C., {et~al.} 2016, \apj, 823, 102,
  \dodoi{10.3847/0004-637X/823/2/102}

\bibitem[{{Ciddor}(1996)}]{1996ApOpt..35.1566C}
{Ciddor}, P.~E. 1996, \ao, 35, 1566, \dodoi{10.1364/AO.35.001566}

\bibitem[{{Cottaar} {et~al.}(2014){Cottaar}, {Covey}, {Meyer}, {Nidever},
  {Stassun}, {Foster}, {Tan}, {Chojnowski}, {da Rio}, {Flaherty}, {Frinchaboy},
  {Skrutskie}, {Majewski}, {Wilson}, \& {Zasowski}}]{2014ApJ...794..125C}
{Cottaar}, M., {Covey}, K.~R., {Meyer}, M.~R., {et~al.} 2014, \apj, 794, 125,
  \dodoi{10.1088/0004-637X/794/2/125}

\bibitem[{{Crepp} {et~al.}(2014){Crepp}, {Johnson}, {Howard}, {Marcy},
  {Brewer}, {Fischer}, {Wright}, \& {Isaacson}}]{2014ApJ...781...29C}
{Crepp}, J.~R., {Johnson}, J.~A., {Howard}, A.~W., {et~al.} 2014, \apj, 781,
  29, \dodoi{10.1088/0004-637X/781/1/29}

\bibitem[{{Crepp} {et~al.}(2015){Crepp}, {Rice}, {Veicht}, {Aguilar}, {Pueyo},
  {Giorla}, {Nilsson}, {Luszcz-Cook}, {Oppenheimer}, {Hinkley}, {Brenner},
  {Vasisht}, {Cady}, {Beichman}, {Hillenbrand}, {Lockhart}, {Matthews},
  {Roberts}, {Sivaramakrishnan}, {Soummer}, \& {Zhai}}]{2015ApJ...798L..43C}
{Crepp}, J.~R., {Rice}, E.~L., {Veicht}, A., {et~al.} 2015, \apjl, 798, L43,
  \dodoi{10.1088/2041-8205/798/2/L43}

\bibitem[{{Cruz} {et~al.}(2007){Cruz}, {Reid}, {Kirkpatrick}, {Burgasser},
  {Liebert}, {Solomon}, {Schmidt}, {Allen}, {Hawley}, \&
  {Covey}}]{2007AJ....133..439C}
{Cruz}, K.~L., {Reid}, I.~N., {Kirkpatrick}, J.~D., {et~al.} 2007, \aj, 133,
  439, \dodoi{10.1086/510132}

\bibitem[{{Cushing} {et~al.}(2004){Cushing}, {Vacca}, \&
  {Rayner}}]{2004PASP..116..362C}
{Cushing}, M.~C., {Vacca}, W.~D., \& {Rayner}, J.~T. 2004, \pasp, 116, 362,
  \dodoi{10.1086/382907}

\bibitem[{{Cutri} \& {et al.}(2014)}]{2014yCat.2328....0C}
{Cutri}, R.~M., \& {et al.} 2014, VizieR Online Data Catalog, 2328

\bibitem[{{Cutri} {et~al.}(2003){Cutri}, {Skrutskie}, {van Dyk}, {Beichman},
  {Carpenter}, {Chester}, {Cambresy}, {Evans}, {Fowler}, {Gizis}, {Howard},
  {Huchra}, {Jarrett}, {Kopan}, {Kirkpatrick}, {Light}, {Marsh}, {McCallon},
  {Schneider}, {Stiening}, {Sykes}, {Weinberg}, {Wheaton}, {Wheelock}, \&
  {Zacarias}}]{2003yCat.2246....0C}
{Cutri}, R.~M., {Skrutskie}, M.~F., {van Dyk}, S., {et~al.} 2003, VizieR Online
  Data Catalog, II/246

\bibitem[{{Czekala} {et~al.}(2015){Czekala}, {Andrews}, {Mandel}, {Hogg}, \&
  {Green}}]{2015ApJ...812..128C}
{Czekala}, I., {Andrews}, S.~M., {Mandel}, K.~S., {Hogg}, D.~W., \& {Green},
  G.~M. 2015, \apj, 812, 128, \dodoi{10.1088/0004-637X/812/2/128}

\bibitem[{{Dahm}(2015)}]{2015ApJ...813..108D}
{Dahm}, S.~E. 2015, \apj, 813, 108, \dodoi{10.1088/0004-637X/813/2/108}

\bibitem[{{Dahn} {et~al.}(2002){Dahn}, {Harris}, {Vrba}, {Guetter}, {Canzian},
  {Henden}, {Levine}, {Luginbuhl}, {Monet}, {Monet}, {Pier}, {Stone}, {Walker},
  {Burgasser}, {Gizis}, {Kirkpatrick}, {Liebert}, \&
  {Reid}}]{2002AJ....124.1170D}
{Dahn}, C.~C., {Harris}, H.~C., {Vrba}, F.~J., {et~al.} 2002, \aj, 124, 1170,
  \dodoi{10.1086/341646}

\bibitem[{{Day-Jones} {et~al.}(2011){Day-Jones}, {Pinfield}, {Ruiz},
  {Beaumont}, {Burningham}, {Gallardo}, {Gianninas}, {Bergeron}, {Napiwotzki},
  {Jenkins}, {Zhang}, {Murray}, {Catal{\'a}n}, \&
  {Gomes}}]{2011MNRAS.410..705D}
{Day-Jones}, A.~C., {Pinfield}, D.~J., {Ruiz}, M.~T., {et~al.} 2011, \mnras,
  410, 705, \dodoi{10.1111/j.1365-2966.2010.17469.x}

\bibitem[{{Day-Jones} {et~al.}(2013){Day-Jones}, {Marocco}, {Pinfield},
  {Zhang}, {Burningham}, {Deacon}, {Ruiz}, {Gallardo}, {Jones}, \&
  {Lucas}}]{2013MNRAS.430.1171D}
{Day-Jones}, A.~C., {Marocco}, F., {Pinfield}, D.~J., {et~al.} 2013, \mnras,
  430, 1171, \dodoi{10.1093/mnras/sts685}

\bibitem[{{Deacon} {et~al.}(2012{\natexlab{a}}){Deacon}, {Liu}, {Magnier},
  {Bowler}, {Redstone}, {Goldman}, {Burgett}, {Chambers}, {Flewelling},
  {Kaiser}, {Morgan}, {Price}, {Sweeney}, {Tonry}, {Wainscoat}, \&
  {Waters}}]{2012ApJ...755...94D}
{Deacon}, N.~R., {Liu}, M.~C., {Magnier}, E.~A., {et~al.} 2012{\natexlab{a}},
  \apj, 755, 94, \dodoi{10.1088/0004-637X/755/2/94}

\bibitem[{{Deacon} {et~al.}(2012{\natexlab{b}}){Deacon}, {Liu}, {Magnier},
  {Bowler}, {Mann}, {Redstone}, {Burgett}, {Chambers}, {Hodapp}, {Kaiser},
  {Kudritzki}, {Morgan}, {Price}, {Tonry}, \&
  {Wainscoat}}]{2012ApJ...757..100D}
---. 2012{\natexlab{b}}, \apj, 757, 100, \dodoi{10.1088/0004-637X/757/1/100}

\bibitem[{{Deacon} {et~al.}(2014){Deacon}, {Liu}, {Magnier}, {Aller}, {Best},
  {Dupuy}, {Bowler}, {Mann}, {Redstone}, {Burgett}, {Chambers}, {Draper},
  {Flewelling}, {Hodapp}, {Kaiser}, {Kudritzki}, {Morgan}, {Metcalfe}, {Price},
  {Tonry}, \& {Wainscoat}}]{2014ApJ...792..119D}
---. 2014, \apj, 792, 119, \dodoi{10.1088/0004-637X/792/2/119}

\bibitem[{{Deacon} {et~al.}(2017){Deacon}, {Magnier}, {Liu}, {Schlieder},
  {Aller}, {Best}, {Bowler}, {Burgett}, {Chambers}, {Draper}, {Flewelling},
  {Hodapp}, {Kaiser}, {Metcalfe}, {Sweeney}, {Wainscoat}, \&
  {Waters}}]{2017MNRAS.467.1126D}
{Deacon}, N.~R., {Magnier}, E.~A., {Liu}, M.~C., {et~al.} 2017, \mnras, 467,
  1126, \dodoi{10.1093/mnras/stx065}

\bibitem[{{Desidera} \& {Barbieri}(2007)}]{2007A&A...462..345D}
{Desidera}, S., \& {Barbieri}, M. 2007, \aap, 462, 345,
  \dodoi{10.1051/0004-6361:20066319}

\bibitem[{{Dhital} {et~al.}(2011){Dhital}, {Burgasser}, {Looper}, \&
  {Stassun}}]{2011AJ....141....7D}
{Dhital}, S., {Burgasser}, A.~J., {Looper}, D.~L., \& {Stassun}, K.~G. 2011,
  \aj, 141, 7, \dodoi{10.1088/0004-6256/141/1/7}

\bibitem[{{Dupuy} \& {Liu}(2012)}]{2012ApJS..201...19D}
{Dupuy}, T.~J., \& {Liu}, M.~C. 2012, \apjs, 201, 19,
  \dodoi{10.1088/0067-0049/201/2/19}

\bibitem[{{Dupuy} \& {Liu}(2017)}]{2017ApJS..231...15D}
---. 2017, \apjs, 231, 15, \dodoi{10.3847/1538-4365/aa5e4c}

\bibitem[{{Dupuy} {et~al.}(2009){Dupuy}, {Liu}, \&
  {Ireland}}]{2009ApJ...699..168D}
{Dupuy}, T.~J., {Liu}, M.~C., \& {Ireland}, M.~J. 2009, \apj, 699, 168,
  \dodoi{10.1088/0004-637X/699/1/168}

\bibitem[{{Dupuy} {et~al.}(2015){Dupuy}, {Liu}, {Leggett}, {Ireland }, {Chiu},
  \& {Golimowski}}]{2015ApJ...805...56D}
{Dupuy}, T.~J., {Liu}, M.~C., {Leggett}, S.~K., {et~al.} 2015, \apj, 805, 56,
  \dodoi{10.1088/0004-637X/805/1/56}

\bibitem[{{Dupuy} {et~al.}(2018){Dupuy}, {Liu}, {Allers}, {Biller}, {Kratter},
  {Mann}, {Shkolnik}, {Kraus}, \& {Best}}]{2018AJ....156...57D}
{Dupuy}, T.~J., {Liu}, M.~C., {Allers}, K.~N., {et~al.} 2018, \aj, 156, 57,
  \dodoi{10.3847/1538-3881/aacbc2}

\bibitem[{{Dupuy} {et~al.}(2019){Dupuy}, {Liu}, {Best}, {Mann}, {Tucker},
  {Zhang}, {Baraffe}, {Chabrier}, {Forveille}, {Metchev}, {Tremblin}, {Do},
  {Payne}, {Shappee}, {Bond}, {Cetre}, {Chun}, {Delorme}, {Jovanovic},
  {Lilley}, {Mawet}, {Ragland }, {Wetherell}, \&
  {Wizinowich}}]{2019AJ....158..174D}
{Dupuy}, T.~J., {Liu}, M.~C., {Best}, W. M.~J., {et~al.} 2019, \aj, 158, 174,
  \dodoi{10.3847/1538-3881/ab3cd1}

\bibitem[{{Dye} {et~al.}(2018){Dye}, {Lawrence}, {Read}, {Fan}, {Kerr},
  {Varricatt}, {Furnell}, {Edge}, {Irwin}, {Hambly}, {Lucas}, {Almaini},
  {Chambers}, {Green}, {Hewett}, {Liu}, {McGreer}, {Best}, {Zhang}, {Sutorius},
  {Froebrich}, {Magnier}, {Hasinger}, {Lederer}, {Bold}, \&
  {Tedds}}]{2018MNRAS.473.5113D}
{Dye}, S., {Lawrence}, A., {Read}, M.~A., {et~al.} 2018, \mnras, 473, 5113,
  \dodoi{10.1093/mnras/stx2622}

\bibitem[{{Edge} {et~al.}(2016){Edge}, {Sutherland}, \& {Viking
  Team}}]{2016yCat.2343....0E}
{Edge}, A., {Sutherland}, W., \& {Viking Team}. 2016, VizieR Online Data
  Catalog, II/343

\bibitem[{{Edwards}(1976)}]{1976AJ.....81..245E}
{Edwards}, T.~W. 1976, \aj, 81, 245, \dodoi{10.1086/111879}

\bibitem[{{Eisenstein} {et~al.}(2006){Eisenstein}, {Liebert}, {Harris},
  {Kleinman}, {Nitta}, {Silvestri}, {Anderson}, {Barentine}, {Brewington},
  {Brinkmann}, {Harvanek}, {Krzesi{\'n}ski}, {Neilsen}, {Long}, {Schneider}, \&
  {Snedden}}]{2006ApJS..167...40E}
{Eisenstein}, D.~J., {Liebert}, J., {Harris}, H.~C., {et~al.} 2006, \apjs, 167,
  40, \dodoi{10.1086/507110}

\bibitem[{{Emerson} {et~al.}(2004){Emerson}, {Sutherland}, {McPherson},
  {Craig}, {Dalton}, \& {Ward}}]{2004Msngr.117...27E}
{Emerson}, J.~P., {Sutherland}, W.~J., {McPherson}, A.~M., {et~al.} 2004, The
  Messenger, 117, 27

\bibitem[{{Evans} {et~al.}(2018){Evans}, {Riello}, {De Angeli}, {Carrasco},
  {Montegriffo}, {Fabricius}, {Jordi}, {Palaversa}, {Diener}, \&
  {Busso}}]{2018AandA...616A...4E}
{Evans}, D.~W., {Riello}, M., {De Angeli}, F., {et~al.} 2018, \aap, 616, A4,
  \dodoi{10.1051/0004-6361/201832756}

\bibitem[{{Faherty} {et~al.}(2014){Faherty}, {Beletsky}, {Burgasser}, {Tinney},
  {Osip}, {Filippazzo}, \& {Simcoe}}]{2014ApJ...790...90F}
{Faherty}, J.~K., {Beletsky}, Y., {Burgasser}, A.~J., {et~al.} 2014, \apj, 790,
  90, \dodoi{10.1088/0004-637X/790/2/90}

\bibitem[{{Faherty} {et~al.}(2010){Faherty}, {Burgasser}, {West}, {Bochanski},
  {Cruz}, {Shara}, \& {Walter}}]{2010AJ....139..176F}
{Faherty}, J.~K., {Burgasser}, A.~J., {West}, A.~A., {et~al.} 2010, \aj, 139,
  176, \dodoi{10.1088/0004-6256/139/1/176}

\bibitem[{{Faherty} {et~al.}(2013){Faherty}, {Rice}, {Cruz}, {Mamajek}, \&
  {N{\'u}{\~n}ez}}]{2013AJ....145....2F}
{Faherty}, J.~K., {Rice}, E.~L., {Cruz}, K.~L., {Mamajek}, E.~E., \&
  {N{\'u}{\~n}ez}, A. 2013, \aj, 145, 2, \dodoi{10.1088/0004-6256/145/1/2}

\bibitem[{{Faherty} {et~al.}(2012){Faherty}, {Burgasser}, {Walter}, {Van der
  Bliek}, {Shara}, {Cruz}, {West}, {Vrba}, \&
  {Anglada-Escud{\'e}}}]{2012ApJ...752...56F}
{Faherty}, J.~K., {Burgasser}, A.~J., {Walter}, F.~M., {et~al.} 2012, \apj,
  752, 56, \dodoi{10.1088/0004-637X/752/1/56}

\bibitem[{{Faherty} {et~al.}(2016){Faherty}, {Riedel}, {Cruz}, {Gagne},
  {Filippazzo}, {Lambrides}, {Fica}, {Weinberger}, {Thorstensen}, {Tinney},
  {Baldassare}, {Lemonier}, \& {Rice}}]{2016ApJS..225...10F}
{Faherty}, J.~K., {Riedel}, A.~R., {Cruz}, K.~L., {et~al.} 2016, \apjs, 225,
  10, \dodoi{10.3847/0067-0049/225/1/10}

\bibitem[{{Feigelson} {et~al.}(2006){Feigelson}, {Lawson}, {Stark}, {Townsley},
  \& {Garmire}}]{2006AJ....131.1730F}
{Feigelson}, E.~D., {Lawson}, W.~A., {Stark}, M., {Townsley}, L., \& {Garmire},
  G.~P. 2006, \aj, 131, 1730, \dodoi{10.1086/499923}

\bibitem[{{Fields} {et~al.}(2016){Fields}, {Farmer}, {Petermann}, {Iliadis}, \&
  {Timmes}}]{2016ApJ...823...46F}
{Fields}, C.~E., {Farmer}, R., {Petermann}, I., {Iliadis}, C., \& {Timmes},
  F.~X. 2016, \apj, 823, 46, \dodoi{10.3847/0004-637X/823/1/46}

\bibitem[{{Filippazzo} {et~al.}(2015){Filippazzo}, {Rice}, {Faherty}, {Cruz},
  {Van Gordon}, \& {Looper}}]{2015ApJ...810..158F}
{Filippazzo}, J.~C., {Rice}, E.~L., {Faherty}, J., {et~al.} 2015, \apj, 810,
  158, \dodoi{10.1088/0004-637X/810/2/158}

\bibitem[{{Fontaine} {et~al.}(2001){Fontaine}, {Brassard}, \&
  {Bergeron}}]{2001PASP..113..409F}
{Fontaine}, G., {Brassard}, P., \& {Bergeron}, P. 2001, \pasp, 113, 409,
  \dodoi{10.1086/319535}

\bibitem[{{Foreman-Mackey} {et~al.}(2013){Foreman-Mackey}, {Hogg}, {Lang}, \&
  {Goodman}}]{2013PASP..125..306F}
{Foreman-Mackey}, D., {Hogg}, D.~W., {Lang}, D., \& {Goodman}, J. 2013, \pasp,
  125, 306, \dodoi{10.1086/670067}

\bibitem[{{Fortney} {et~al.}(2008){Fortney}, {Marley}, {Saumon}, \&
  {Lodders}}]{2008ApJ...683.1104F}
{Fortney}, J.~J., {Marley}, M.~S., {Saumon}, D., \& {Lodders}, K. 2008, \apj,
  683, 1104, \dodoi{10.1086/589942}

\bibitem[{{Gagn{\'e}} {et~al.}(2018){Gagn{\'e}}, {Allers}, {Theissen},
  {Faherty}, {Bardalez Gagliuffi}, \& {Artigau}}]{2018ApJ...854L..27G}
{Gagn{\'e}}, J., {Allers}, K.~N., {Theissen}, C.~A., {et~al.} 2018, \apjl, 854,
  L27, \dodoi{10.3847/2041-8213/aaacfd}

\bibitem[{{Gagn{\'e}} {et~al.}(2015){Gagn{\'e}}, {Burgasser}, {Faherty},
  {Lafreni{\'e}re}, {Doyon}, {Filippazzo}, {Bowsher}, \&
  {Nicholls}}]{2015ApJ...808L..20G}
{Gagn{\'e}}, J., {Burgasser}, A.~J., {Faherty}, J.~K., {et~al.} 2015, \apjl,
  808, L20, \dodoi{10.1088/2041-8205/808/1/L20}

\bibitem[{{Gagn{\'e}} {et~al.}(2017){Gagn{\'e}}, {Faherty}, {Burgasser},
  {Artigau}, {Bouchard}, {Albert}, {Lafreni{\`e}re}, {Doyon}, \& {Bardalez
  Gagliuffi}}]{2017ApJ...841L...1G}
{Gagn{\'e}}, J., {Faherty}, J.~K., {Burgasser}, A.~J., {et~al.} 2017, \apjl,
  841, L1, \dodoi{10.3847/2041-8213/aa70e2}

\bibitem[{{Gaia Collaboration} {et~al.}(2016){Gaia Collaboration}, {Prusti},
  {de Bruijne}, {Brown}, {Vallenari}, {Babusiaux}, {Bailer-Jones}, {Bastian},
  {Biermann}, {Evans}, {Eyer}, {Jansen}, {Jordi}, {Klioner}, {Lammers},
  {Lindegren}, {Luri}, {Mignard}, {Milligan}, {Panem}, {Poinsignon},
  {Pourbaix}, {Randich}, {Sarri}, {Sartoretti}, {Siddiqui}, {Soubiran},
  {Valette}, {van Leeuwen}, {Walton}, {Aerts}, {Arenou}, {Cropper}, {Drimmel},
  {H{\o}g}, {Katz}, {Lattanzi}, {O'Mullane}, {Grebel}, {Holland}, {Huc},
  {Passot}, {Bramante}, {Cacciari}, {Casta{\~n}eda}, {Chaoul}, {Cheek}, {De
  Angeli}, {Fabricius}, {Guerra}, {Hern{\'a}ndez}, {Jean-Antoine-Piccolo},
  {Masana}, {Messineo}, {Mowlavi}, {Nienartowicz}, {Ord{\'o}{\~n}ez-Blanco},
  {Panuzzo}, {Portell}, {Richards}, {Riello}, {Seabroke}, {Tanga},
  {Th{\'e}venin}, {Torra}, {Els}, {Gracia-Abril}, {Comoretto},
  {Garcia-Reinaldos}, {Lock}, {Mercier}, {Altmann}, {Andrae}, {Astraatmadja},
  {Bellas-Velidis}, {Benson}, {Berthier}, {Blomme}, {Busso}, {Carry},
  {Cellino}, {Clementini}, {Cowell}, {Creevey}, {Cuypers}, {Davidson}, {De
  Ridder}, {de Torres}, {Delchambre}, {Dell'Oro}, {Ducourant}, {Fr{\'e}mat},
  {Garc{\'\i}a-Torres}, {Gosset}, {Halbwachs}, {Hambly}, {Harrison}, {Hauser},
  {Hestroffer}, {Hodgkin}, {Huckle}, {Hutton}, {Jasniewicz}, {Jordan},
  {Kontizas}, {Korn}, {Lanzafame}, {Manteiga}, {Moitinho}, {Muinonen},
  {Osinde}, {Pancino}, {Pauwels}, {Petit}, {Recio-Blanco}, {Robin}, {Sarro},
  {Siopis}, {Smith}, {Smith}, {Sozzetti}, {Thuillot}, {van Reeven}, {Viala},
  {Abbas}, {Abreu Aramburu}, {Accart}, {Aguado}, {Allan}, {Allasia},
  {Altavilla}, {{\'A}lvarez}, {Alves}, {Anderson}, {Andrei}, {Anglada Varela},
  {Antiche}, {Antoja}, {Ant{\'o}n}, {Arcay}, {Atzei}, {Ayache}, {Bach},
  {Baker}, {Balaguer-N{\'u}{\~n}ez}, {Barache}, {Barata}, {Barbier}, {Barblan},
  {Baroni}, {Barrado y Navascu{\'e}s}, {Barros}, {Barstow}, {Becciani},
  {Bellazzini}, {Bellei}, {Bello Garc{\'\i}a}, {Belokurov}, {Bendjoya},
  {Berihuete}, {Bianchi}, {Bienaym{\'e}}, {Billebaud}, {Blagorodnova},
  {Blanco-Cuaresma}, {Boch}, {Bombrun}, {Borrachero}, {Bouquillon}, {Bourda},
  {Bouy}, {Bragaglia}, {Breddels}, {Brouillet}, {Br{\"u}semeister},
  {Bucciarelli}, {Budnik}, {Burgess}, {Burgon}, {Burlacu}, {Busonero}, {Buzzi},
  {Caffau}, {Cambras}, {Campbell}, {Cancelliere}, {Cantat-Gaudin}, {Carlucci},
  {Carrasco}, {Castellani}, {Charlot}, {Charnas}, {Charvet}, {Chassat},
  {Chiavassa}, {Clotet}, {Cocozza}, {Collins}, {Collins}, {Costigan}, {Crifo},
  {Cross}, {Crosta}, {Crowley}, {Dafonte}, {Damerdji}, {Dapergolas}, {David},
  {David}, {De Cat}, {de Felice}, {de Laverny}, {De Luise}, {De March}, {de
  Martino}, {de Souza}, {Debosscher}, {del Pozo}, {Delbo}, {Delgado},
  {Delgado}, {di Marco}, {Di Matteo}, {Diakite}, {Distefano}, {Dolding}, {Dos
  Anjos}, {Drazinos}, {Dur{\'a}n}, {Dzigan}, {Ecale}, {Edvardsson}, {Enke},
  {Erdmann}, {Escolar}, {Espina}, {Evans}, {Eynard Bontemps}, {Fabre},
  {Fabrizio}, {Faigler}, {Falc{\~a}o}, {Farr{\`a}s Casas}, {Faye}, {Federici},
  {Fedorets}, {Fern{\'a}ndez-Hern{\'a}ndez}, {Fernique}, {Fienga}, {Figueras},
  {Filippi}, {Findeisen}, {Fonti}, {Fouesneau}, {Fraile}, {Fraser}, {Fuchs},
  {Furnell}, {Gai}, {Galleti}, {Galluccio}, {Garabato}, {Garc{\'\i}a-Sedano},
  {Gar{\'e}}, {Garofalo}, {Garralda}, {Gavras}, {Gerssen}, {Geyer}, {Gilmore},
  {Girona}, {Giuffrida}, {Gomes}, {Gonz{\'a}lez-Marcos},
  {Gonz{\'a}lez-N{\'u}{\~n}ez}, {Gonz{\'a}lez-Vidal}, {Granvik}, {Guerrier},
  {Guillout}, {Guiraud}, {G{\'u}rpide}, {Guti{\'e}rrez-S{\'a}nchez}, {Guy},
  {Haigron}, {Hatzidimitriou}, {Haywood}, {Heiter}, {Helmi}, {Hobbs},
  {Hofmann}, {Holl}, {Holland }, {Hunt}, {Hypki}, {Icardi}, {Irwin}, {Jevardat
  de Fombelle}, {Jofr{\'e}}, {Jonker}, {Jorissen}, {Julbe}, {Karampelas},
  {Kochoska}, {Kohley}, {Kolenberg}, {Kontizas}, {Koposov}, {Kordopatis},
  {Koubsky}, {Kowalczyk}, {Krone-Martins}, {Kudryashova}, {Kull}, {Bachchan},
  {Lacoste-Seris}, {Lanza}, {Lavigne}, {Le Poncin-Lafitte}, {Lebreton},
  {Lebzelter}, {Leccia}, {Leclerc}, {Lecoeur-Taibi}, {Lemaitre}, {Lenhardt},
  {Leroux}, {Liao}, {Licata}, {Lindstr{\o}m}, {Lister}, {Livanou}, {Lobel},
  {L{\"o}ffler}, {L{\'o}pez}, {Lopez-Lozano}, {Lorenz}, {Loureiro},
  {MacDonald}, {Magalh{\~a}es Fernandes}, {Managau}, {Mann}, {Mantelet},
  {Marchal}, {Marchant}, {Marconi}, {Marie}, {Marinoni}, {Marrese},
  {Marschalk{\'o}}, {Marshall}, {Mart{\'\i}n-Fleitas}, {Martino}, {Mary},
  {Matijevi{\v{c}}}, {Mazeh}, {McMillan}, {Messina}, {Mestre}, {Michalik},
  {Millar}, {Miranda}, {Molina}, {Molinaro}, {Molinaro}, {Moln{\'a}r},
  {Moniez}, {Montegriffo}, {Monteiro}, {Mor}, {Mora}, {Morbidelli}, {Morel},
  {Morgenthaler}, {Morley}, {Morris}, {Mulone}, {Muraveva}, {Musella},
  {Narbonne}, {Nelemans}, {Nicastro}, {Noval}, {Ord{\'e}novic},
  {Ordieres-Mer{\'e}}, {Osborne}, {Pagani}, {Pagano}, {Pailler}, {Palacin},
  {Palaversa}, {Parsons}, {Paulsen}, {Pecoraro}, {Pedrosa}, {Pentik{\"a}inen},
  {Pereira}, {Pichon}, {Piersimoni}, {Pineau}, {Plachy}, {Plum}, {Poujoulet},
  {Pr{\v{s}}a}, {Pulone}, {Ragaini}, {Rago}, {Rambaux}, {Ramos-Lerate},
  {Ranalli}, {Rauw}, {Read}, {Regibo}, {Renk}, {Reyl{\'e}}, {Ribeiro},
  {Rimoldini}, {Ripepi}, {Riva}, {Rixon}, {Roelens}, {Romero-G{\'o}mez},
  {Rowell}, {Royer}, {Rudolph}, {Ruiz-Dern}, {Sadowski}, {Sagrist{\`a}
  Sell{\'e}s}, {Sahlmann}, {Salgado}, {Salguero}, {Sarasso}, {Savietto},
  {Schnorhk}, {Schultheis}, {Sciacca}, {Segol}, {Segovia}, {Segransan},
  {Serpell}, {Shih}, {Smareglia}, {Smart}, {Smith}, {Solano}, {Solitro},
  {Sordo}, {Soria Nieto}, {Souchay}, {Spagna}, {Spoto}, {Stampa}, {Steele},
  {Steidelm{\"u}ller}, {Stephenson}, {Stoev}, {Suess}, {S{\"u}veges}, {Surdej},
  {Szabados}, {Szegedi-Elek}, {Tapiador}, {Taris}, {Tauran}, {Taylor},
  {Teixeira}, {Terrett}, {Tingley}, {Trager}, {Turon}, {Ulla}, {Utrilla},
  {Valentini}, {van Elteren}, {Van Hemelryck}, {van Leeuwen}, {Varadi},
  {Vecchiato}, {Veljanoski}, {Via}, {Vicente}, {Vogt}, {Voss}, {Votruba},
  {Voutsinas}, {Walmsley}, {Weiler}, {Weingrill}, {Werner}, {Wevers},
  {Whitehead}, {Wyrzykowski}, {Yoldas}, {{\v{Z}}erjal}, {Zucker}, {Zurbach},
  {Zwitter}, {Alecu}, {Allen}, {Allende Prieto}, {Amorim},
  {Anglada-Escud{\'e}}, {Arsenijevic}, {Azaz}, {Balm}, {Beck}, {Bernstein},
  {Bigot}, {Bijaoui}, {Blasco}, {Bonfigli}, {Bono}, {Boudreault}, {Bressan},
  {Brown}, {Brunet}, {Bunclark}, {Buonanno}, {Butkevich}, {Carret}, {Carrion},
  {Chemin}, {Ch{\'e}reau}, {Corcione}, {Darmigny}, {de Boer}, {de Teodoro}, {de
  Zeeuw}, {Delle Luche}, {Domingues}, {Dubath}, {Fodor}, {Fr{\'e}zouls},
  {Fries}, {Fustes}, {Fyfe}, {Gallardo}, {Gallegos}, {Gardiol}, {Gebran},
  {Gomboc}, {G{\'o}mez}, {Grux}, {Gueguen}, {Heyrovsky}, {Hoar}, {Iannicola},
  {Isasi Parache}, {Janotto}, {Joliet}, {Jonckheere}, {Keil}, {Kim},
  {Klagyivik}, {Klar}, {Knude}, {Kochukhov}, {Kolka}, {Kos}, {Kutka}, {Lainey},
  {LeBouquin}, {Liu}, {Loreggia}, {Makarov}, {Marseille}, {Martayan},
  {Martinez-Rubi}, {Massart}, {Meynadier}, {Mignot}, {Munari}, {Nguyen},
  {Nordlander}, {Ocvirk}, {O'Flaherty}, {Olias Sanz}, {Ortiz}, {Osorio},
  {Oszkiewicz}, {Ouzounis}, {Palmer}, {Park}, {Pasquato}, {Peltzer}, {Peralta},
  {P{\'e}turaud}, {Pieniluoma}, {Pigozzi}, {Poels}, {Prat}, {Prod'homme},
  {Raison}, {Rebordao}, {Risquez}, {Rocca-Volmerange}, {Rosen}, {Ruiz-Fuertes},
  {Russo}, {Sembay}, {Serraller Vizcaino}, {Short}, {Siebert}, {Silva},
  {Sinachopoulos}, {Slezak}, {Soffel}, {Sosnowska}, {Strai{\v{z}}ys}, {ter
  Linden}, {Terrell}, {Theil}, {Tiede}, {Troisi}, {Tsalmantza}, {Tur},
  {Vaccari}, {Vachier}, {Valles}, {Van Hamme}, {Veltz}, {Virtanen}, {Wallut},
  {Wichmann}, {Wilkinson}, {Ziaeepour}, \& {Zschocke}}]{2016AandA...595A...1G}
{Gaia Collaboration}, {Prusti}, T., {de Bruijne}, J.~H.~J., {et~al.} 2016,
  \aap, 595, A1, \dodoi{10.1051/0004-6361/201629272}

\bibitem[{{Gaia Collaboration} {et~al.}(2018){Gaia Collaboration}, {Brown},
  {Vallenari}, {Prusti}, {de Bruijne}, {Babusiaux}, {Bailer-Jones}, {Biermann},
  {Evans}, {Eyer}, \& et~al.}]{2018AandA...616A...1G}
{Gaia Collaboration}, {Brown}, A.~G.~A., {Vallenari}, A., {et~al.} 2018, \aap,
  616, A1, \dodoi{10.1051/0004-6361/201833051}

\bibitem[{{Garcia} {et~al.}(2017){Garcia}, {Ammons}, {Salama}, {Crossfield},
  {Bendek}, {Chilcote}, {Garrel}, {Graham}, {Kalas}, {Konopacky}, {Lu},
  {Macintosh}, {Marin}, {Marois}, {Nielsen}, {Neichel}, {Pham}, {De Rosa},
  {Ryan}, {Service}, \& {Sivo}}]{2017ApJ...846...97G}
{Garcia}, E.~V., {Ammons}, S.~M., {Salama}, M., {et~al.} 2017, \apj, 846, 97,
  \dodoi{10.3847/1538-4357/aa844f}

\bibitem[{{Gauza} {et~al.}(2015){Gauza}, {B{\'e}jar}, {P{\'e}rez-Garrido},
  {Zapatero Osorio}, {Lodieu}, {Rebolo}, {Pall{\'e}}, \&
  {Nowak}}]{2015ApJ...804...96G}
{Gauza}, B., {B{\'e}jar}, V. J.~S., {P{\'e}rez-Garrido}, A., {et~al.} 2015,
  \apj, 804, 96, \dodoi{10.1088/0004-637X/804/2/96}

\bibitem[{{Gauza} {et~al.}(2019){Gauza}, {B{\'e}jar}, {P{\'e}rez-Garrido},
  {Lodieu}, {Rebolo}, {Zapatero Osorio}, {Pantoja}, {Velasco}, \&
  {Jenkins}}]{2019MNRAS.487.1149G}
{Gauza}, B., {B{\'e}jar}, V.~J.~S., {P{\'e}rez-Garrido}, A., {et~al.} 2019,
  \mnras, 487, 1149, \dodoi{10.1093/mnras/stz1284}

\bibitem[{{Geballe} {et~al.}(2002){Geballe}, {Knapp}, {Leggett}, {Fan},
  {Golimowski}, {Anderson}, {Brinkmann}, {Csabai}, {Gunn}, {Hawley},
  {Hennessy}, {Henry}, {Hill}, {Hindsley}, {Ivezi{\'c}}, {Lupton}, {McDaniel},
  {Munn}, {Narayanan}, {Peng}, {Pier}, {Rockosi}, {Schneider}, {Smith},
  {Strauss}, {Tsvetanov}, {Uomoto}, {York}, \& {Zheng}}]{2002ApJ...564..466G}
{Geballe}, T.~R., {Knapp}, G.~R., {Leggett}, S.~K., {et~al.} 2002, \apj, 564,
  466, \dodoi{10.1086/324078}

\bibitem[{{Gentile Fusillo} {et~al.}(2019){Gentile Fusillo}, {Tremblay},
  {G{\"a}nsicke}, {Manser}, {Cunningham}, {Cukanovaite}, {Hollands}, {Marsh},
  {Raddi}, \& {Jordan}}]{2019MNRAS.482.4570G}
{Gentile Fusillo}, N.~P., {Tremblay}, P.-E., {G{\"a}nsicke}, B.~T., {et~al.}
  2019, \mnras, 482, 4570, \dodoi{10.1093/mnras/sty3016}

\bibitem[{{Giammichele} {et~al.}(2012){Giammichele}, {Bergeron}, \&
  {Dufour}}]{2012ApJS..199...29G}
{Giammichele}, N., {Bergeron}, P., \& {Dufour}, P. 2012, \apjs, 199, 29,
  \dodoi{10.1088/0067-0049/199/2/29}

\bibitem[{{Gizis}(1998)}]{1998AJ....115.2053G}
{Gizis}, J.~E. 1998, \aj, 115, 2053, \dodoi{10.1086/300325}

\bibitem[{{Gizis} {et~al.}(2015){Gizis}, {Allers}, {Liu}, {Harris}, {Faherty},
  {Burgasser}, \& {Kirkpatrick}}]{2015ApJ...799..203G}
{Gizis}, J.~E., {Allers}, K.~N., {Liu}, M.~C., {et~al.} 2015, \apj, 799, 203,
  \dodoi{10.1088/0004-637X/799/2/203}

\bibitem[{{Gizis} {et~al.}(2003){Gizis}, {Reid}, {Knapp}, {Liebert},
  {Kirkpatrick}, {Koerner}, \& {Burgasser}}]{2003AJ....125.3302G}
{Gizis}, J.~E., {Reid}, I.~N., {Knapp}, G.~R., {et~al.} 2003, \aj, 125, 3302,
  \dodoi{10.1086/374991}

\bibitem[{{Gizis} {et~al.}(2012){Gizis}, {Faherty}, {Liu}, {Castro}, {Shaw},
  {Vrba}, {Harris}, {Aller}, \& {Deacon}}]{2012AJ....144...94G}
{Gizis}, J.~E., {Faherty}, J.~K., {Liu}, M.~C., {et~al.} 2012, \aj, 144, 94,
  \dodoi{10.1088/0004-6256/144/4/94}

\bibitem[{{Golimowski} {et~al.}(2004){Golimowski}, {Leggett}, {Marley}, {Fan},
  {Geballe}, {Knapp}, {Vrba}, {Henden}, {Luginbuhl}, {Guetter}, {Munn},
  {Canzian}, {Zheng}, {Tsvetanov}, {Chiu}, {Glazebrook}, {Hoversten},
  {Schneider}, \& {Brinkmann}}]{2004AJ....127.3516G}
{Golimowski}, D.~A., {Leggett}, S.~K., {Marley}, M.~S., {et~al.} 2004, \aj,
  127, 3516, \dodoi{10.1086/420709}

\bibitem[{{Gonz{\'a}lez-Fern{\'a}ndez}
  {et~al.}(2018){Gonz{\'a}lez-Fern{\'a}ndez}, {Hodgkin}, {Irwin},
  {Gonz{\'a}lez-Solares}, {Koposov}, {Lewis}, {Emerson}, {Hewett},
  {Yolda{\textcommabelow s}}, \& {Riello}}]{2018MNRAS.474.5459G}
{Gonz{\'a}lez-Fern{\'a}ndez}, C., {Hodgkin}, S.~T., {Irwin}, M.~J., {et~al.}
  2018, \mnras, 474, 5459, \dodoi{10.1093/mnras/stx3073}

\bibitem[{{Goodman} \& {Weare}(2010)}]{2010CAMCS...5...65G}
{Goodman}, J., \& {Weare}, J. 2010, Communications in Applied Mathematics and
  Computational Science, 5, 65, \dodoi{10.2140/camcos.2010.5.65}

\bibitem[{{Gray} {et~al.}(2006){Gray}, {Corbally}, {Garrison}, {McFadden},
  {Bubar}, {McGahee}, {O'Donoghue}, \& {Knox}}]{2006AJ....132..161G}
{Gray}, R.~O., {Corbally}, C.~J., {Garrison}, R.~F., {et~al.} 2006, \aj, 132,
  161, \dodoi{10.1086/504637}

\bibitem[{{Gray} {et~al.}(2003){Gray}, {Corbally}, {Garrison}, {McFadden}, \&
  {Robinson}}]{2003AJ....126.2048G}
{Gray}, R.~O., {Corbally}, C.~J., {Garrison}, R.~F., {McFadden}, M.~T., \&
  {Robinson}, P.~E. 2003, \aj, 126, 2048, \dodoi{10.1086/378365}

\bibitem[{{Gray} {et~al.}(2001){Gray}, {Napier}, \&
  {Winkler}}]{2001AJ....121.2148G}
{Gray}, R.~O., {Napier}, M.~G., \& {Winkler}, L.~I. 2001, \aj, 121, 2148,
  \dodoi{10.1086/319956}

\bibitem[{{Greenbaum} {et~al.}(2018){Greenbaum}, {Pueyo}, {Ruffio}, {Wang}, {De
  Rosa}, {Aguilar}, {Rameau}, {Barman}, {Marois}, {Marley}, {Konopacky},
  {Rajan}, {Macintosh}, {Ansdell}, {Arriaga}, {Bailey}, {Bulger}, {Burrows},
  {Chilcote}, {Cotten}, {Doyon}, {Duch{\^e}ne}, {Fitzgerald}, {Follette},
  {Gerard}, {Goodsell}, {Graham}, {Hibon}, {Hung}, {Ingraham}, {Kalas},
  {Larkin}, {Maire}, {Marchis}, {Metchev}, {Millar-Blanchaer}, {Nielsen},
  {Norton}, {Oppenheimer}, {Palmer}, {Patience}, {Perrin}, {Poyneer},
  {Rantakyr{\"o}}, {Savransky}, {Schneider}, {Sivaramakrishnan}, {Song},
  {Soummer}, {Thomas}, {Wallace}, {Ward-Duong}, {Wiktorowicz}, \&
  {Wolff}}]{2018AJ....155..226G}
{Greenbaum}, A.~Z., {Pueyo}, L., {Ruffio}, J.-B., {et~al.} 2018, \aj, 155, 226,
  \dodoi{10.3847/1538-3881/aabcb8}

\bibitem[{{Hewett} {et~al.}(2006){Hewett}, {Warren}, {Leggett}, \&
  {Hodgkin}}]{2006MNRAS.367..454H}
{Hewett}, P.~C., {Warren}, S.~J., {Leggett}, S.~K., \& {Hodgkin}, S.~T. 2006,
  \mnras, 367, 454, \dodoi{10.1111/j.1365-2966.2005.09969.x}

\bibitem[{{Hiranaka} {et~al.}(2016){Hiranaka}, {Cruz}, {Douglas}, {Marley}, \&
  {Baldassare}}]{2016ApJ...830...96H}
{Hiranaka}, K., {Cruz}, K.~L., {Douglas}, S.~T., {Marley}, M.~S., \&
  {Baldassare}, V.~F. 2016, \apj, 830, 96, \dodoi{10.3847/0004-637X/830/2/96}

\bibitem[{{Hollands} {et~al.}(2018){Hollands}, {Tremblay}, {G{\"a}nsicke},
  {Gentile-Fusillo}, \& {Toonen}}]{2018MNRAS.480.3942H}
{Hollands}, M.~A., {Tremblay}, P.~E., {G{\"a}nsicke}, B.~T., {Gentile-Fusillo},
  N.~P., \& {Toonen}, S. 2018, \mnras, 480, 3942, \dodoi{10.1093/mnras/sty2057}

\bibitem[{{Houk}(1978)}]{1978mcts.book.....H}
{Houk}, N. 1978, {Michigan catalogue of two-dimensional spectral types for the
  HD stars}

\bibitem[{{Houk} \& {Smith-Moore}(1988)}]{1988mcts.book.....H}
{Houk}, N., \& {Smith-Moore}, M. 1988, {Michigan Catalogue of Two-dimensional
  Spectral Types for the HD Stars. Volume 4.}, Vol.~4

\bibitem[{{Houk} \& {Swift}(1999)}]{1999MSS...C05....0H}
{Houk}, N., \& {Swift}, C. 1999, Michigan Spectral Survey, 5, 0

\bibitem[{Hunter(2007)}]{Hunter:2007}
Hunter, J.~D. 2007, Computing in Science \& Engineering, 9, 90,
  \dodoi{10.1109/MCSE.2007.55}

\bibitem[{{James}(1964)}]{1964ApJ...140..552J}
{James}, R.~A. 1964, \apj, 140, 552, \dodoi{10.1086/147949}

\bibitem[{{Janson} {et~al.}(2012){Janson}, {Hormuth}, {Bergfors}, {Brand ner},
  {Hippler}, {Daemgen}, {Kudryavtseva}, {Schmalzl}, {Schnupp}, \&
  {Henning}}]{2012ApJ...754...44J}
{Janson}, M., {Hormuth}, F., {Bergfors}, C., {et~al.} 2012, \apj, 754, 44,
  \dodoi{10.1088/0004-637X/754/1/44}

\bibitem[{{Janson} {et~al.}(2014){Janson}, {Bergfors}, {Brandner}, {Bonnefoy},
  {Schlieder}, {K{\"o}hler}, {Hormuth}, {Henning}, \&
  {Hippler}}]{2014ApJS..214...17J}
{Janson}, M., {Bergfors}, C., {Brandner}, W., {et~al.} 2014, \apjs, 214, 17,
  \dodoi{10.1088/0067-0049/214/2/17}

\bibitem[{{Jarrett} {et~al.}(2011){Jarrett}, {Cohen}, {Masci}, {Wright},
  {Stern}, {Benford}, {Blain}, {Carey}, {Cutri}, \&
  {Eisenhardt}}]{2011ApJ...735..112J}
{Jarrett}, T.~H., {Cohen}, M., {Masci}, F., {et~al.} 2011, \apj, 735, 112,
  \dodoi{10.1088/0004-637X/735/2/112}

\bibitem[{{Jaschek}(1978)}]{1978BICDS..15..121J}
{Jaschek}, M. 1978, Bulletin d'Information du Centre de Donnees Stellaires, 15,
  121

\bibitem[{{Jeffries} \& {Stevens}(1996)}]{1996MNRAS.279..180J}
{Jeffries}, R.~D., \& {Stevens}, I.~R. 1996, \mnras, 279, 180,
  \dodoi{10.1093/mnras/279.1.180}

\bibitem[{Jones {et~al.}(2001)Jones, Oliphant, Peterson, {et~al.}}]{scipy}
Jones, E., Oliphant, T., Peterson, P., {et~al.} 2001, {SciPy}: Open source
  scientific tools for {Python}.
\newblock \url{http://www.scipy.org/}

\bibitem[{{Kalirai} {et~al.}(2008){Kalirai}, {Hansen}, {Kelson}, {Reitzel},
  {Rich}, \& {Richer}}]{2008ApJ...676..594K}
{Kalirai}, J.~S., {Hansen}, B. M.~S., {Kelson}, D.~D., {et~al.} 2008, The
  Astrophysical Journal, 676, 594, \dodoi{10.1086/527028}

\bibitem[{{Karovska} {et~al.}(1997){Karovska}, {Hack}, {Raymond}, \&
  {Guinan}}]{1997ApJ...482L.175K}
{Karovska}, M., {Hack}, W., {Raymond}, J., \& {Guinan}, E. 1997, \apjl, 482,
  L175, \dodoi{10.1086/310704}

\bibitem[{{Kendall} {et~al.}(2004){Kendall}, {Delfosse}, {Mart{\'\i}n}, \&
  {Forveille}}]{2004AandA...416L..17K}
{Kendall}, T.~R., {Delfosse}, X., {Mart{\'\i}n}, E.~L., \& {Forveille}, T.
  2004, \aap, 416, L17, \dodoi{10.1051/0004-6361:20040046}

\bibitem[{{King} {et~al.}(2010){King}, {McCaughrean}, {Homeier}, {Allard},
  {Scholz}, \& {Lodieu}}]{2010AandA...510A..99K}
{King}, R.~R., {McCaughrean}, M.~J., {Homeier}, D., {et~al.} 2010, \aap, 510,
  A99, \dodoi{10.1051/0004-6361/200912981}

\bibitem[{{Kirkpatrick}(2005)}]{2005ARA&A..43..195K}
{Kirkpatrick}, J.~D. 2005, \araa, 43, 195,
  \dodoi{10.1146/annurev.astro.42.053102.134017}

\bibitem[{{Kirkpatrick} {et~al.}(1999){Kirkpatrick}, {Allard}, {Bida},
  {Zuckerman}, {Becklin}, {Chabrier}, \& {Baraffe}}]{1999ApJ...519..834K}
{Kirkpatrick}, J.~D., {Allard}, F., {Bida}, T., {et~al.} 1999, \apj, 519, 834,
  \dodoi{10.1086/307380}

\bibitem[{{Kirkpatrick} {et~al.}(2001){Kirkpatrick}, {Dahn}, {Monet}, {Reid},
  {Gizis}, {Liebert}, \& {Burgasser}}]{2001AJ....121.3235K}
{Kirkpatrick}, J.~D., {Dahn}, C.~C., {Monet}, D.~G., {et~al.} 2001, \aj, 121,
  3235, \dodoi{10.1086/321085}

\bibitem[{{Kirkpatrick} {et~al.}(2000){Kirkpatrick}, {Reid}, {Liebert},
  {Gizis}, {Burgasser}, {Monet}, {Dahn}, {Nelson}, \&
  {Williams}}]{2000AJ....120..447K}
{Kirkpatrick}, J.~D., {Reid}, I.~N., {Liebert}, J., {et~al.} 2000, \aj, 120,
  447, \dodoi{10.1086/301427}

\bibitem[{{Kirkpatrick} {et~al.}(2008){Kirkpatrick}, {Cruz}, {Barman},
  {Burgasser}, {Looper}, {Tinney}, {Gelino}, {Lowrance}, {Liebert},
  {Carpenter}, {Hillenbrand}, \& {Stauffer}}]{2008ApJ...689.1295K}
{Kirkpatrick}, J.~D., {Cruz}, K.~L., {Barman}, T.~S., {et~al.} 2008, \apj, 689,
  1295, \dodoi{10.1086/592768}

\bibitem[{{Kirkpatrick} {et~al.}(2010){Kirkpatrick}, {Looper}, {Burgasser},
  {Schurr}, {Cutri}, {Cushing}, {Cruz}, {Sweet}, {Knapp}, {Barman},
  {Bochanski}, {Roellig}, {McLean}, {McGovern}, \&
  {Rice}}]{2010ApJS..190..100K}
{Kirkpatrick}, J.~D., {Looper}, D.~L., {Burgasser}, A.~J., {et~al.} 2010,
  \apjs, 190, 100, \dodoi{10.1088/0067-0049/190/1/100}

\bibitem[{{Knapp} {et~al.}(2004){Knapp}, {Leggett}, {Fan}, {Marley}, {Geballe},
  {Golimowski}, {Finkbeiner}, {Gunn}, {Hennawi}, {Ivezi{\'c}}, {Lupton},
  {Schlegel}, {Strauss}, {Tsvetanov}, {Chiu}, {Hoversten}, {Glazebrook},
  {Zheng}, {Hendrickson}, {Williams}, {Uomoto}, {Vrba}, {Henden}, {Luginbuhl},
  {Guetter}, {Munn}, {Canzian}, {Schneider}, \&
  {Brinkmann}}]{2004AJ....127.3553K}
{Knapp}, G.~R., {Leggett}, S.~K., {Fan}, X., {et~al.} 2004, \aj, 127, 3553,
  \dodoi{10.1086/420707}

\bibitem[{{Koen} {et~al.}(2010){Koen}, {Kilkenny}, {van Wyk}, \&
  {Marang}}]{2010MNRAS.403.1949K}
{Koen}, C., {Kilkenny}, D., {van Wyk}, F., \& {Marang}, F. 2010, \mnras, 403,
  1949, \dodoi{10.1111/j.1365-2966.2009.16182.x}

\bibitem[{{Lagrange} {et~al.}(2006){Lagrange}, {Beust}, {Udry}, {Chauvin}, \&
  {Mayor}}]{2006A&A...459..955L}
{Lagrange}, A.~M., {Beust}, H., {Udry}, S., {Chauvin}, G., \& {Mayor}, M. 2006,
  \aap, 459, 955, \dodoi{10.1051/0004-6361:20054710}

\bibitem[{{Lantz} {et~al.}(2004){Lantz}, {Aldering}, {Antilogus}, {Bonnaud},
  {Capoani}, {Castera}, {Copin}, {Dubet}, {Gangler}, \&
  {Henault}}]{2004SPIE.5249..146L}
{Lantz}, B., {Aldering}, G., {Antilogus}, P., {et~al.} 2004, in Society of
  Photo-Optical Instrumentation Engineers (SPIE) Conference Series, Vol. 5249,
  Optical Design and Engineering, ed. L.~{Mazuray}, P.~J. {Rogers}, \&
  R.~{Wartmann}, 146--155

\bibitem[{{Lawrence} {et~al.}(2007){Lawrence}, {Warren}, {Almaini}, {Edge},
  {Hambly}, {Jameson}, {Lucas}, {Casali}, {Adamson}, {Dye}, {Emerson},
  {Foucaud}, {Hewett}, {Hirst}, {Hodgkin}, {Irwin}, {Lodieu}, {McMahon},
  {Simpson}, {Smail}, {Mortlock}, \& {Folger}}]{2007MNRAS.379.1599L}
{Lawrence}, A., {Warren}, S.~J., {Almaini}, O., {et~al.} 2007, \mnras, 379,
  1599, \dodoi{10.1111/j.1365-2966.2007.12040.x}

\bibitem[{{Lawrence} {et~al.}(2012){Lawrence}, {Warren}, {Almaini}, {Edge},
  {Hambly}, {Jameson}, {Lucas}, {Casali}, {Adamson}, {Dye}, {Emerson},
  {Foucaud}, {Hewett}, {Hirst}, {Hodgkin}, {Irwin}, {Lodieu}, {McMahon},
  {Simpson}, {Smail}, {Mortlock}, \& {Folger}}]{2012yCat.2314....0L}
---. 2012, VizieR Online Data Catalog, II/314

\bibitem[{{Leggett} {et~al.}(2015){Leggett}, {Morley}, {Marley}, \&
  {Saumon}}]{2015ApJ...799...37L}
{Leggett}, S.~K., {Morley}, C.~V., {Marley}, M.~S., \& {Saumon}, D. 2015, \apj,
  799, 37, \dodoi{10.1088/0004-637X/799/1/37}

\bibitem[{{Liebert} {et~al.}(2005){Liebert}, {Bergeron}, \&
  {Holberg}}]{2005ApJS..156...47L}
{Liebert}, J., {Bergeron}, P., \& {Holberg}, J.~B. 2005, \apjs, 156, 47,
  \dodoi{10.1086/425738}

\bibitem[{{Lindegren}(2018)}]{Lindegren2018}
{Lindegren}, L. 2018, Gaia Technical Note: GAIA-C3-TN-LU-LL-124-01

\bibitem[{{Lindegren} {et~al.}(2018){Lindegren}, {Hern{\'a}ndez}, {Bombrun},
  {Klioner}, {Bastian}, {Ramos-Lerate}, {de Torres}, {Steidelm{\"u}ller},
  {Stephenson}, {Hobbs}, {Lammers}, {Biermann}, {Geyer}, {Hilger}, {Michalik},
  {Stampa}, {McMillan}, {Casta{\~n}eda}, {Clotet}, {Comoretto}, {Davidson},
  {Fabricius}, {Gracia}, {Hambly}, {Hutton}, {Mora}, {Portell}, {van Leeuwen},
  {Abbas}, {Abreu}, {Altmann}, {Andrei}, {Anglada}, {Balaguer-N{\'u}{\~n}ez},
  {Barache}, {Becciani}, {Bertone}, {Bianchi}, {Bouquillon}, {Bourda},
  {Br{\"u}semeister}, {Bucciarelli}, {Busonero}, {Buzzi}, {Cancelliere},
  {Carlucci}, {Charlot}, {Cheek}, {Crosta}, {Crowley}, {de Bruijne}, {de
  Felice}, {Drimmel}, {Esquej}, {Fienga}, {Fraile}, {Gai}, {Garralda},
  {Gonz{\'a}lez-Vidal}, {Guerra}, {Hauser}, {Hofmann}, {Holl}, {Jordan},
  {Lattanzi}, {Lenhardt}, {Liao}, {Licata}, {Lister}, {L{\"o}ffler},
  {Marchant}, {Martin-Fleitas}, {Messineo}, {Mignard}, {Morbidelli}, {Poggio},
  {Riva}, {Rowell}, {Salguero}, {Sarasso}, {Sciacca}, {Siddiqui}, {Smart},
  {Spagna}, {Steele}, {Taris}, {Torra}, {van Elteren}, {van Reeven}, \&
  {Vecchiato}}]{2018A&A...616A...2L}
{Lindegren}, L., {Hern{\'a}ndez}, J., {Bombrun}, A., {et~al.} 2018, \aap, 616,
  A2, \dodoi{10.1051/0004-6361/201832727}

\bibitem[{{Liu} {et~al.}(2016){Liu}, {Dupuy}, \&
  {Allers}}]{2016ApJ...833...96L}
{Liu}, M.~C., {Dupuy}, T.~J., \& {Allers}, K.~N. 2016, \apj, 833, 96,
  \dodoi{10.3847/1538-4357/833/1/96}

\bibitem[{{Liu} {et~al.}(2008){Liu}, {Dupuy}, \&
  {Ireland}}]{2008ApJ...689..436L}
{Liu}, M.~C., {Dupuy}, T.~J., \& {Ireland}, M.~J. 2008, \apj, 689, 436,
  \dodoi{10.1086/591837}

\bibitem[{{Liu} {et~al.}(2010){Liu}, {Dupuy}, \&
  {Leggett}}]{2010ApJ...722..311L}
{Liu}, M.~C., {Dupuy}, T.~J., \& {Leggett}, S.~K. 2010, \apj, 722, 311,
  \dodoi{10.1088/0004-637X/722/1/311}

\bibitem[{{Liu} {et~al.}(2007){Liu}, {Leggett}, \&
  {Chiu}}]{2007ApJ...660.1507L}
{Liu}, M.~C., {Leggett}, S.~K., \& {Chiu}, K. 2007, \apj, 660, 1507,
  \dodoi{10.1086/512662}

\bibitem[{{Liu} {et~al.}(2006){Liu}, {Leggett}, {Golimowski}, {Chiu}, {Fan},
  {Geballe}, {Schneider}, \& {Brinkmann}}]{2006ApJ...647.1393L}
{Liu}, M.~C., {Leggett}, S.~K., {Golimowski}, D.~A., {et~al.} 2006, \apj, 647,
  1393, \dodoi{10.1086/505561}

\bibitem[{{Liu} {et~al.}(2013){Liu}, {Magnier}, {Deacon}, {Allers}, {Dupuy},
  {Kotson}, {Aller}, {Burgett}, {Chambers}, {Draper}, {Hodapp}, {Jedicke},
  {Kaiser}, {Kudritzki}, {Metcalfe}, {Morgan}, {Price}, {Tonry}, \&
  {Wainscoat}}]{2013ApJ...777L..20L}
{Liu}, M.~C., {Magnier}, E.~A., {Deacon}, N.~R., {et~al.} 2013, \apjl, 777,
  L20, \dodoi{10.1088/2041-8205/777/2/L20}

\bibitem[{{Lodieu} {et~al.}(2014){Lodieu}, {P{\'e}rez-Garrido}, {B{\'e}jar},
  {Gauza}, {Ruiz}, {Rebolo}, {Pinfield}, \&
  {Mart{\'\i}n}}]{2014AandA...569A.120L}
{Lodieu}, N., {P{\'e}rez-Garrido}, A., {B{\'e}jar}, V.~J.~S., {et~al.} 2014,
  \aap, 569, A120, \dodoi{10.1051/0004-6361/201424210}

\bibitem[{{Looper} {et~al.}(2008{\natexlab{a}}){Looper}, {Gelino}, {Burgasser},
  \& {Kirkpatrick}}]{2008ApJ...685.1183L}
{Looper}, D.~L., {Gelino}, C.~R., {Burgasser}, A.~J., \& {Kirkpatrick}, J.~D.
  2008{\natexlab{a}}, \apj, 685, 1183, \dodoi{10.1086/590382}

\bibitem[{{Looper} {et~al.}(2007){Looper}, {Kirkpatrick}, \&
  {Burgasser}}]{2007AJ....134.1162L}
{Looper}, D.~L., {Kirkpatrick}, J.~D., \& {Burgasser}, A.~J. 2007, \aj, 134,
  1162, \dodoi{10.1086/520645}

\bibitem[{{Looper} {et~al.}(2008{\natexlab{b}}){Looper}, {Kirkpatrick},
  {Cutri}, {Barman}, {Burgasser}, {Cushing}, {Roellig}, {McGovern}, {McLean},
  {Rice}, {Swift}, \& {Schurr}}]{2008ApJ...686..528L}
{Looper}, D.~L., {Kirkpatrick}, J.~D., {Cutri}, R.~M., {et~al.}
  2008{\natexlab{b}}, \apj, 686, 528, \dodoi{10.1086/591025}

\bibitem[{{Loutrel} {et~al.}(2011){Loutrel}, {Luhman}, {Lowrance}, \&
  {Bochanski}}]{2011ApJ...739...81L}
{Loutrel}, N.~P., {Luhman}, K.~L., {Lowrance}, P.~J., \& {Bochanski}, J.~J.
  2011, \apj, 739, 81, \dodoi{10.1088/0004-637X/739/2/81}

\bibitem[{{Luhman}(2013)}]{2013ApJ...767L...1L}
{Luhman}, K.~L. 2013, \apjl, 767, L1, \dodoi{10.1088/2041-8205/767/1/L1}

\bibitem[{{Luhman} {et~al.}(2011){Luhman}, {Burgasser}, \&
  {Bochanski}}]{2011ApJ...730L...9L}
{Luhman}, K.~L., {Burgasser}, A.~J., \& {Bochanski}, J.~J. 2011, \apjl, 730,
  L9, \dodoi{10.1088/2041-8205/730/1/L9}

\bibitem[{{Luhman} {et~al.}(2012){Luhman}, {Burgasser}, {Labb{\'e}}, {Saumon},
  {Marley}, {Bochanski}, {Monson}, \& {Persson}}]{2012ApJ...744..135L}
{Luhman}, K.~L., {Burgasser}, A.~J., {Labb{\'e}}, I., {et~al.} 2012, \apj, 744,
  135, \dodoi{10.1088/0004-637X/744/2/135}

\bibitem[{{Luhman} {et~al.}(2007){Luhman}, {Patten}, {Marengo}, {Schuster},
  {Hora}, {Ellis}, {Stauffer}, {Sonnett}, {Winston}, {Gutermuth}, {Megeath},
  {Backman}, {Henry}, {Werner}, \& {Fazio}}]{2007ApJ...654..570L}
{Luhman}, K.~L., {Patten}, B.~M., {Marengo}, M., {et~al.} 2007, \apj, 654, 570,
  \dodoi{10.1086/509073}

\bibitem[{{Mace} {et~al.}(2013){Mace}, {Kirkpatrick}, {Cushing}, {Gelino},
  {McLean}, {Logsdon}, {Wright}, {Skrutskie}, {Beichman}, {Eisenhardt}, \&
  {Kulas}}]{2013ApJ...777...36M}
{Mace}, G.~N., {Kirkpatrick}, J.~D., {Cushing}, M.~C., {et~al.} 2013, \apj,
  777, 36, \dodoi{10.1088/0004-637X/777/1/36}

\bibitem[{{Mace} {et~al.}(2018){Mace}, {Mann}, {Skiff}, {Sneden},
  {Kirkpatrick}, {Schneider}, {Kidder}, {Gosnell}, {Kim}, \&
  {Mulligan}}]{2018ApJ...854..145M}
{Mace}, G.~N., {Mann}, A.~W., {Skiff}, B.~A., {et~al.} 2018, \apj, 854, 145,
  \dodoi{10.3847/1538-4357/aaa8dd}

\bibitem[{{Macintosh} {et~al.}(2015){Macintosh}, {Graham}, {Barman}, {De Rosa},
  {Konopacky}, {Marley}, {Marois}, {Nielsen}, {Pueyo}, {Rajan}, {Rameau},
  {Saumon}, {Wang}, {Patience}, {Ammons}, {Arriaga}, {Artigau}, {Beckwith},
  {Brewster}, {Bruzzone}, {Bulger}, {Burningham}, {Burrows}, {Chen}, {Chiang},
  {Chilcote}, {Dawson}, {Dong}, {Doyon}, {Draper}, {Duch{\^e}ne}, {Esposito},
  {Fabrycky}, {Fitzgerald}, {Follette}, {Fortney}, {Gerard}, {Goodsell},
  {Greenbaum}, {Hibon}, {Hinkley}, {Cotten}, {Hung}, {Ingraham},
  {Johnson-Groh}, {Kalas}, {Lafreniere}, {Larkin}, {Lee}, {Line}, {Long},
  {Maire}, {Marchis}, {Matthews}, {Max}, {Metchev}, {Millar-Blanchaer},
  {Mittal}, {Morley}, {Morzinski}, {Murray-Clay}, {Oppenheimer}, {Palmer},
  {Patel}, {Perrin}, {Poyneer}, {Rafikov}, {Rantakyr{\"o}}, {Rice}, {Rojo},
  {Rudy}, {Ruffio}, {Ruiz}, {Sadakuni}, {Saddlemyer}, {Salama}, {Savransky},
  {Schneider}, {Sivaramakrishnan}, {Song}, {Soummer}, {Thomas}, {Vasisht},
  {Wallace}, {Ward-Duong}, {Wiktorowicz}, {Wolff}, \&
  {Zuckerman}}]{2015Sci...350...64M}
{Macintosh}, B., {Graham}, J.~R., {Barman}, T., {et~al.} 2015, Science, 350,
  64, \dodoi{10.1126/science.aac5891}

\bibitem[{{Magnier} \& {Cuillandre}(2004)}]{2004PASP..116..449M}
{Magnier}, E.~A., \& {Cuillandre}, J.-C. 2004, \pasp, 116, 449,
  \dodoi{10.1086/420756}

\bibitem[{{Magnier} {et~al.}(2016){Magnier}, {Schlafly}, {Finkbeiner}, {Tonry},
  {Goldman}, {R{\"o}ser}, {Schilbach}, {Chambers}, {Flewelling}, {Huber},
  {Price}, {Sweeney}, {Waters}, {Denneau}, {Draper}, {Hodapp}, {Jedicke},
  {Kudritzki}, {Metcalfe}, {Stubbs}, \& {Wainscoast}}]{2016arXiv161205242M}
{Magnier}, E.~A., {Schlafly}, E.~F., {Finkbeiner}, D.~P., {et~al.} 2016, arXiv
  e-prints.
\newblock \doarXiv{1612.05242}

\bibitem[{{Marley} {et~al.}(2013){Marley}, {Ackerman}, {Cuzzi}, \&
  {Kitzmann}}]{2013cctp.book..367M}
{Marley}, M.~S., {Ackerman}, A.~S., {Cuzzi}, J.~N., \& {Kitzmann}, D. 2013,
  {Clouds and Hazes in Exoplanet Atmospheres}, ed. S.~J. {Mackwell}, A.~A.
  {Simon-Miller}, J.~W. {Harder}, \& M.~A. {Bullock}, 367

\bibitem[{{Marley} \& {Robinson}(2015)}]{2015ARA&A..53..279M}
{Marley}, M.~S., \& {Robinson}, T.~D. 2015, \araa, 53, 279,
  \dodoi{10.1146/annurev-astro-082214-122522}

\bibitem[{{Marley} {et~al.}(2012){Marley}, {Saumon}, {Cushing}, {Ackerman},
  {Fortney}, \& {Freedman}}]{2012ApJ...754..135M}
{Marley}, M.~S., {Saumon}, D., {Cushing}, M., {et~al.} 2012, \apj, 754, 135,
  \dodoi{10.1088/0004-637X/754/2/135}

\bibitem[{{Marley} {et~al.}(2017){Marley}, {Saumon}, {Fortney}, {Morley},
  {Lupu}, {Freedman}, \& {Visscher}}]{2017AAS...23031507M}
{Marley}, M.~S., {Saumon}, D., {Fortney}, J.~J., {et~al.} 2017, in American
  Astronomical Society Meeting Abstracts, Vol. 230, American Astronomical
  Society Meeting Abstracts \#230, 315.07

\bibitem[{{Marley} {et~al.}(2010){Marley}, {Saumon}, \&
  {Goldblatt}}]{2010ApJ...723L.117M}
{Marley}, M.~S., {Saumon}, D., \& {Goldblatt}, C. 2010, \apjl, 723, L117,
  \dodoi{10.1088/2041-8205/723/1/L117}

\bibitem[{{Marley} {et~al.}(2002){Marley}, {Seager}, {Saumon}, {Lodders},
  {Ackerman}, {Freedman}, \& {Fan}}]{2002ApJ...568..335M}
{Marley}, M.~S., {Seager}, S., {Saumon}, D., {et~al.} 2002, \apj, 568, 335,
  \dodoi{10.1086/338800}

\bibitem[{{Marley} \& {Sengupta}(2011)}]{2011MNRAS.417.2874M}
{Marley}, M.~S., \& {Sengupta}, S. 2011, \mnras, 417, 2874,
  \dodoi{10.1111/j.1365-2966.2011.19448.x}

\bibitem[{{Marocco} {et~al.}(2013){Marocco}, {Andrei}, {Smart}, {Jones},
  {Pinfield}, {Day-Jones}, {Clarke}, {Sozzetti}, {Lucas}, {Bucciarelli}, \&
  {Penna}}]{2013AJ....146..161M}
{Marocco}, F., {Andrei}, A.~H., {Smart}, R.~L., {et~al.} 2013, \aj, 146, 161,
  \dodoi{10.1088/0004-6256/146/6/161}

\bibitem[{{Marocco} {et~al.}(2015){Marocco}, {Jones}, {Day-Jones}, {Pinfield},
  {Lucas}, {Burningham}, {Zhang}, {Smart}, {Gomes}, \&
  {Smith}}]{2015MNRAS.449.3651M}
{Marocco}, F., {Jones}, H.~R.~A., {Day-Jones}, A.~C., {et~al.} 2015, \mnras,
  449, 3651, \dodoi{10.1093/mnras/stv530}

\bibitem[{{Marois} {et~al.}(2008){Marois}, {Macintosh}, {Barman}, {Zuckerman},
  {Song}, {Patience}, {Lafreni{\`e}re}, \& {Doyon}}]{2008Sci...322.1348M}
{Marois}, C., {Macintosh}, B., {Barman}, T., {et~al.} 2008, Science, 322, 1348,
  \dodoi{10.1126/science.1166585}

\bibitem[{{Marois} {et~al.}(2010){Marois}, {Zuckerman}, {Konopacky},
  {Macintosh}, \& {Barman}}]{2010Natur.468.1080M}
{Marois}, C., {Zuckerman}, B., {Konopacky}, Q.~M., {Macintosh}, B., \&
  {Barman}, T. 2010, \nat, 468, 1080, \dodoi{10.1038/nature09684}

\bibitem[{{Mason}(1996)}]{1996AJ....112.2260M}
{Mason}, B.~D. 1996, \aj, 112, 2260, \dodoi{10.1086/118179}

\bibitem[{{Matthews} {et~al.}(2014){Matthews}, {Crepp}, {Skemer}, {Hinz},
  {Gianninas}, {Kilic}, {Skrutskie}, {Bailey}, {Defrere}, {Leisenring},
  {Esposito}, \& {Puglisi}}]{2014ApJ...783L..25M}
{Matthews}, C.~T., {Crepp}, J.~R., {Skemer}, A., {et~al.} 2014, \apjl, 783,
  L25, \dodoi{10.1088/2041-8205/783/2/L25}

\bibitem[{{McCaughrean} {et~al.}(2004){McCaughrean}, {Close}, {Scholz},
  {Lenzen}, {Biller}, {Brandner}, {Hartung}, \&
  {Lodieu}}]{2004AandA...413.1029M}
{McCaughrean}, M.~J., {Close}, L.~M., {Scholz}, R.~D., {et~al.} 2004, \aap,
  413, 1029, \dodoi{10.1051/0004-6361:20034292}

\bibitem[{{McMahon} {et~al.}(2013){McMahon}, {Banerji}, {Gonzalez}, {Koposov},
  {Bejar}, {Lodieu}, {Rebolo}, \& {VHS Collaboration}}]{2013Msngr.154...35M}
{McMahon}, R.~G., {Banerji}, M., {Gonzalez}, E., {et~al.} 2013, The Messenger,
  154, 35

\bibitem[{{Melis} {et~al.}(2014){Melis}, {Reid}, {Mioduszewski}, {Stauffer}, \&
  {Bower}}]{2014Sci...345.1029M}
{Melis}, C., {Reid}, M.~J., {Mioduszewski}, A.~J., {Stauffer}, J.~R., \&
  {Bower}, G.~C. 2014, Science, 345, 1029, \dodoi{10.1126/science.1256101}

\bibitem[{{Metchev} \& {Hillenbrand}(2006)}]{2006ApJ...651.1166M}
{Metchev}, S.~A., \& {Hillenbrand}, L.~A. 2006, \apj, 651, 1166,
  \dodoi{10.1086/507836}

\bibitem[{{Miles-P{\'a}ez} {et~al.}(2017){Miles-P{\'a}ez}, {Metchev}, {Luhman},
  {Marengo}, \& {Hulsebus}}]{2017AJ....154..262M}
{Miles-P{\'a}ez}, P.~A., {Metchev}, S., {Luhman}, K.~L., {Marengo}, M., \&
  {Hulsebus}, A. 2017, \aj, 154, 262, \dodoi{10.3847/1538-3881/aa9711}

\bibitem[{{Minniti} {et~al.}(2010){Minniti}, {Lucas}, {Emerson}, {Saito},
  {Hempel}, {Pietrukowicz}, {Ahumada}, {Alonso}, {Alonso-Garcia}, {Arias},
  {Bandyopadhyay}, {Barb{\'a}}, {Barbuy}, {Bedin}, {Bica}, {Borissova},
  {Bronfman}, {Carraro}, {Catelan}, {Clari{\'a}}, {Cross}, {de Grijs},
  {D{\'e}k{\'a}ny}, {Drew}, {Fari{\~n}a}, {Feinstein}, {Fern{\'a}ndez
  Laj{\'u}s}, {Gamen}, {Geisler}, {Gieren}, {Goldman}, {Gonzalez}, {Gunthardt},
  {Gurovich}, {Hambly}, {Irwin}, {Ivanov}, {Jord{\'a}n}, {Kerins}, {Kinemuchi},
  {Kurtev}, {L{\'o}pez-Corredoira}, {Maccarone}, {Masetti}, {Merlo},
  {Messineo}, {Mirabel}, {Monaco}, {Morelli}, {Padilla}, {Palma}, {Parisi},
  {Pignata}, {Rejkuba}, {Roman-Lopes}, {Sale}, {Schreiber}, {Schr{\"o}der},
  {Smith}, {}, {Soto}, {Tamura}, {Tappert}, {Thompson}, {Toledo}, {Zoccali}, \&
  {Pietrzynski}}]{2010NewA...15..433M}
{Minniti}, D., {Lucas}, P.~W., {Emerson}, J.~P., {et~al.} 2010, \na, 15, 433,
  \dodoi{10.1016/j.newast.2009.12.002}

\bibitem[{{Morley} {et~al.}(2012){Morley}, {Fortney}, {Marley}, {Visscher},
  {Saumon}, \& {Leggett}}]{2012ApJ...756..172M}
{Morley}, C.~V., {Fortney}, J.~J., {Marley}, M.~S., {et~al.} 2012, \apj, 756,
  172, \dodoi{10.1088/0004-637X/756/2/172}

\bibitem[{{Murray} {et~al.}(2011){Murray}, {Burningham}, {Jones}, {Pinfield},
  {Lucas}, {Leggett}, {Tinney}, {Day-Jones}, {Weights}, {Lodieu}, {P{\'e}rez
  Prieto}, {Nickson}, {Zhang}, {Clarke}, {Jenkins}, \&
  {Tamura}}]{2011MNRAS.414..575M}
{Murray}, D.~N., {Burningham}, B., {Jones}, H.~R.~A., {et~al.} 2011, \mnras,
  414, 575, \dodoi{10.1111/j.1365-2966.2011.18424.x}

\bibitem[{{Naud} {et~al.}(2014){Naud}, {Artigau}, {Malo}, {Albert}, {Doyon},
  {Lafreni{\`e}re}, {Gagn{\'e}}, {Saumon}, {Morley}, {Allard}, {Homeier},
  {Beichman}, {Gelino}, \& {Boucher}}]{2014ApJ...787....5N}
{Naud}, M.-E., {Artigau}, {\'E}., {Malo}, L., {et~al.} 2014, \apj, 787, 5,
  \dodoi{10.1088/0004-637X/787/1/5}

\bibitem[{{Nielsen} {et~al.}(2019){Nielsen}, {De Rosa}, {Macintosh}, {Wang},
  {Ruffio}, {Chiang}, {Marley}, {Saumon}, {Savransky}, {Ammons}, {Bailey},
  {Barman}, {Blain}, {Bulger}, {Chilcote}, {Cotten}, {Czekala}, {Doyon},
  {Duchene}, {Esposito}, {Fabrycky}, {Fitzgerald}, {Follette}, {Fortney},
  {Gerard}, {Goodsell}, {Graham}, {Greenbaum}, {Hibon}, {Hinkley}, {Hirsch},
  {Hom}, {Hung}, {Dawson}, {Ingraham}, {Kalas}, {Konopacky}, {Larkin}, {Lee},
  {Lin}, {Maire}, {Marchis}, {Marois}, {Metchev}, {Millar-Blanchaer},
  {Morzinski}, {Oppenheimer}, {Palmer}, {Patience}, {Perrin}, {Poyneer},
  {Pueyo}, {Rafikov}, {Rajan}, {Rameau}, {Rantakyr o}, {Ren}, {Schneider},
  {Sivaramakrishnan}, {Song}, {Soummer}, {Tallis}, {Thomas}, {Ward-Duong}, \&
  {Wolff}}]{2019arXiv190405358N}
{Nielsen}, E.~L., {De Rosa}, R.~J., {Macintosh}, B., {et~al.} 2019, arXiv
  e-prints.
\newblock \doarXiv{1904.05358}

\bibitem[{Oliphant(2006)}]{numpy}
Oliphant, T. 2006, {NumPy}: A guide to {NumPy}, USA: Trelgol Publishing.
\newblock \url{http://www.numpy.org/}

\bibitem[{P\'erez \& Granger(2007)}]{PER-GRA:2007}
P\'erez, F., \& Granger, B.~E. 2007, Computing in Science and Engineering, 9,
  21, \dodoi{10.1109/MCSE.2007.53}

\bibitem[{{Pinfield} {et~al.}(2006){Pinfield}, {Jones}, {Lucas}, {Kendall},
  {Folkes}, {Day-Jones}, {Chappelle}, \& {Steele}}]{2006MNRAS.368.1281P}
{Pinfield}, D.~J., {Jones}, H.~R.~A., {Lucas}, P.~W., {et~al.} 2006, \mnras,
  368, 1281, \dodoi{10.1111/j.1365-2966.2006.10213.x}

\bibitem[{{Press} {et~al.}(1992){Press}, {Teukolsky}, {Vetterling}, \&
  {Flannery}}]{1992nrfa.book.....P}
{Press}, W.~H., {Teukolsky}, S.~A., {Vetterling}, W.~T., \& {Flannery}, B.~P.
  1992, {Numerical recipes in FORTRAN. The art of scientific computing}

\bibitem[{{Rajan} {et~al.}(2017){Rajan}, {Rameau}, {De Rosa}, {Marley},
  {Graham}, {Macintosh}, {Marois}, {Morley}, {Patience}, {Pueyo}, {Saumon},
  {Ward-Duong}, {Ammons}, {Arriaga}, {Bailey}, {Barman}, {Bulger}, {Burrows},
  {Chilcote}, {Cotten}, {Czekala}, {Doyon}, {Duch{\^e}ne}, {Esposito},
  {Fitzgerald}, {Follette}, {Fortney}, {Goodsell}, {Greenbaum}, {Hibon},
  {Hung}, {Ingraham}, {Johnson-Groh}, {Kalas}, {Konopacky}, {Lafreni{\`e}re},
  {Larkin}, {Maire}, {Marchis}, {Metchev}, {Millar-Blanchaer}, {Morzinski},
  {Nielsen}, {Oppenheimer}, {Palmer}, {Patel}, {Perrin}, {Poyneer},
  {Rantakyr{\"o}}, {Ruffio}, {Savransky}, {Schneider}, {Sivaramakrishnan},
  {Song}, {Soummer}, {Thomas}, {Vasisht}, {Wallace}, {Wang}, {Wiktorowicz}, \&
  {Wolff}}]{2017AJ....154...10R}
{Rajan}, A., {Rameau}, J., {De Rosa}, R.~J., {et~al.} 2017, \aj, 154, 10,
  \dodoi{10.3847/1538-3881/aa74db}

\bibitem[{{Rayner} {et~al.}(2003){Rayner}, {Toomey}, {Onaka}, {Denault},
  {Stahlberger}, {Vacca}, {Cushing}, \& {Wang}}]{2003PASP..115..362R}
{Rayner}, J.~T., {Toomey}, D.~W., {Onaka}, P.~M., {et~al.} 2003, \pasp, 115,
  362, \dodoi{10.1086/367745}

\bibitem[{{Reid} {et~al.}(2008){Reid}, {Cruz}, {Kirkpatrick}, {Allen},
  {Mungall}, {Liebert}, {Lowrance}, \& {Sweet}}]{2008AJ....136.1290R}
{Reid}, I.~N., {Cruz}, K.~L., {Kirkpatrick}, J.~D., {et~al.} 2008, \aj, 136,
  1290, \dodoi{10.1088/0004-6256/136/3/1290}

\bibitem[{{Reid} {et~al.}(2006{\natexlab{a}}){Reid}, {Lewitus}, {Allen},
  {Cruz}, \& {Burgasser}}]{2006AJ....132..891R}
{Reid}, I.~N., {Lewitus}, E., {Allen}, P.~R., {Cruz}, K.~L., \& {Burgasser},
  A.~J. 2006{\natexlab{a}}, \aj, 132, 891, \dodoi{10.1086/505626}

\bibitem[{{Reid} {et~al.}(2006{\natexlab{b}}){Reid}, {Lewitus}, {Burgasser}, \&
  {Cruz}}]{2006ApJ...639.1114R}
{Reid}, I.~N., {Lewitus}, E., {Burgasser}, A.~J., \& {Cruz}, K.~L.
  2006{\natexlab{b}}, \apj, 639, 1114, \dodoi{10.1086/499484}

\bibitem[{{Reid} \& {Walkowicz}(2006)}]{2006PASP..118..671R}
{Reid}, I.~N., \& {Walkowicz}, L.~M. 2006, \pasp, 118, 671,
  \dodoi{10.1086/503446}

\bibitem[{{Reyl{\'e}} {et~al.}(2010){Reyl{\'e}}, {Delorme}, {Willott},
  {Albert}, {Delfosse}, {Forveille}, {Artigau}, {Malo}, {Hill}, \&
  {Doyon}}]{2010AandA...522A.112R}
{Reyl{\'e}}, C., {Delorme}, P., {Willott}, C.~J., {et~al.} 2010, \aap, 522,
  A112, \dodoi{10.1051/0004-6361/200913234}

\bibitem[{{Reyl{\'e}} {et~al.}(2014){Reyl{\'e}}, {Delorme}, {Artigau},
  {Delfosse}, {Albert}, {Forveille}, {Rajpurohit}, {Allard}, {Homeier}, \&
  {Robin}}]{2014AandA...561A..66R}
{Reyl{\'e}}, C., {Delorme}, P., {Artigau}, E., {et~al.} 2014, \aap, 561, A66,
  \dodoi{10.1051/0004-6361/201322107}

\bibitem[{{Riaz} {et~al.}(2006){Riaz}, {Gizis}, \&
  {Harvin}}]{2006AJ....132..866R}
{Riaz}, B., {Gizis}, J.~E., \& {Harvin}, J. 2006, \aj, 132, 866,
  \dodoi{10.1086/505632}

\bibitem[{{Richichi} {et~al.}(2000){Richichi}, {Ragland}, {Calamai}, {Richter},
  \& {Stecklum}}]{2000AandA...361..594R}
{Richichi}, A., {Ragland}, S., {Calamai}, G., {Richter}, S., \& {Stecklum}, B.
  2000, \aap, 361, 594

\bibitem[{{Sahlmann} \& {Lazorenko}(2015)}]{2015MNRAS.453L.103S}
{Sahlmann}, J., \& {Lazorenko}, P.~F. 2015, \mnras, 453, L103,
  \dodoi{10.1093/mnrasl/slv113}

\bibitem[{{Salaris} {et~al.}(2009){Salaris}, {Serenelli}, {Weiss}, \& {Miller
  Bertolami}}]{2009ApJ...692.1013S}
{Salaris}, M., {Serenelli}, A., {Weiss}, A., \& {Miller Bertolami}, M. 2009,
  \apj, 692, 1013, \dodoi{10.1088/0004-637X/692/2/1013}

\bibitem[{{Samland} {et~al.}(2017){Samland}, {Molli{\`e}re}, {Bonnefoy},
  {Maire}, {Cantalloube}, {Cheetham}, {Mesa}, {Gratton}, {Biller}, \&
  {Wahhaj}}]{2017A&A...603A..57S}
{Samland}, M., {Molli{\`e}re}, P., {Bonnefoy}, M., {et~al.} 2017, \aap, 603,
  A57, \dodoi{10.1051/0004-6361/201629767}

\bibitem[{{Saumon} \& {Marley}(2008)}]{2008ApJ...689.1327S}
{Saumon}, D., \& {Marley}, M.~S. 2008, \apj, 689, 1327, \dodoi{10.1086/592734}

\bibitem[{{Saumon} {et~al.}(2006){Saumon}, {Marley}, {Cushing}, {Leggett},
  {Roellig}, {Lodders}, \& {Freedman}}]{2006ApJ...647..552S}
{Saumon}, D., {Marley}, M.~S., {Cushing}, M.~C., {et~al.} 2006, \apj, 647, 552,
  \dodoi{10.1086/505419}

\bibitem[{{Scholz}(2010)}]{2010AandA...515A..92S}
{Scholz}, R.-D. 2010, \aap, 515, A92, \dodoi{10.1051/0004-6361/201014264}

\bibitem[{{Scholz}(2014)}]{2014AandA...561A.113S}
{Scholz}, R.~D. 2014, \aap, 561, A113, \dodoi{10.1051/0004-6361/201323015}

\bibitem[{{Scholz} {et~al.}(2003){Scholz}, {McCaughrean}, {Lodieu}, \&
  {Kuhlbrodt}}]{2003AandA...398L..29S}
{Scholz}, R.~D., {McCaughrean}, M.~J., {Lodieu}, N., \& {Kuhlbrodt}, B. 2003,
  \aap, 398, L29, \dodoi{10.1051/0004-6361:20021847}

\bibitem[{{S{\'e}gransan} {et~al.}(2011){S{\'e}gransan}, {Mayor}, {Udry},
  {Lovis}, {Benz}, {Bouchy}, {Lo Curto}, {Mordasini}, {Moutou}, {Naef}, {Pepe},
  {Queloz}, \& {Santos}}]{2011AandA...535A..54S}
{S{\'e}gransan}, D., {Mayor}, M., {Udry}, S., {et~al.} 2011, \aap, 535, A54,
  \dodoi{10.1051/0004-6361/200913580}

\bibitem[{{Shkolnik} {et~al.}(2009){Shkolnik}, {Liu}, \&
  {Reid}}]{2009ApJ...699..649S}
{Shkolnik}, E., {Liu}, M.~C., \& {Reid}, I.~N. 2009, \apj, 699, 649,
  \dodoi{10.1088/0004-637X/699/1/649}

\bibitem[{{Siegler} {et~al.}(2007){Siegler}, {Close}, {Burgasser}, {Cruz},
  {Marois}, {Macintosh}, \& {Barman}}]{2007AJ....133.2320S}
{Siegler}, N., {Close}, L.~M., {Burgasser}, A.~J., {et~al.} 2007, \aj, 133,
  2320, \dodoi{10.1086/513273}

\bibitem[{{Skrutskie} {et~al.}(2006){Skrutskie}, {Cutri}, {Stiening},
  {Weinberg}, {Schneider}, {Carpenter}, {Beichman}, {Capps}, {Chester},
  {Elias}, {Huchra}, {Liebert}, {Lonsdale}, {Monet}, {Price}, {Seitzer},
  {Jarrett}, {Kirkpatrick}, {Gizis}, {Howard}, {Evans}, {Fowler}, {Fullmer},
  {Hurt}, {Light}, {Kopan}, {Marsh}, {McCallon}, {Tam}, {Van Dyk}, \&
  {Wheelock}}]{2006AJ....131.1163S}
{Skrutskie}, M.~F., {Cutri}, R.~M., {Stiening}, R., {et~al.} 2006, \aj, 131,
  1163, \dodoi{10.1086/498708}

\bibitem[{{Smith} {et~al.}(2018){Smith}, {Lucas}, {Kurtev}, {Smart}, {Minniti},
  {Borissova}, {Jones}, {Zhang}, {Marocco}, {Contreras Pe{\~n}a}, {Gromadzki},
  {Kuhn}, {Drew}, {Pinfield}, \& {Bedin}}]{2018MNRAS.474.1826S}
{Smith}, L.~C., {Lucas}, P.~W., {Kurtev}, R., {et~al.} 2018, \mnras, 474, 1826,
  \dodoi{10.1093/mnras/stx2789}

\bibitem[{{Steele} {et~al.}(2009){Steele}, {Burleigh}, {Farihi},
  {G{\"a}nsicke}, {Jameson}, {Dobbie}, \& {Barstow}}]{2009AandA...500.1207S}
{Steele}, P.~R., {Burleigh}, M.~R., {Farihi}, J., {et~al.} 2009, \aap, 500,
  1207, \dodoi{10.1051/0004-6361/200911694}

\bibitem[{{Stephens} {et~al.}(2009){Stephens}, {Leggett}, {Cushing}, {Marley},
  {Saumon}, {Geballe}, {Golimowski}, {Fan}, \& {Noll}}]{2009ApJ...702..154S}
{Stephens}, D.~C., {Leggett}, S.~K., {Cushing}, M.~C., {et~al.} 2009, \apj,
  702, 154, \dodoi{10.1088/0004-637X/702/1/154}

\bibitem[{{Stone} {et~al.}(2016){Stone}, {Skemer}, {Kratter}, {Dupuy}, {Close},
  {Eisner}, {Fortney}, {Hinz}, {Males}, {Morley}, {Morzinski}, \&
  {Ward-Duong}}]{2016ApJ...818L..12S}
{Stone}, J.~M., {Skemer}, A.~J., {Kratter}, K.~M., {et~al.} 2016, \apjl, 818,
  L12, \dodoi{10.3847/2041-8205/818/1/L12}

\bibitem[{{Struve} \& {Franklin}(1955)}]{1955ApJ...121..337S}
{Struve}, O., \& {Franklin}, K.~L. 1955, \apj, 121, 337, \dodoi{10.1086/145993}

\bibitem[{{Sutherland} {et~al.}(2015){Sutherland}, {Emerson}, {Dalton},
  {Atad-Ettedgui}, {Beard}, {Bennett}, {Bezawada}, {Born}, {Caldwell}, {Clark},
  {Craig}, {Henry}, {Jeffers}, {Little}, {McPherson}, {Murray}, {Stewart},
  {Stobie}, {Terrett}, {Ward}, {Whalley}, \& {Woodhouse}}]{2015A&A...575A..25S}
{Sutherland}, W., {Emerson}, J., {Dalton}, G., {et~al.} 2015, \aap, 575, A25,
  \dodoi{10.1051/0004-6361/201424973}

\bibitem[{{Taylor}(2005)}]{2005ASPC..347...29T}
{Taylor}, M.~B. 2005, Astronomical Society of the Pacific Conference Series,
  Vol. 347, {TOPCAT \&amp; STIL: Starlink Table/VOTable Processing Software},
  ed. P.~{Shopbell}, M.~{Britton}, \& R.~{Ebert}, 29

\bibitem[{{Tinney} {et~al.}(2003){Tinney}, {Burgasser}, \&
  {Kirkpatrick}}]{2003AJ....126..975T}
{Tinney}, C.~G., {Burgasser}, A.~J., \& {Kirkpatrick}, J.~D. 2003, \aj, 126,
  975, \dodoi{10.1086/376481}

\bibitem[{{Tonry} {et~al.}(2012){Tonry}, {Stubbs}, {Lykke}, {Doherty},
  {Shivvers}, {Burgett}, {Chambers}, {Hodapp}, {Kaiser}, \&
  {Kudritzki}}]{2012ApJ...750...99T}
{Tonry}, J.~L., {Stubbs}, C.~W., {Lykke}, K.~R., {et~al.} 2012, \apj, 750, 99,
  \dodoi{10.1088/0004-637X/750/2/99}

\bibitem[{{Torres} {et~al.}(2006){Torres}, {Quast}, {da Silva}, {de La Reza},
  {Melo}, \& {Sterzik}}]{2006AandA...460..695T}
{Torres}, C.~A.~O., {Quast}, G.~R., {da Silva}, L., {et~al.} 2006, \aap, 460,
  695, \dodoi{10.1051/0004-6361:20065602}

\bibitem[{{Tremblay} {et~al.}(2011){Tremblay}, {Bergeron}, \&
  {Gianninas}}]{2011ApJ...730..128T}
{Tremblay}, P.~E., {Bergeron}, P., \& {Gianninas}, A. 2011, \apj, 730, 128,
  \dodoi{10.1088/0004-637X/730/2/128}

\bibitem[{{Tremblin} {et~al.}(2016){Tremblin}, {Amundsen}, {Chabrier},
  {Baraffe}, {Drummond}, {Hinkley}, {Mourier}, \&
  {Venot}}]{2016ApJ...817L..19T}
{Tremblin}, P., {Amundsen}, D.~S., {Chabrier}, G., {et~al.} 2016, \apjl, 817,
  L19, \dodoi{10.3847/2041-8205/817/2/L19}

\bibitem[{{Tremblin} {et~al.}(2015){Tremblin}, {Amundsen}, {Mourier},
  {Baraffe}, {Chabrier}, {Drummond}, {Homeier}, \&
  {Venot}}]{2015ApJ...804L..17T}
{Tremblin}, P., {Amundsen}, D.~S., {Mourier}, P., {et~al.} 2015, \apjl, 804,
  L17, \dodoi{10.1088/2041-8205/804/1/L17}

\bibitem[{{Tremblin} {et~al.}(2017){Tremblin}, {Chabrier}, {Baraffe}, {Liu},
  {Magnier}, {Lagage}, {Alves de Oliveira}, {Burgasser}, {Amundsen}, \&
  {Drummond}}]{2017ApJ...850...46T}
{Tremblin}, P., {Chabrier}, G., {Baraffe}, I., {et~al.} 2017, \apj, 850, 46,
  \dodoi{10.3847/1538-4357/aa9214}

\bibitem[{{Tsuji} {et~al.}(1999){Tsuji}, {Ohnaka}, \&
  {Aoki}}]{1999ApJ...520L.119T}
{Tsuji}, T., {Ohnaka}, K., \& {Aoki}, W. 1999, \apj, 520, L119,
  \dodoi{10.1086/312161}

\bibitem[{{Valenti} \& {Fischer}(2005)}]{2005ApJS..159..141V}
{Valenti}, J.~A., \& {Fischer}, D.~A. 2005, \apjs, 159, 141,
  \dodoi{10.1086/430500}

\bibitem[{{van Leeuwen}(2007)}]{2007AandA...474..653V}
{van Leeuwen}, F. 2007, \aap, 474, 653, \dodoi{10.1051/0004-6361:20078357}

\bibitem[{{Vos} {et~al.}(2019){Vos}, {Biller}, {Bonavita}, {Eriksson}, {Liu},
  {Best}, {Metchev}, {Radigan}, {Allers}, {Janson}, {Buenzli}, {Dupuy},
  {Bonnefoy}, {Manjavacas}, {Brand ner}, {Crossfield}, {Deacon}, {Henning},
  {Homeier}, {Kopytova}, \& {Schlieder}}]{2019MNRAS.483..480V}
{Vos}, J.~M., {Biller}, B.~A., {Bonavita}, M., {et~al.} 2019, \mnras, 483, 480,
  \dodoi{10.1093/mnras/sty3123}

\bibitem[{{Vrba} {et~al.}(2004){Vrba}, {Henden}, {Luginbuhl}, {Guetter},
  {Munn}, {Canzian}, {Burgasser}, {Kirkpatrick}, {Fan}, {Geballe},
  {Golimowski}, {Knapp}, {Leggett}, {Schneider}, \&
  {Brinkmann}}]{2004AJ....127.2948V}
{Vrba}, F.~J., {Henden}, A.~A., {Luginbuhl}, C.~B., {et~al.} 2004, \aj, 127,
  2948, \dodoi{10.1086/383554}

\bibitem[{{Williams} {et~al.}(2009){Williams}, {Bolte}, \&
  {Koester}}]{2009ApJ...693..355W}
{Williams}, K.~A., {Bolte}, M., \& {Koester}, D. 2009, The Astrophysical
  Journal, 693, 355, \dodoi{10.1088/0004-637X/693/1/355}

\bibitem[{{Wilson} {et~al.}(2001){Wilson}, {Kirkpatrick}, {Gizis}, {Skrutskie},
  {Monet}, \& {Houck}}]{2001AJ....122.1989W}
{Wilson}, J.~C., {Kirkpatrick}, J.~D., {Gizis}, J.~E., {et~al.} 2001, \aj, 122,
  1989, \dodoi{10.1086/323134}

\bibitem[{{Wright} {et~al.}(2010){Wright}, {Eisenhardt}, {Mainzer}, {Ressler},
  {Cutri}, {Jarrett}, {Kirkpatrick}, {Padgett}, {McMillan}, {Skrutskie},
  {Stanford}, {Cohen}, {Walker}, {Mather}, {Leisawitz}, {Gautier}, {McLean},
  {Benford}, {Lonsdale}, {Blain}, {Mendez}, {Irace}, {Duval}, {Liu}, {Royer},
  {Heinrichsen}, {Howard}, {Shannon}, {Kendall}, {Walsh}, {Larsen}, {Cardon},
  {Schick}, {Schwalm}, {Abid}, {Fabinsky}, {Naes}, \&
  {Tsai}}]{2010AJ....140.1868W}
{Wright}, E.~L., {Eisenhardt}, P.~R.~M., {Mainzer}, A.~K., {et~al.} 2010, \aj,
  140, 1868, \dodoi{10.1088/0004-6256/140/6/1868}

\bibitem[{{Zahnle} \& {Marley}(2014)}]{2014ApJ...797...41Z}
{Zahnle}, K.~J., \& {Marley}, M.~S. 2014, \apj, 797, 41,
  \dodoi{10.1088/0004-637X/797/1/41}

\bibitem[{{Zapatero Osorio} {et~al.}(2018){Zapatero Osorio}, {B{\'e}jar},
  {Lodieu}, \& {Manjavacas}}]{2018MNRAS.475..139Z}
{Zapatero Osorio}, M.~R., {B{\'e}jar}, V.~J.~S., {Lodieu}, N., \& {Manjavacas},
  E. 2018, \mnras, 475, 139, \dodoi{10.1093/mnras/stx3154}

\bibitem[{{Zapatero Osorio} {et~al.}(2014){Zapatero Osorio}, {G{\'a}lvez
  Ortiz}, {Bihain}, {Bailer-Jones}, {Rebolo}, {Henning}, {Boudreault},
  {B{\'e}jar}, {Goldman}, {Mundt}, \& {Caballero}}]{2014AandA...568A..77Z}
{Zapatero Osorio}, M.~R., {G{\'a}lvez Ortiz}, M.~C., {Bihain}, G., {et~al.}
  2014, \aap, 568, A77, \dodoi{10.1051/0004-6361/201423848}

\bibitem[{{Zuckerman} {et~al.}(2006){Zuckerman}, {Bessell}, {Song}, \&
  {Kim}}]{2006ApJ...649L.115Z}
{Zuckerman}, B., {Bessell}, M.~S., {Song}, I., \& {Kim}, S. 2006, \apjl, 649,
  L115, \dodoi{10.1086/508060}

\bibitem[{{Zuckerman} {et~al.}(2011){Zuckerman}, {Rhee}, {Song}, \&
  {Bessell}}]{2011ApJ...732...61Z}
{Zuckerman}, B., {Rhee}, J.~H., {Song}, I., \& {Bessell}, M.~S. 2011, \apj,
  732, 61, \dodoi{10.1088/0004-637X/732/2/61}

\bibitem[{{Zuckerman} {et~al.}(2001){Zuckerman}, {Song}, {Bessell}, \&
  {Webb}}]{2001ApJ...562L..87Z}
{Zuckerman}, B., {Song}, I., {Bessell}, M.~S., \& {Webb}, R.~A. 2001, \apjl,
  562, L87, \dodoi{10.1086/337968}

\bibitem[{{Zurlo} {et~al.}(2013){Zurlo}, {Vigan}, {Hagelberg}, {Desidera},
  {Chauvin}, {Almenara}, {Biazzo}, {Bonnefoy}, {Carson}, {Covino}, {Delorme},
  {D'Orazi}, {Gratton}, {Mesa}, {Messina}, {Moutou}, {Segransan}, {Turatto},
  {Udry}, \& {Wildi}}]{2013A&A...554A..21Z}
{Zurlo}, A., {Vigan}, A., {Hagelberg}, J., {et~al.} 2013, \aap, 554, A21,
  \dodoi{10.1051/0004-6361/201321179}

\end{thebibliography}

\clearpage

\clearpage

\clearpage

\clearpage

\tabletypesize{\scriptsize}

\clearpage
\begin{deluxetable*}{lccccccccc}
\tablewidth{0pc}
\tablecaption{Properties of COCONUTS-1 \label{tab:info}}
\tablehead{
\multicolumn{1}{l}{Properties}    &&&&\colhead{COCONUTS-1A}  &&&&    &\colhead{COCONUTS-1B}  }
\startdata
Spectral Type &&&& DA &&&& & T4$\pm0.5$    \\
Age (Gyr) &&&& $7.3^{+2.8}_{-1.6}$ &&&& & --  \\
\hline
\multicolumn{10}{c}{{\it Gaia}~DR2 Astrometry} \\
\hline
Designation &&&& Gaia DR2 244214799689691904 &&&& & --    \\
Coordinates ($\alpha$, $\delta$; hms, dms) &&&& 03:55:56.53, +45:25:06.2 &&&& & --    \\
visibility\_periods\_used   &&&& $14$ &&&& & --       \\
astrometric excess noise (mas)   &&&& $0.12\ (0.35\sigma)$ &&&& & --       \\
Renormalised Unit Weight Error   &&&& $1.05$ &&&& & --       \\
\enddata
\tablenotetext{a}{The spectrophotometric distance of the companion.}
\tablenotetext{b}{The PS1 tangential velocities are computed using the white dwarf's {\it Gaia}~DR2 distance and uncertainties.}
\tablenotetext{c}{The $H_{\rm MKO}$ and $K_{\rm MKO}$ photometry of COCONUTS-1AB are synthesized using $J_{\rm MKO}$ from their near-infrared spectra. We do not use these photometry to (1) conduct the SED analysis of COCONUTS-1A (Section~\ref{sec:WD_analysis}), and (2) flux-calibrate the near-infrared spectrum of COCONUTS-1B for the atmospheric model analysis (Section~\ref{subsec:atm}). }
\tablenotetext{d}{The $K_{\rm 2MASS}$ magnitude of COCONUTS-1A and the W3 and W4 magnitudes of COCONUTS-1B are $95\%$ upper limits.}
\end{deluxetable*}
\clearpage

\clearpage
\begin{deluxetable*}{lcccccc}
\tablewidth{0pc}
\tablecaption{Model-Derived Physical Properties of COCONUTS-1B \label{tab:atmevo}}
\tablehead{
\multicolumn{1}{l}{Properties}    &\colhead{Atmospheric Models\tablenotemark{a}}  &&&&    &\colhead{Evolutionary Models}  }
\startdata
Bolometric Luminosity ($\log\ (L_{\rm bol}/L_{\odot})$) & -- &&&& & $-4.832 \pm 0.007$    \\
Effective Temperature ($T_{\rm eff}$; K) & $1304^{+3}_{-6}\ (\pm 50)$ &&&& & $1255^{+6}_{-8}$    \\
Logarithmic Surface Gravity ($\log\ g$; cgs) & $4.86^{+0.03}_{-0.04}\ (\pm 0.13)$ &&&& & $5.44^{+0.02}_{-0.03}$    \\
Metallicity ($Z$) & $0.002^{+0.004}_{-0.002}\ (\pm 0.250)$ &&&& & $0$     \\
Radial Velocity ($v_{r}$; km/s) & $498^{+40}_{-40}\ (\pm 200)$ &&&& & --    \\
Projected Rotational Velocity ($v\ \sin i$; km/s) & $76^{+50}_{-52}$ &&&& & --   \\
Logarithmic Solid Angle ($\log\ \Omega$) & $-20.542^{+0.010}_{-0.006}\ (\pm 0.067)$ &&&& & $-20.474^{+0.013}_{-0.006}$    \\
Radius ($R$; $R_{\rm Jup}$) & $0.730^{+0.009}_{-0.006}\ (\pm 0.056)$ &&&& & $0.789^{+0.011}_{-0.005}$    \\
Mass ($M$; $M_{\rm Jup}$) & $15.4^{+0.9}_{-0.8}\ (\pm 5.2)$ &&&& & $69.3^{+1.6}_{-3.4}$    \\
\hline
\enddata
\tablenotetext{a}{Values inside parentheses are the final adopted uncertainties as discussed in Section~\ref{subsubsec:atm_res}.}
\end{deluxetable*}
\clearpage

\global\pdfpageattr\expandafter{\the\pdfpageattr/Rotate 90} 
\begin{longrotatetable} 
\begin{deluxetable}{lcccccccccr} 
\tablewidth{0pc} 
\setlength{\tabcolsep}{0.05in} 
\tablecaption{L/T Transition Benchmarks: Spectral Type, Photometry, Parallax, and Age \label{tab:benchmarks_sptphotparage}} 
\tablehead{ \multicolumn{1}{l}{Object}  &  \multicolumn{1}{c}{Type\tablenotemark{a}}  &  \multicolumn{1}{c}{SpT\tablenotemark{b}}  &  \multicolumn{1}{c}{$J_{\rm MKO}$}  &  \multicolumn{1}{c}{$H_{\rm MKO}$}  &  \multicolumn{1}{c}{$K_{\rm MKO}$}  &  \multicolumn{1}{c}{Parallax}  &  \multicolumn{1}{c}{Age}  &  \multicolumn{1}{c}{Primary SpT}  &  \multicolumn{1}{c}{Separation\tablenotemark{c}}  &  \multicolumn{1}{r}{Reference\tablenotemark{d}}   \\ 
\multicolumn{1}{l}{}  &  \multicolumn{1}{c}{}  &  \multicolumn{1}{c}{}  &  \multicolumn{1}{c}{(mag)}  &  \multicolumn{1}{c}{(mag)}  &  \multicolumn{1}{c}{(mag)}  &  \multicolumn{1}{c}{(mas)}  &  \multicolumn{1}{c}{(Gyr)}  &  \multicolumn{1}{c}{}  &  \multicolumn{1}{c}{($''$)}  &  \multicolumn{1}{r}{Disc.;\ SpT;\ Phot.;\ $\pi$;\ Age;\ P.SpT}   } 
\startdata 
\multicolumn{11}{c}{Young Benchmarks ($\leqslant 0.3$~Gyr)} \\ 
\cline{1-11} 
HIP~65426~b  &  A\tablenotemark{e}+C  &  L$6\pm1$  &  $19.39 \pm 0.40$\tablenotemark{f}  &  --  &  --  &  $9.16 \pm 0.06$  &  $0.014\pm0.004$  &  A2~V  &  $0.83$  &  109;\ \ 109;\ \ 1,109;\ \ 105,118;\ \ 109;\ \ 4   \\ 
LP~$261-75$B  &  C  &  L$6\pm0.5$  &  $17.10 \pm 0.21$  &  $15.94 \pm 0.14$  &  $15.08 \pm 0.13$  &  $29.45 \pm 0.14$  &  $0.1-0.2$  &  M4.5~V  &  $13.6$  &  11,37;\ \ 11;\ \ 75;\ \ 105,118;\ \ 37;\ \ 37   \\ 
2MASS~J$2244316+204343$  &  A\tablenotemark{e}  &  L$6.5\pm0.5$  &  $16.33 \pm 0.03$  &  $15.06 \pm 0.03$  &  $13.90 \pm 0.03$  &  $58.70 \pm 1.00$  &  $0.149^{+0.051}_{-0.019}$  &  --  &  --  &  17,104;\ \ 49;\ \ 106;\ \ 106;\ \ 92;\ \ ...   \\ 
Calar~21  &  A  &  L$6.5\pm0.5$  &  $20.23 \pm 0.08$  &  $18.74 \pm 0.03$  &  $17.58 \pm 0.07$  &  $7.34 \pm 0.06$  &  $0.112 \pm 0.005$  &  --  &  --  &  90;\ \ 121;\ \ 45,90;\ \ 86;\ \ 97;\ \ ...   \\ 
Calar~22  &  A  &  L$6.5\pm0.5$  &  $20.29 \pm 0.07$  &  $18.92 \pm 0.02$  &  $17.80 \pm 0.09$  &  $7.34 \pm 0.06$  &  $0.112 \pm 0.005$  &  --  &  --  &  90;\ \ 121;\ \ 45,90;\ \ 86;\ \ 97;\ \ ...   \\ 
PSO~J$318.5338-22.8603$  &  A\tablenotemark{e}  &  L$7\pm1$  &  $17.15 \pm 0.04$  &  $15.68 \pm 0.02$  &  $14.41 \pm 0.02$  &  $45.14 \pm 1.73$  &  $0.024 \pm 0.003$  &  --  &  --  &  82;\ \ 82;\ \ 82;\ \ 106;\ \ 92;\ \ ...   \\ 
WISE~J$004701.06+680352.1$  &  A\tablenotemark{e}  &  L$7\pm1$ pec  &  $15.49 \pm 0.07$  &  $14.04 \pm 0.04$  &  $13.01 \pm 0.03$  &  $82.30 \pm 1.80$  &  $0.149^{+0.051}_{-0.019}$  &  --  &  --  &  76,98;\ \ 98;\ \ 106;\ \ 106;\ \ 92;\ \ ...   \\
\enddata
\end{deluxetable} 
\end{longrotatetable}

\clearpage 
\global\pdfpageattr\expandafter{\the\pdfpageattr/Rotate 90} 
\begin{longrotatetable} 
\begin{deluxetable}{lccccccccccccr} 
\tablewidth{0pc} 
\setlength{\tabcolsep}{0.05in} 
\tablecaption{L/T Transition Benchmarks: Bolometric Luminosity, Effective Temperature, Surface Gravity, and Mass \label{tab:benchmarks_lbolteff}} 
\tablehead{ \multicolumn{1}{l}{}  &  \multicolumn{1}{c}{}  &  \multicolumn{1}{c}{}  &  \multicolumn{5}{c}{Literature}  &  \multicolumn{1}{c}{}  &  \multicolumn{5}{c}{Adopted}   \\ 
\cline{4-8} \cline{10-14} 
\multicolumn{1}{l}{Object}  &  \multicolumn{1}{c}{SpT}  &  \multicolumn{1}{c}{}  &  \multicolumn{1}{c}{$\log\ (L_{\rm bol}/L_{\odot})$}  &  \multicolumn{1}{c}{$L_{\rm bol}$ Method\tablenotemark{a}}  &  \multicolumn{1}{c}{$T_{\rm eff}$}  &  \multicolumn{1}{c}{$T_{\rm eff}$ Method\tablenotemark{b}}  &  \multicolumn{1}{c}{Reference}  &  \multicolumn{1}{c}{}  &  \multicolumn{1}{c}{$\log\ (L_{\rm bol}/L_{\odot})$}  &  \multicolumn{1}{c}{$T_{\rm eff}$}  &  \multicolumn{1}{c}{$\log\ g$}  &  \multicolumn{1}{c}{$M$}  &  \multicolumn{1}{r}{Reference\tablenotemark{c}}   \\ 
\multicolumn{1}{l}{}  &  \multicolumn{1}{c}{}  &  \multicolumn{1}{c}{}  &  \multicolumn{1}{c}{(dex)}  &  \multicolumn{1}{c}{}  &  \multicolumn{1}{c}{(K)}  &  \multicolumn{1}{c}{}  &  \multicolumn{1}{c}{}  &  \multicolumn{1}{c}{}  &  \multicolumn{1}{c}{(dex)}  &  \multicolumn{1}{c}{(K)}  &  \multicolumn{1}{c}{(dex)}  &  \multicolumn{1}{c}{($M_{\rm Jup}$)}  &  \multicolumn{1}{r}{$L_{\rm bol}$; Phys.}   } 
\startdata 
\multicolumn{14}{c}{Young Benchmarks ($\leqslant 0.3$~Gyr)} \\ 
\cline{1-14} 
HIP~65426~b  &  L$6\pm1$  &  &  $-4.06 \pm 0.10$  &  BC$-$SpT (34)  &  $1500^{+100}_{-200}$  &  Evo$-$Age (4,7)  &  39  &  &  $-4.136^{+0.177}_{-0.177}$  &  $1359^{+126}_{-105}$  &  $3.99^{+0.11}_{-0.13}$  &  $8.7^{+2.7}_{-2.5}$  &  1; 1   \\ 
LP~$261-75$B  &  L$6\pm0.5$  &  &  $-4.43 \pm 0.09$  &  SED  &  --  &  --  &  28  &  &  $-4.430 \pm 0.090$  &  $1264^{+75}_{-49}$  &  $4.44^{+0.13}_{-0.09}$  &  $16.3^{+5.4}_{-2.2}$  &  28; 1   \\ 
2MASS~J$2244316+204343$  &  L$6.5\pm0.5$  &  &  $-4.503 \pm 0.007$  &  SED  &  $1184 \pm 10$\tablenotemark{d}  &  Evo$-$Age (11)  &  37  &  &  $-4.503 \pm 0.007$  &  $1228^{+24}_{-10}$  &  $4.42^{+0.13}_{-0.04}$  &  $15.5^{+3.9}_{-1.1}$  &  37; 1   \\ 
Calar~21  &  L$6.5\pm0.5$  &  &  $-4.29 \pm 0.10$  &  SED  &  $1350^{+100}_{-80}$  &  Evo$-$Age (30,33)  &  48  &  &  $-4.290 \pm 0.100$  &  $1352^{+97}_{-72}$  &  $4.44^{+0.16}_{-0.08}$  &  $17.9^{+6.7}_{-3.1}$  &  48; 1   \\ 
Calar~22  &  L$6.5\pm0.5$  &  &  $-4.33 \pm 0.10$  &  SED  &  $1350^{+100}_{-80}$  &  Evo$-$Age (30,33)  &  48  &  &  $-4.330 \pm 0.100$  &  $1322^{+69}_{-68}$  &  $4.40^{+0.11}_{-0.06}$  &  $16.9^{+3.8}_{-3.0}$  &  48; 1   \\ 
PSO~J$318.5338-22.8603$  &  L$7\pm1$  &  &  $-4.42 \pm 0.06$  &  SED  &  $1160^{+30}_{-40}$\tablenotemark{d}  &  Evo$-$Age (11)  &  29  &  &  $-4.420 \pm 0.060$  &  $1210^{+36}_{-37}$  &  $4.00^{+0.04}_{-0.04}$  &  $7.5^{+0.7}_{-0.6}$  &  29; 1   \\ 
WISE~J$004701.06+680352.1$  &  L$7\pm1$ pec  &  &  $-4.45 \pm 0.04$  &  SED  &  $1270-1300$\tablenotemark{d}  &  Radius (3,4)  &  35  &  &  $-4.450 \pm 0.040$  &  $1256^{+39}_{-19}$  &  $4.42^{+0.12}_{-0.06}$  &  $15.8^{+4.1}_{-1.4}$  &  35; 1   \\
\enddata
\end{deluxetable} 
\end{longrotatetable}

\global\pdfpageattr\expandafter{\the\pdfpageattr/Rotate 90} 
\clearpage
\begin{rotatetable}
\begin{deluxetable}{llllllllllll}
\tablewidth{0pc}
\tablecaption{Directly-Imaged L, T, Y Companions to White Dwarfs\label{tab:known_WD}}
\tablehead{
\multicolumn{5}{c}{Companion}                  &&&&\multicolumn{4}{c}{White Dwarf Primary}           \\
\cline{1-5} \cline{9-12}
\multicolumn{1}{l}{Companion }                    &\multicolumn{1}{l}{Other }                    &\multicolumn{1}{l}{Spectral }      & \multicolumn{1}{l}{Projected }  &\multicolumn{1}{l}{Discovery}       &&&&\multicolumn{1}{l}{Spectral} &\multicolumn{1}{l}{{\it Gaia}~DR2}      &\multicolumn{1}{l}{Estimated}                 &\multicolumn{1}{l}{Age}                             \\
\multicolumn{1}{l}{Name}                    &\multicolumn{1}{l}{Name}                    &\multicolumn{1}{l}{Type}      & \multicolumn{1}{l}{Separation}  &\multicolumn{1}{l}{Reference}       &&&&\multicolumn{1}{l}{Type} &\multicolumn{1}{l}{Distance}      &\multicolumn{1}{l}{Age}                 &\multicolumn{1}{l}{Reference}                             \\
\multicolumn{1}{l}{}    & \multicolumn{1}{l}{}                   &\multicolumn{1}{l}{}      &\multicolumn{1}{l}{($''$)} &\multicolumn{1}{l}{}       &&&&\multicolumn{1}{l}{} &\multicolumn{1}{l}{(pc)}      &\multicolumn{1}{l}{(Gyr)}                 &\multicolumn{1}{l}{}               }
\startdata
LSPM~J0241+2553B & 2MASS~J02415153+2553452 & L1 & 31.2 & (1) &&&& --\tablenotemark{a} & $72.7\pm1.0$  & $<10$\tablenotemark{b} & (1)  \\
GD~165B & 2MASS~J14243909+0917104 & L4 & 4 & (2) &&&& DA & $33.41\pm0.06$  & $1.2-5.5$ & (3) \\
PHL~5038B & SDSS~J222030.68-004107.3B & L8 & 0.94 & (4) &&&& DA & $72.9\pm1.2$  & $1.9-2.7$ & (4) \\
COCONUTS-1B & PSO~J058.9869+45.4296 & T4 & 40.61 & This Work  &&&& DA & $31.51\pm0.09$  & $7.3^{+2.8}_{-1.6}$ & This Work \\
LSPM~J1459+0857B & ULAS~J145935.25+085751.2 & T4.5 & 385 & (5) &&&& DA & $72.5\pm1.8$  & $>4.8$ & (5) \\
Wolf~1130C\tablenotemark{c} & WISE~J200520.38+542433.9  & T8 & 188.5 & (6) &&&& --\tablenotemark{a} & $16.559\pm0.009$  & $>10$\tablenotemark{d} & (7) \\
WD~0806-661B & L~97$-$3B & Y1\tablenotemark{e} & 130.2 & (8) &&&& DQ & $19.255\pm0.007$  & $2\pm0.5$ & (9) \\
\enddata
\tablecomments{The estimated age of the white dwarfs are derived from the cooling age and the evolutionary time of their progenitors unless otherwise noted. }
\tablenotetext{a}{\cite{2014ApJ...792..119D} and \cite{2018ApJ...854..145M} did not have spectra of the white dwarf LSPM~J0241+2553B and Wolf~1130C, respectively.}
\tablenotetext{b}{The maximum age of the white dwarf is estimated based on its disk-like kinematics.}
\tablenotetext{c}{This companion is in a multiple system with the primary being a binary composed of a M3 subdwarf and a white dwarf \citep{1998AJ....115.2053G}.}
\tablenotetext{d}{The white dwarf age is estimated using the metallicity and kinematics of the M3 subdwarf Wolf~1130A, which forms a binary with the white dwarf Wolf~1130B. The white dwarf is not detected in $0.2-2.5$~$\mu$m, providing an age constraint as $>3.4$~Gyr \citep{2018ApJ...854..145M}. }
\tablenotetext{e}{The companion's spectral type is assigned by \cite{2015ApJ...799...37L} based on photometry.}
\tablerefs{(1) \cite{2014ApJ...792..119D}, (2) \cite{1988Natur.336..656B}, (3) \cite{1999ApJ...519..834K}, (4) \cite{2009AandA...500.1207S}, (5) \cite{2011MNRAS.410..705D}, (6) \cite{2013ApJ...777...36M}, (7) \cite{2018ApJ...854..145M}, (8) \cite{2011ApJ...730L...9L}, (9) \cite{2012ApJ...744..135L}  }
\end{deluxetable}
\end{rotatetable}
\clearpage

\vfill
\eject
\end{document}